%% file: main.tex
\documentclass[a4paper,reprint,superscriptaddress,notitlepage,floatfix]{revtex4-2}

\input{preamble}

\graphicspath{{figures/}}

\begin{document}
    
\begin{abstract}
    Living systems display complex behaviors driven by physical forces as well as decision-making.
    Hydrodynamic theories hold promise for simplified universal descriptions of socially-generated collective behaviors. 
    However, the construction of such theories is often divorced from the data they should describe.
    Here, we develop and apply a data-driven pipeline that links micromotives to macrobehavior by augmenting hydrodynamics with individual preferences that guide motion.
    We illustrate this pipeline on a case study of residential dynamics in the United States, for which census and sociological data is available.
    Guided by Census data, sociological surveys, and neural network analysis, we systematically assess standard hydrodynamic assumptions to construct a sociohydrodynamic model.
    Solving our simple hydrodynamic model, calibrated using statistical inference, qualitatively captures key features of residential dynamics at the level of individual US counties.
    We highlight that a social memory, akin to hysteresis in magnets, emerges in the segregation-integration transition even with memory-less agents. This suggests an explanation for the phenomenon of neighborhood tipping, whereby a small change in a neighborhood’s population leads to a rapid demographic shift.
    Beyond residential segregation, our work paves the way for systematic investigations of decision-guided motility in real space, from micro-organisms to humans, as well as fitness-mediated motion in more abstract genomic spaces.
\end{abstract}

\title{Sociohydrodynamics: data-driven modelling of social behavior}

\author{Daniel S. Seara}
\affiliation{University of Chicago, James Franck Institute, 929 E 57th Street, Chicago, IL 60637}

\author{Jonathan Colen}
\affiliation{University of Chicago, James Franck Institute, 929 E 57th Street, Chicago, IL 60637}
\affiliation{University of Chicago, Department of Physics, 5720 S Ellis Ave, Chicago, IL 60637}
\affiliation{Old Dominion University, Joint Institute on Advanced Computing for Environmental Studies, 1070 University Blvd, Portsmouth, VA, 23703}

\author{Michel Fruchart}
\affiliation{University of Chicago, James Franck Institute, 929 E 57th Street, Chicago, IL 60637}
\affiliation{University of Chicago, Department of Physics, 5720 S Ellis Ave, Chicago, IL 60637}
\affiliation{ESPCI, Laboratoire Gulliver, 10 rue Vauquelin, 75231 Paris cedex 05}

\author{Yael Avni}
\affiliation{University of Chicago, James Franck Institute, 929 E 57th Street, Chicago, IL 60637}

\author{David Martin}
\affiliation{University of Chicago, Kadanoff Center for Theoretical Physics, 933 E 56th St, Chicago, IL 60637}
\affiliation{University of Chicago, Enrico Fermi Institute, 933 E 56th St, Chicago, IL 60637}

\author{Vincenzo Vitelli}
\affiliation{University of Chicago, James Franck Institute, 929 E 57th Street, Chicago, IL 60637}
\affiliation{University of Chicago, Department of Physics, 5720 S Ellis Ave, Chicago, IL 60637}
\affiliation{University of Chicago, Kadanoff Center for Theoretical Physics, 933 E 56th St, Chicago, IL 60637}

\date{\today}
\maketitle


Individual social organisms, from bacteria to ants to humans, display complex behaviors shaped by their interactions with each other and their environment.
Groups of such organisms often form large, coherent patterns across space and time~\cite{Reichenbach2007,Couzin2009,Ouellette2022,Vicsek2012,Bettencourt2013}.
This regularity suggests that aspects of social behavior may be captured using generalized \textit{hydrodynamic theories} that account for individual choices.
Hydrodynamic theories are mathematical descriptions of the time evolution of spatially extended systems that involve only a small number of slowly-varying fields~\cite{Kadanoff1963,Landau1987,Anderson1972,Saarloos2023}.
This approach, originating in fluid mechanics, has since been applied to living systems ranging from microbial suspensions~\cite{Sokolov2007,Wensink2012,Wioland2013,Li2018,Copenhagen2020,Curatolo2020} and cellular tissues~\cite{Mertz2012,Saw2017,PerezGonzalez2018,Streichan2018,Alert2019,Boocock2020,Yousafzai2022,ArmengolCollado2023} to insect swarms~\cite{Cavagna2023,Gorbonos2024} and human crowds~\cite{Hughes2003,Bain2019,Gu2025}.
In these systems, active mechanical~\cite{Marchetti2013,Vrugt2024} or \enquote{social} forces~\cite{Helbing1995,Ballerini2008,Corbetta2023} between individuals drive the dynamics of the hydrodynamic variables, such as density and polarization.
However, we lack a principled way to incorporate cognitive decision-making into hydrodynamic models.

Here, we develop a data-driven pipeline to capture the physical manifestations of non-mechanical choices within a hydrodynamic theory. 
We take inspiration from microeconomics to codify individual preferences (micromotives) into utility functions~\cite{Neumann2007,Osborne2006}, and then we incorporate them into a \enquote{sociohydrodynamic} theory that can account for collective behavior (macrobehavior). 
We illustrate our approach on the case study of human residential dynamics, focusing on segregation between non-Hispanic White and non-Hispanic Black residents in the United States, for which both sociological research~\cite{Williams2001,Pager2008,Reardon2014,Alexander2017,Charles2003,Hwang2022,DuBois2018,Massey1990,Jargowsky1998,Taeuber2008} and data~\cite{Manson2022} are available.
Theoretical explorations have examined the connection between micromotives and macrobehavior~\cite{Schelling1971,Schelling1980,Vinkovic2006,Grauwin2009,Grauwin2012,Clark2008,Fossett2006,Gauvin2009}, including recently-proposed hydrodynamic theories~\cite{Zakine2024,GarnierBrun2024a,Becharat2024}.
In addition, recent work suggests that statistical methods can forecast local trends in segregation observed in US census data~\cite{Chen2020,Kinkhabwala2021,Barron2022}.
Our analysis combines these two perspectives to forecast demographic distributions using a hydrodynamic theory constructed directly from data.

We demonstrate that, for a period of four decades, both local and global aspects of the dynamics of US population distributions can be described by an intelligible, analytical model constructed from data.
Our model shows that the segregation-integration transition is history-dependent, and suggests a possible mechanism for a phenomenon dubbed \enquote{neighborhood tipping}, whereby a small change in a neighborhood's population leads to a rapid demographic shift~\cite{Card2008,Zhang2011,Gualdi2015}.

\section{The sociohydrodynamic pipeline}
\begin{figure}
        \centering
        \includegraphics[width=\linewidth]{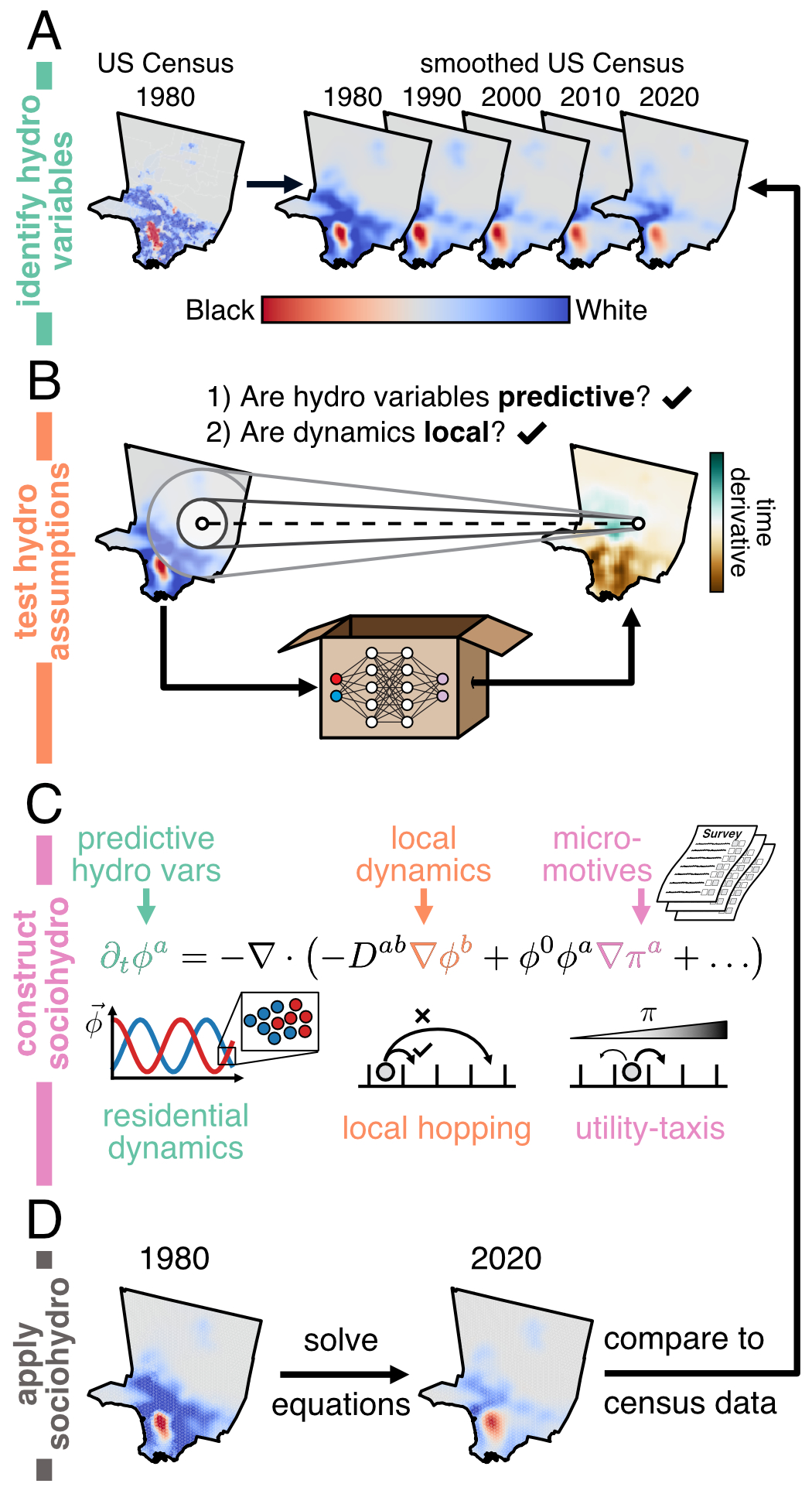}
        \caption{
        \textbf{How to construct a sociohydrodynamic model from data.}
        (A) 1980 Census data of Los Angeles County, California, colored by the difference in relative population density, $\phi^W - \phi^B$, where $\phi^a = \rho^a / \rho_\mathrm{max}$ (left). 
        Right shows the result of smoothing the census data on a regular square lattice for decennial Census data from 1980, 1990, 2000, 2010, and 2020.
        (B) We test the assumptions made in simple hydrodynamic theories using a model-free, data-driven approach. Namely, we test whether population distributions are predictive of their own future, and if they evolve locally. The prior is determined by the accuracy of the neural network predictions, and the latter is determined by the network's saliency (see Section~\ref{sec:nn_results} and Methods).
        (C) (left) Sociological surveys ask individuals of different groups (red and blue) to give opinions on communities with different compositions of the groups. These choices are discretized into binary decisions, denoting a positive ($\checkmark$) or negative ($\mathbf{\times}$) opinion of the proposed community. (right) We estimate a utility function as a function of the community structure based on the proportion of respondents of each type that give a positive assessment of that community.
        (D) Constructing our sociohydrodynamic model based on dynamics of $\vphi$ (green), with local movements (orange) that are biased according to spatial gradients of utility (purple).
        (E) We verify our sociohydrodynamic equations by solving them on individual US counties. The coefficients are learned from Census data for each county, and then simulated with the initial condition set as the 1980 US Census data. The results in 2020 are then compared to the 2020 US Census.}
        \label{fig:pipeline}
    \end{figure}

    Figure~\ref{fig:pipeline} summarizes our sociohydrodynamic pipeline.
    First, we identify candidate hydrodynamic variables that evolve slowly both in time and space (Fig.~\ref{fig:pipeline}A).
    To be useful, these variables must contain enough information to forecast their own future values -- they must be \emph{self-predictive}. 
    To test this self-predictability in a model-agnostic way, we train a neural network to forecast the evolution of the candidate hydrodynamic variables and examine whether the forecasting is possible.
    Examining further the trained neural network allows us to assess whether the dynamics are local, i.e. whether hydrodynamic variables are only instantaneously influenced by their immediate surroundings (Fig.~\ref{fig:pipeline}B).
    This locality, when it holds, considerably simplifies the models we have to consider next.
    We then construct an analytical, phenomenological model that relates micromotives to macrobehavior using a combination of physics and economic theory (Fig.~\ref{fig:pipeline}C).
    Finally, we apply this model to predict real data by first inferring the equation's coefficients and then checking their numerical solution against experimentally measured trajectories (Fig.~\ref{fig:pipeline}D).
    
    In the sections below, we detail how we apply this pipeline to the case of human residential dynamics.
    There, we make precise the variables we use, how we check predictability and locality, and the specific model we build to describe the data.

    \subsection{Identifying hydrodynamic variables in social behavior}
        Before building a hydrodynamic theory, we must first find suitable hydrodynamic variables.
        When the collective variables are not easy to guess, data-driven techniques can be used to propose candidates~\cite{Schmitt2023,Schmitt2023b,Lefebvre2023}.
        In our example of residential dynamics, a reasonable guess for collective variables are the populations of different groups. 
        Here, we focus on non-Hispanic White and non-Hispanic Black residents.
        Data from the decennial US Census~\cite{Manson2022} gives us access to population densities of each group $a$ at position $\mathbf{r}$ and time $t$, $\rho^a(\mathbf{r}, t)$.
        From these densities, we define a proxy for the local housing capacity $h(\mathbf{r})$ by finding the highest density of people who have lived at each location within a county over time.
        In this work, we take fill fractions as our hydrodynamic variables, defined as the population densities normalized by the maximum housing availability in the county -- $\phi^a(\mathbf{r}, t) = \rho^a(\mathbf{r}, t) / \mathrm{max}(h)$ (Methods).

        In order to apply the usual tools of hydrodynamics, we must check that the collective variables are slowly varying both in space and time, compared to the space and time scales we are able to resolve.
        Qualitatively, we observe that the main features of the population distributions appear unaffected by spatial smoothing over the length-scale of single counties.
        Quantitatively, fill fractions are correlated over distances four to seven times larger than the typical census tract $\ell$ in the region (Methods, Fig.~\ref{fig:slowVars}A,B).
        In the following analysis, we smooth the fill fractions using a Gaussian filter with a kernel width of 3 km (SI Fig.~1).
        
        We also require hydrodynamic variables to vary slowly in time. 
        We assess the temporal change in populations via measurements of segregation using the entropy index, which measures how local demographic distribution differs from the global composition~\cite{Duncan1955,Reardon2004,Roberto2015} (Methods).
        Qualitatively, maps of the entropy index appear similar over a span of 40 years.
        Quantitatively, its overall magnitude decreases slowly between the years of 1980-2020, as measured previously~\cite{Hwang2022} (Fig.~\ref{fig:slowVars}C,D).
        
        Together, these two results indicate that human residential dynamics may indeed be described by hydrodynamic variables that evolve slowly over the scale of individual counties and over a time-scale of decades.

    \subsection{Testing hydrodynamic assumptions with neural networks}\label{sec:nn_results}
        
        \begin{figure}[t!]
            \centering
            \includegraphics[width=\linewidth]{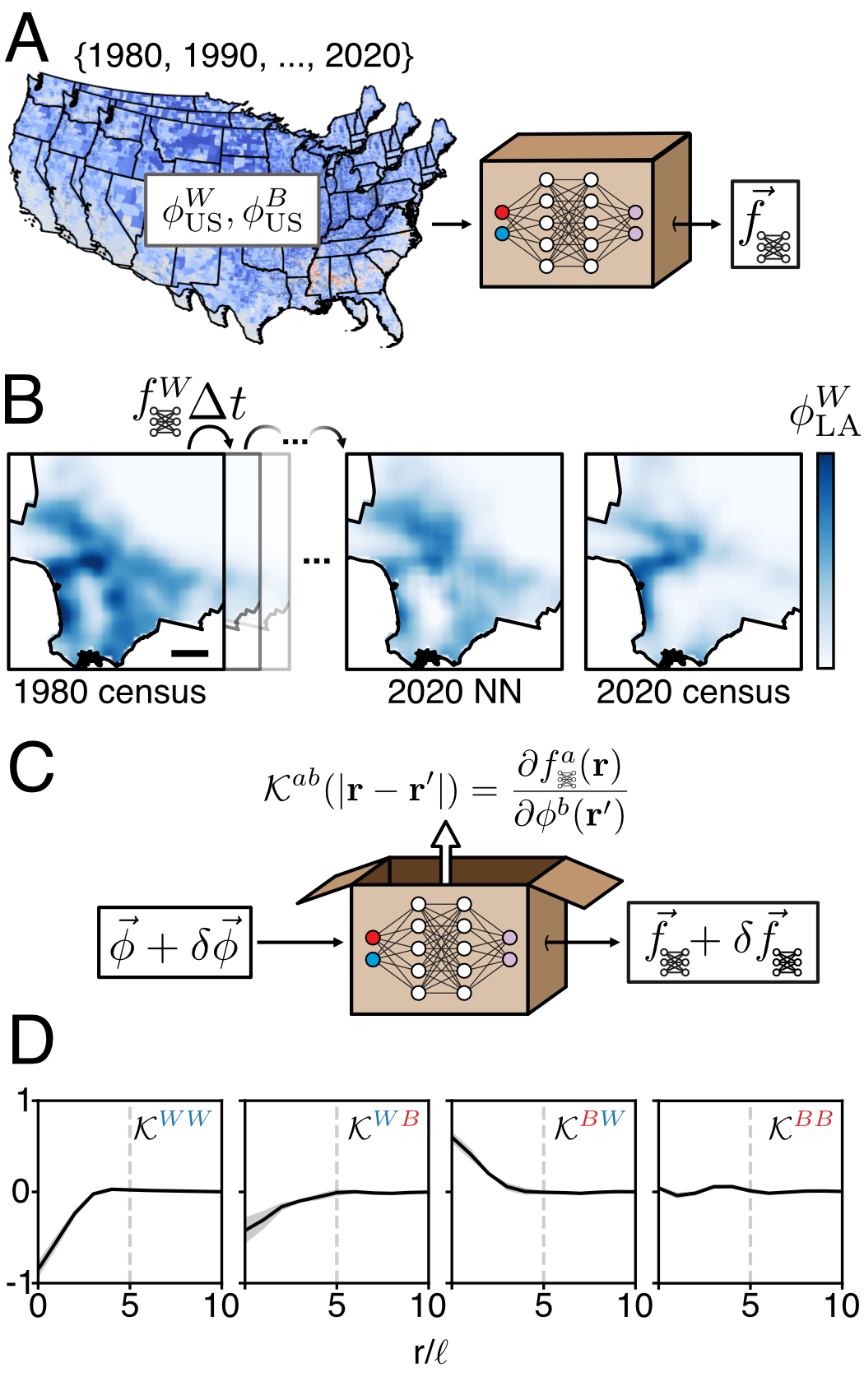}
            \caption{
                \textbf{Human residential dynamics are predictable and local.}
                (A) Illustration of neural network training procedure.
                A convolutional neural network takes processed US Census data, $\vphi$ as input, and produces a functional $\vec{f}_\NN$. This functional is optimized to provide an accurate prediction of 2020 census data when $\vphi(1980)$ is evolved by $\vec{f}_\NN$. 
                (B) Example of evolution by neural network for the White population in Los Angeles County.
                Starting with the 1980 census as an initial condition, populations within individual counties are evolved forward in time using $\vec{f}_{\NN}$ to 2020, and then compared with the 2020 census.
                Scale bar indicates a length of 5~$\ell = $~11.5~km.
                (C) Schematic illustration of saliency $\mathcal{K}^{ab}$ as the linear response of the neural network.
                (D) Saliency measured radially for four regions, Cook County IL, Fulton County GA, Harris County TX, and Los Angeles County CA. Mean $\pm$ standard deviation are shown by black line and gray shading, respectively.
            }
            \label{fig:nn_results}
        \end{figure}
        
        \subsubsection{Self-predictability}
        We next test how well the candidate hydrodynamic variables $\vec{\phi}$ identified in the previous section can predict their own future states (Fig.~\ref{fig:pipeline}B).
        Formally, we check whether $\partial_t \phi^a(\mathbf{r}, t)$ depends only on $\vphi(\mathbf{r'}, t)$ (for all $\mathbf{r'}$).
        Testing the self-predictability of collective variables requires a reliable rule for dynamics, which we do not have \textit{a priori}.
        To resolve this, we train a neural network on US Census data to map the fields $\phi^a$ to the time derivatives of the fields $\partial_t \phi^a$ (Methods). 
        Our dynamical system
        \begin{equation}
            \label{nndynamics}
            {\partial}_t \phi^a(\mathbf{r}, t) = f_\NN^a \left[ \vec{\phi}; \, h \right](\mathbf{r}),
        \end{equation}
        uses a machine-learned dynamical rule $\vec{f}_\NN$ for $\vphi$ that depends on both $\vphi$ and the local housing approximation $h$ introduced earlier (Fig.~\ref{fig:nn_results}A). 

        The predicted population dynamics, obtained by integrating \eqref{nndynamics}, capture qualitative large-scale features of the population distributions of US counties.
        Fig.~\ref{fig:nn_results}B shows an example of the predicted White population in Los Angeles County, CA.
        Similar results are seen for both White and Black populations in 3 additional counties dispersed throughout the US (SI Fig.~2).
        In the SI, we also quantify the prediction accuracy and show that the neural network out-performs several alternative dynamical rules, including no dynamics, linear growth, and exponential growth (SI Fig.~3).

        \subsubsection{Locality}
        Then, we evaluate whether the dynamics are generated locally (Fig.~\ref{fig:pipeline}B).
        Locality allows us to approximate $\partial_t \phi^a(\mathbf{r}, t)$ as a function of $\vphi$ and its gradients, $f^a[\vphi] = f^a(\vphi, \nabla \vphi, ...)$ evaluated at $\mathbf{r}$.
        To evaluate locality, we consider the saliency
        \begin{equation}
            \mathcal{K}^{ab} \left( r \right) = \bigg\langle \dfrac{ \partial f^a_\NN(\mathbf{r})}{\partial \phi^b(\mathbf{r'})} \bigg\rangle_{|\mathbf{r} - \mathbf{r'}| = r}
        \end{equation}
        of the machine-learned dynamics.   
        In essence, $\mathcal{K}^{ab}(r)$ measures how strongly the predicted dynamics of $\phi^a$ at a point $\mathbf{r}$ depends on the value of the field $\phi^b$ at another point a distance $r$ away (Fig.~\ref{fig:nn_results}C, see Methods)~\cite{Simonyan2014,Schmitt2023}. A $\mathcal{K}^{ab}(r)$ that decays rapidly to zero indicates that $f_{\NN}$ uses information within a very narrow region to generate its dynamics. 
        Note that correlation functions would not be sufficient to assess whether the dynamics is local as they do not directly address the evolution rules of the system.

        Indeed, the trained neural network identifies a local rule (Fig.~\ref{fig:nn_results}D).
        $\mathcal{K}^{ab}$ exhibits a similar structure for all predicted regions and is narrowly peaked at $r < 5\ell$, where $\ell$ is the typical size of a census tract within that county (Fig.~\ref{fig:nn_results}D).
        Furthermore, the cross-saliencies ($a \neq b$) may echo the preferences underlying residential decision-making. 
        Below, we codify these preferences in \enquote{utility functions}.
        We find that saliencies extracted from a neural network trained on agent-based simulations based on the Schelling model~\cite{Schelling1971} support this claim.
        Namely, the sign of the saliencies at $r=0$ directly reflect the slope of the utility functions input to the simulations (SI Fig.~4).

        Microscopically, nonlocality can both arise from the fact that agents are influenced by what happens away from them and from the fact that they can move far away.
        However, microscopic sources of nonlocality do not necessarily lead to a nonlocal effective description.
        %
        For example, a variant of the Ising model with local interactions but infinite-range displacements~\cite{Conti2002} is described by a hydrodynamic theory with local motility at a sufficiently coarse-grained level (SI Fig.~5).
        Indeed, human residential dynamics feature non-local displacements at the individual level: according to the US Census Bureau, 53\% of moves were within the same county in 2022, down from 64\% in 2012~\cite{censusMigration2023}.

    \subsection{Constructing a hydrodynamic theory}
    Using the insights and elements extracted from data in the initial steps of our pipeline, we now develop a hydrodynamic model designed to capture universal features of residential dynamics found across US cities. 
    
    \subsubsection{General theory}
    Having identified the collective variables $\phi^a(\mathbf{r}, t)$ and verified that they are predictive of their own future, we write a general equation of motion for these variables
    \begin{subequations}
    \label{eq:sociohydro}
        \begin{equation}
            \label{eq:flux+source}
            \partial_t \phi^a(\mathbf{r}, t) = - \nabla \cdot \mathbf{J}^a + S^a
        \end{equation}
        in which we have separated the dynamics into two parts: the divergence of a flux $\mathbf{J}^a$ that redistributes $\phi^a$ in space, and a source term $S^a$ that changes the total populations. The self-predictability and locality of the dynamics for $\vphi$ (Fig.~\ref{fig:nn_results}) motivates us to write $\mathbf{J}^a = \mathbf{J}^a(\vphi)$ and $S^a = S^a(\vphi)$.

        The source term $S^a$ in \eqref{eq:flux+source} describes how $\phi^a$ changes locally due to, for example, reproduction or immigration (SI Fig.~6).
        In addition to human populations, these processes play a crucial role in contexts such as microbiology and ecology~\cite{Fisher1937,Hallatschek2023}.
        Growth can be related to a local evolutionary \enquote{fitness function}, $f^a$, that describes to what extent the environment promotes the growth of a certain group~\cite{Crow1970,Kimura1983,Tsimring1996,Mustonen2009}.
        Although the fitness and utility functions share similarities, they have no reason to be identical~\cite{Cremer2019}.

        We write the flux $\mathbf{J}^a$ using a gradient expansion as
        \begin{equation}
        \label{eq:flux}
            \mathbf{J}^a(\vphi) = \phi^a \mathbf{v}^a - D^{ab} \nabla \phi^b + \Gamma^{ab} \nabla^3 \phi^b
        \end{equation}
        in which we have assumed isotropy, and where $D^{ab}[\vphi]$ accounts for diffusion of $\phi^a$ down the gradients of $\phi^b$ and $\Gamma^{ab}[\vphi]$ acts like a surface tension, penalizing spatial gradients in $\phi^a$~\cite{Cahn1958,Hohenberg1977}. Higher order terms in the gradients have been neglected.
        The first term in \eqref{eq:flux} describes advection of $\phi^a$ at a velocity $\mathbf{v}^a$.
        We assume that the velocity is proportional to the gradient of a \textit{utility function} $\pi^a$,
        \begin{equation}
        \label{eq:utility-taxis}
            \mathbf{v}^a \propto \nabla \pi^a.
        \end{equation}
        The proportionality factor will be determined below based on a microscopic, agent-based model.
        The utility function quantifies the preference of an $a$ individual for the location $\mathbf{r}$ at time $t$,
        providing a link between motility and socioeconomical incentives.
        The gradient reflects the propensity of individuals to move towards regions they prefer, i.e. up gradients in their utility.
        We call this behavior \enquote{utility-taxis}, in reference to other guided navigation strategies such as chemotaxis~\cite{Keller1971,Adler1975,Berg1975} or infotaxis~\cite{Vergassola2007}.
        \end{subequations}

        \subsubsection{Utility functions}
        The question remains -- what is $\pi^a$?
        Although many socioeconomic and personal factors may contribute~\cite{Charles2003,Hwang2022}, here we focus on the impact of neighborhood demographic preferences in driving residential dynamics.
        In other words, we seek a utility function written as
        \begin{equation}
            \pi^a(\mathbf{r}, t) = \pi^a\left(\vphi(\mathbf{r}, t) \right).
        \end{equation}
        This is the key feature of sociohydrodynamics: it establishes a feedback loop between the slowly-evolving hydrodynamic variables and the decision-making processes that lead to motility in the first place.
        
\begin{figure}
        \centering
        \includegraphics[width=0.45\textwidth]{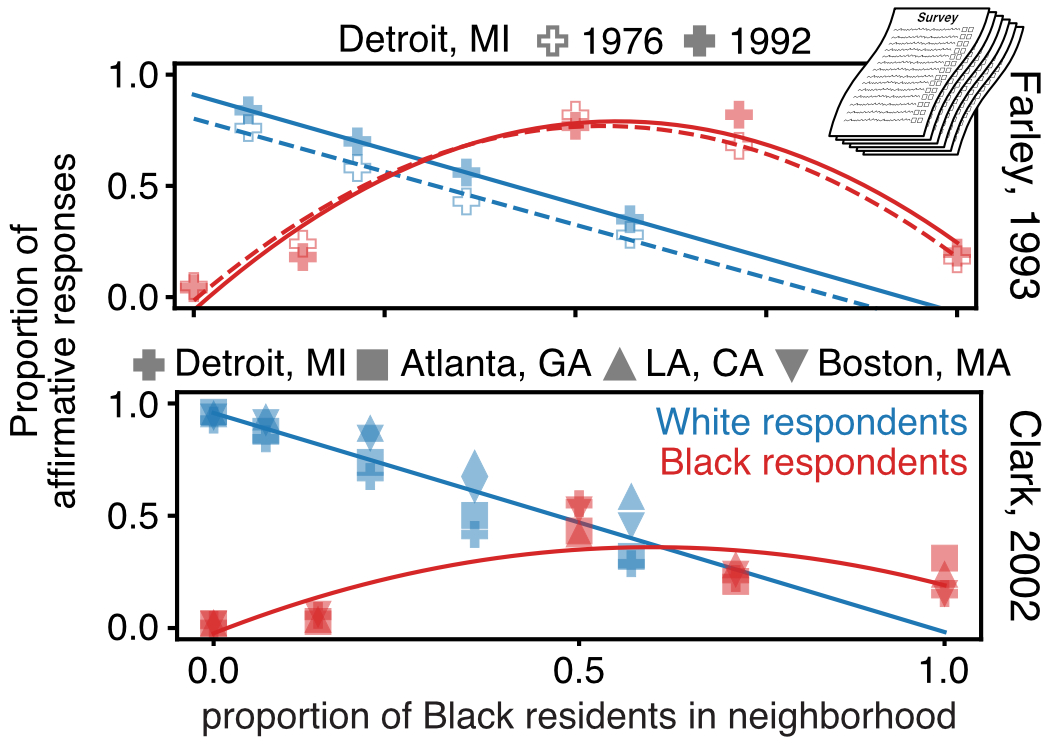}
        \caption{\textbf{Persistent residential preferences from sociological surveys.}
        (top) Reproduction of survey results shown in Figs. 8 \& 10 in Ref.~\cite{Farley1993}. Survey's taken around Detroit, MI ($+$). Top plot shows proportion of White respondents \enquote{indicating they would be feel comfortable in the neighborhood} with the proportion of Black residents shown on the x-axis. Bottom shows proportion of Black residents ranking a neighborhood with the given proportion of Black residents as either their first or second choice among five options. Hollow circles show results from surveys taken in 1976, and filled circles show results from surveys taken in 1992. Dashed (1976) and solid (1992) lines are linear (top) and quadratic (bottom) fits to each set of data.
        (bottom) Reproduction of Figs.4 \& 7 in Ref.~\cite{Clark2002}, which itself accumulates data from Ref.~\cite{Bobo1996}. White respondents were asked if they would move into a hypothetical neighborhood with the given proportion of Black residents. Black respondents, similarly to Ref.~\cite{Farley1993}, were asked to rank hypothetical neighborhoods and the results for the proportion of respondents ranking each neighborhood as their first choice is shown. Surveys were taken in Detroit, MI ($+$), Atlanta, GA ($\square$), Los Angeles, CA ($\triangle$), and Boston, MA ($\triangledown$). Solid lines are linear (top) and quadratic (bottom) fits to each set of data. Precise wording of survey questions are reproduced from the respective references in the SI.}
        \label{fig:survey_results}
    \end{figure}

        To model these utility functions in the case at hand, we turn to the social science literature measuring neighborhood demographic preferences (Fig.~\ref{fig:survey_results}).
        Social scientists have found that residential preferences remained consistent across time over a span of 16 years between 1976 and 1992 in the Detroit metropolitan area~\cite{Farley1993}, and across space over several major US metropolitan areas~\cite{Clark2002} (Fig.~\ref{fig:survey_results}).
        White residents show a monotonic decrease in their preference of neighborhoods with increasing proportion of Black residents. 
        On the other hand, Black residents show a marked preference for mixed neighborhoods, which remains consistent between the two surveys.
        Qualitatively similar results were obtained in other US-based surveys~\cite{Farley1978,Clark1991,Farley1993,Clark2002,Clark2008} (SI Fig.~7).
        Thus, we assume that each group $a$ has a distinct, time-invariant utility function written as a non-linear function of $\vphi$.
    
    \subsubsection{Coarse-graining the Schelling model}
    To further constrain our equations of motion, we construct and coarse-grain an agent-based model for residential dynamics, based on the Schelling model~
    \cite{Schelling1971, Schelling1980, Card2008, Vinkovic2006, Grauwin2009, Grauwin2012, Zhang2011, Clark2008, Fossett2006, Gauvin2009, Zakine2024}.
    The same approach has been independently developed in Refs.~\cite{Zakine2024,GarnierBrun2024a}, resulting in hydrodynamic equations similar to ours.
    In short, it models agents on a lattice that randomly move to adjacent sites with a bias towards increasing their utility (SI).
    Within a mean-field approximation, coarse-graining this agent-based model leads to \eqref{eq:flux+source} with $S^a = 0$ and 
    \begin{subequations}
    \begin{align}
        \mathbf{v}^{a}(\vphi) &= \vac \nabla \pi^a(\vphi) \\
        D^{ab}(\vphi) &= T^a(\phi^a + \delta^{ab} \vac) \\
        \Gamma^{ab}(\vphi) &= \Gamma^a \delta^{ab} \phi^b \vac.
    \end{align}
    \end{subequations}
    In the above, $\vac = 1 - \sum_b \phi^b$ is a vacancy fraction that arises because each lattice site has a maximum carrying capacity~\cite{Giacomin1996}, mimicking the availability of housing.
    The parameters $T^a$ and $\Gamma^a$ control the rate of hopping and penalize spatial gradients of $\vphi(x)$, respectively (see SI for discussion of the origin of the $\Gamma$ term, SI Fig.~8).

    \subsection{Applying the sociohydrodynamic model}

        \begin{figure*}[t!]
            \centering
            \includegraphics{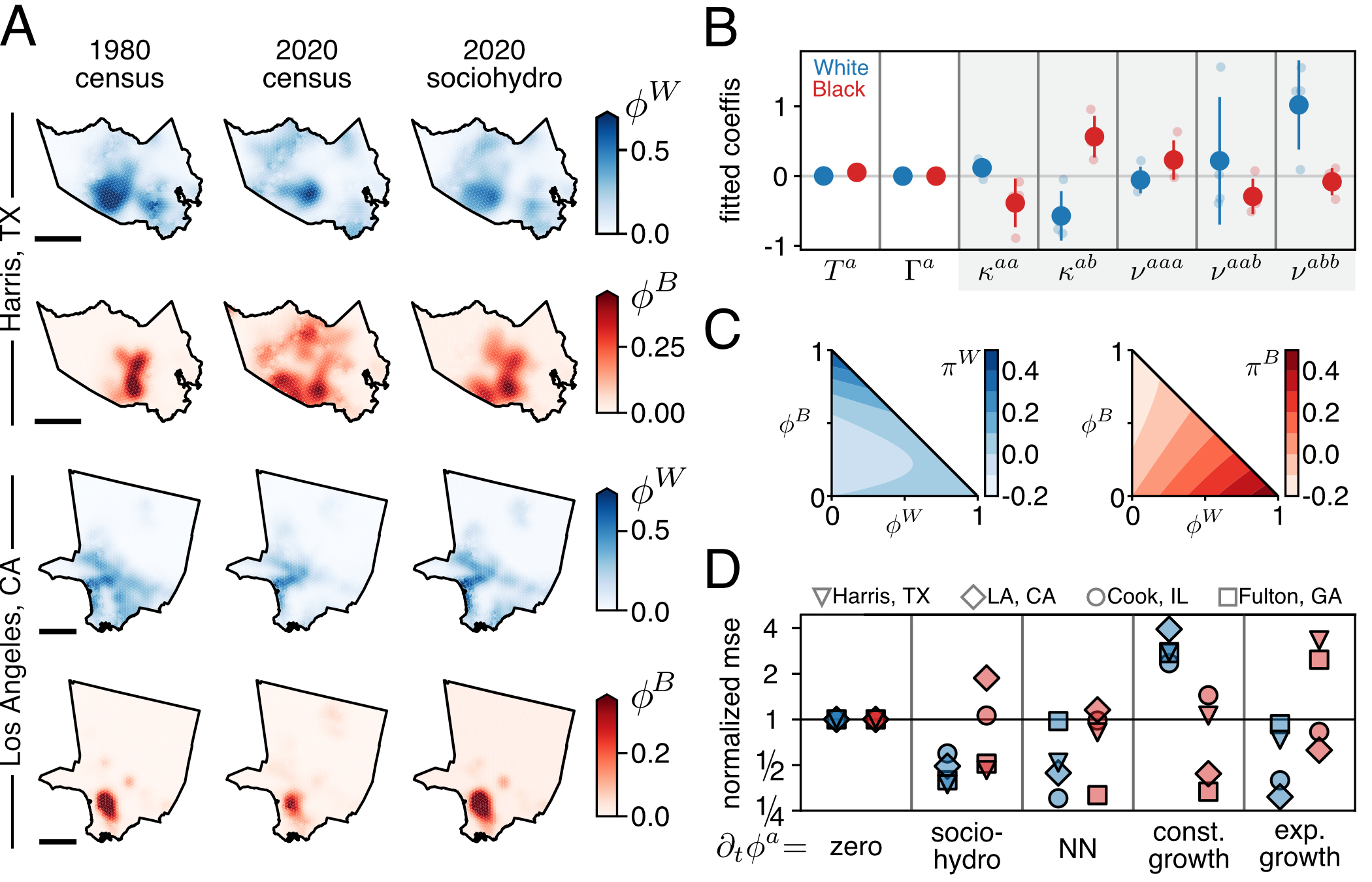}
            \caption{
            \textbf{Sociohydrodynamics predicts real population dynamics.}
            (A) (top) Sociohydrodynamic predictions for Harris County, TX. The left column shows fill fractions, $\phi^W$ and $\phi^B$ the 1980 Census data for White and Black populations (first and third rows, respectively), that serve as initial conditions for the simulations.
            The second column shows the 2020 Census data for White and Black populations, in addition to the total change from the 1980 census, (second and fourth row).
            The third column show the neural network's predictions for 2020 fill fractions starting from 1980.
            The fourth column shows the result from solving \eqref{eq:sociohydro} with our learned coefficients starting from 1980 and continuing to 2020.
            (bottom) Similar to left, but for Fulton County, GA.
            Scale bars indicate 25 km for each county.
            See SI for results on other counties.
            (B) Learned coefficients resulting from linear regression on US Census data. Coefficients are scaled by the standard deviation of their corresponding feature (e.g. $T_i$ is normalized by the standard deviation of $\nabla^2 \phi^i$ measured from the data). Dots show results from fitting on the four counties in Fig.~\ref{fig:sociosim}B, and circles and bars show mean $\pm$ standard deviation. Red and Blue symbols show results for Black and White populations, respectively.
            (C) Utility functions learned by the linear regression, highlighting the presence of non-reciprocity between the two groups.
            (D) Mean-square errors of various model predictions of human residential dynamics for four US counties; Harris TX ($\triangledown$), Los Angeles CA ($\diamond$), Cook IL($\circ$), and Fulton GA ($\square$).
            All values are normalized to the error measured assuming that there is no change in the population from 1980 to 2020.
            The second column shows the results from numerically solving \eqref{eq:sociohydro}.
            The third column shows the errors from the neural network, whose results are shown in Fig.~\ref{fig:nn_results}.
            The fourth column shows the results from assuming linear growth, defined as an extrapolation from 1990 - 2020 after a linear spline interpolation between 1980-1990.
            The final column shows the results from assuming only exponential growth.}
            \label{fig:sociosim}
        \end{figure*}

        To validate our model, we simulate \eqref{eq:sociohydro} and compare against the observed evolution of human populations in the US (Fig.~\ref{fig:sociosim}).
        To do so, we use linear regression to infer the coefficients of our model focusing on the same metropolitan areas as used to train the neural network in Sec.~\ref{sec:nn_results} (SI Fig.~9).
        
        In light of the responses to the social surveys in Fig.~\ref{fig:survey_results}, we parameterize $\pi^a$ as
        \begin{equation}
        \label{eq:quadUtility}
            \pi^a(\vphi) = \sum_{b, c} \kappa^{ab} \phi^b + \nu^{abc}\phi^b \phi^c.
        \end{equation}
        This is a Taylor expansion of a non-constant $\pi^a$ around $\phi^a = 0$.
        Therefore, we fit 5 coefficients for the utility function of each field $\left(\kappa^{aW}, \kappa^{aB}, \nu^{aWW},\nu^{aWB},\nu^{aBB}\right)$, as well as the diffusion constant $T^a$ and the surface tension $\Gamma^a$.
        
        The growth term $S^a$ in \eqref{eq:sociohydro} is fixed by fitting the total population change assuming an exponential growth model, $S^a =  r^a \phi^a$.
        This simple model captures the roughly exponential growth rates we empirically measure across US counties (SI Fig.~6), although it neglects growth that may occur from immigration outside of the region.
        With this growth rate in hand, we fit the mass-conserving portion of \eqref{eq:sociohydro}, $-\nabla\cdot {\bf J}^a = \partial_t \phi^a - S^a$.
        
        We then numerically solve \eqref{eq:sociohydro} using the learned coefficients.
        We take $\vphi(\mathbf{r}, 1980)$ as our initial condition and evolve in time until 2020.
        As shown in Fig.~\ref{fig:sociosim}A, our simulations capture several qualitative features of observed population dynamics.
        For example, we predict the increase of the Black in the southwest of Harris county, as well as a decrease in the White population in the southeaster portion of Los Angeles county (see SI Fig.~3 for results from other counties).
        
        The fitted coefficients are given in Figs.~\ref{fig:sociosim}B-C.
        Across multiple counties, we observed that the fitted diffusion constant of the Black population is larger than that of the White population, $T^B > T^W$ (Fig.~\ref{fig:sociosim}B), reflecting that Black families are more likely to move. Analysis by the US Census Bureau indeed found that $\sim 10\%$ of Black residents moved, compared to $\sim 8\%$ of White residents between the years of 2019 and 2020~\cite{censusMigrationByRace2020,Desmond2015}.
        The coefficients of the utility functions are also consistent across counties (Fig.~\ref{fig:sociosim}B).
        The linear coefficients $\kappa^{ab}$ have opposite signs for the two groups, signaling that White and Black residents have incompatible residential preferences.
        More formally, the utilities $\pi^W$ and $\pi^B$ are incompatible in the sense that
        \begin{equation}
            \frac{\partial \pi^W}{\partial \phi^B} \neq \frac{\partial \pi^B}{\partial \phi^W}.
            \label{incomptible_utilities}
        \end{equation}
        When this incompatibility condition is met, our equations of motion are non-equilibrium in the sense that they cannot be derived by optimizing a potential function~\cite{Grauwin2009,Lemoy2011}.
        These signs are consistent with survey results showing Black residents will move towards areas of higher White populations, while White residents will move away from areas of higher Black population (Fig.~\ref{fig:survey_results}).
        However, the nonlinear terms for the utility of White residents $\nu^{Wbc}$ additionally indicate a preference for neighborhoods with low White populations and high Black populations (Fig.~\ref{fig:sociosim}C).
        Such behavior may originate from gentrification, where wealthy residents move into lower-income neighborhoods.
        Future work could test this hypothesis by supplementing our model with income or housing cost data.

        Figure~\ref{fig:sociosim}D compares the sociohydrodynamic model's performance against other time-evolution models, including the neural networks in Fig.~\ref{fig:nn_results}.
        Despite the simplicity of our sociohydrodynamic equations compared to $f^a_\NN$, they achieve comparable accuracy to the model-free estimate from the neural networks (see SI Fig.~4 for results from other counties).

    \section{A phase diagram for social behavior}

        \begin{figure*}[t!]
            \centering
            \includegraphics{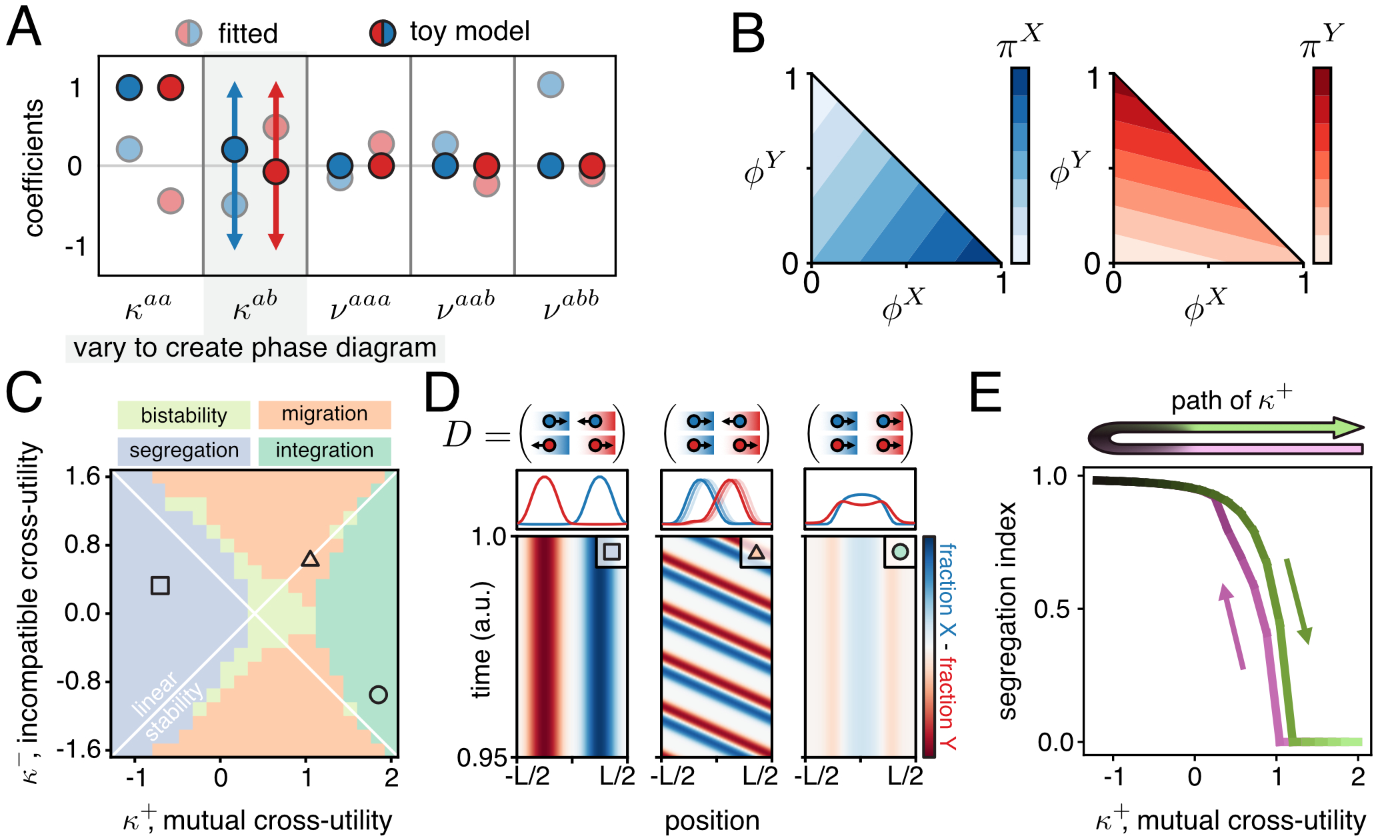}
            \caption{
            \textbf{Phase diagram for toy sociohydrodynamic model.}
            (A) Toy model utility function coefficients (solid circles) compared to the learned coefficients from Fig.~\ref{fig:sociosim} (transparent circles).
            We consider $\nu^{abc} = 0$ for $\kappa^{aa} = 1$ $\forall a, b, c$, where $a \in \lbrace X, Y \rbrace$.
            We then vary the cross-utilities $\kappa^{ab}$ to create our phase diagram.
            It will prove useful to parameterize our phase space in terms of $\kappa^\pm = \kappa^{YX} \pm \kappa^{XY}$.
            Other coefficients are $T^X = T^Y = 0.1$ and $\Gamma^X = \Gamma^Y = 1$.
            (B) Illustration of the linear utility functions for the two groups shown in the left and right subpanels, respectively. Plots show mutual cross-utility coefficient $\kappa^+ = \kappa^{XY} + \kappa^{YX} = -0.5$, incompatible cross-utility coefficient $\kappa^- = \kappa^{YX} - \kappa^{XY} = 1$.
            (C) Phase diagram showing different possible steady state dynamics depending on mutual and incompatible cross-utility coefficients, $\kappa^\pm$. We see 4 phases: segregation (purple), migration (orange), integration (green), and bistability of two phases (yellow).
            The white lines indicate results from a linear stability analysis for the onset of migratory states (details in SI).
            (D) Kymographs for steady state dynamics in the segregated phase, $(\kappa^+, \kappa^-) = (-0.07, 0.02)$ (left), migrating phase, $(\kappa^+, \kappa^-) = (0.02, 0.04)$ (middle), and integrated phase,  $(\kappa^+, \kappa^-) = (0.09, -0.06)$ (right). Plots on the top show the state of the system at the final time. Schematics of the resulting diffusion matrix shown on top of each phase. We set $T = 0.1$ and the average fill fraction of both types to be $\langle \phi^X \rangle = \langle \phi^Y \rangle = 0.25$.
            (e) The emergence of memory, in the form of hysteresis, in our simple model. Keeping $\kappa^-=0$, simulations are run starting at $\kappa^+ = 2$ until a steady state is reached, and then $\kappa^+$ is slightly decreased. This process is repeated until $\kappa^+=-1.2$, and then the process is reversed. The difference in the entropy index on the backward and forward pass of $\kappa^+$.
            }
            \label{fig:model_phases}
        \end{figure*}
        
        Inspired by the above results, we simplify our residential dynamics equations to a toy model in order to gain a better mathematical understanding of the possible sociohydrodynamic behaviors.
        We consider two groups $a=X,Y$ whose utility functions $\pi^a$ are linear in the fields $\phi^a$ 
        ($\nu^{abc}=0$, see Fig.~\ref{fig:model_phases}A,B).

        The flux in \eqref{eq:flux} becomes
        \begin{equation*}
            J^a(\vphi) = \sum_b ( -D^{ab} + \phi^a \vac \kappa^{ab} ) \nabla \phi^b + \Gamma^{ab} \nabla^3 \phi^b.
        \end{equation*}
        This results in diffusion of $\phi^a$ either up or down gradients of $\phi^b$, depending on the symmetries of the matrix $\kappa^{ab}$. 
        Based on known results on the Schelling model~\cite{Schelling1971,Schelling1980,Card2008,Vinkovic2006,Grauwin2009,Grauwin2012,Zhang2011}, we expect to observe segregation (where the two groups concentrate at different places) and integration (where the two groups occupy mainly the same place), depending on the utility functions.
        
        For every utility function, we solve \eqref{eq:sociohydro} numerically, and then report the resulting dynamical phase of the system in Fig.~\ref{fig:model_phases}C (see SI for criteria used in the categorization).
        We indeed observe the expected static states of segregation and integration (purple and green regions).
        In addition, we observe a time-dependent steady-state in which both groups continuously migrate (orange region, see also Fig.~\ref{fig:model_phases}D), in agreement with the results of Ref.~\cite{Zakine2024}.
        Finally, there are bistable regions (yellow) where multiple phases are observed depending on the initial state of the system (SI Fig.~10).
        
        The mechanism leading to the migratory states in Fig.~\ref{fig:model_phases}C,D stems from the incompatibility of the utilities $\pi^X$ and $\pi^Y$ in the sense of \eqref{incomptible_utilities}.
        For compatible utilities, meaning ${\partial \pi^Y}/{\partial \phi^X} = {\partial \pi^X}/{\partial \phi^Y}$, one can cast the dynamics as a gradient descent in a high-dimensional space (SI and Refs.~\cite{Lemoy2011,Grauwin2009,Fruchart2021,Dinelli2023}), thereby excluding time-dependent steady states like traveling waves.
        When the compatibility condition is violated, the corresponding interactions are non-reciprocal, a common ingredient to induce time-dependent steady-states~\cite{Fruchart2021,You2020,Saha2020,Ivlev2015,Hargus2021,Baek2018,FrohoffHuelsmann2021,Brauns2024,Dinelli2023,Scheibner2020,Fruchart2023,Banerjee2022}.
        More precisely, Fig.~\ref{fig:model_phases}B shows that the utility of $X$ increases when $\phi^Y$ \emph{decreases}, whereas the utility of $Y$ increases when $\phi^X$ \emph{increases}.
        In other words, $Y$ tends to move towards $X$ while $X$ tends to move away from $Y$. 
        When this tendency is strong enough, a time-dependent steady-state can emerge where the populations $X$ and $Y$ continuously chase or run away from each other.
        
        To predict the behavior of our simplified model, we perform a linear stability analysis around an initially spatially uniform state for $\vphi$.
        We find excellent agreement for the onset of pattern formation, when the uniform $\vphi$ becomes unstable.
        The onset of the traveling states is not captured by the linear stability analysis due to the non-linearities that play a role in their propagation.
        We see that no migration occurs when the overlap between the populations is below a threshold value, even when linear stability would predict migratory states (SI Fig.~11), in agreement with Ref.~\cite{Zakine2024}.
    
        The coexistence regions in Fig.~\ref{fig:model_phases}C illustrate that a single set of preferences can support multiple states of segregation.
        Which state is selected depends on the history of the system.
        To demonstrate this with our measured utility functions, we slowly change preferences of both groups over time. 
        We implement a cycle in the mutual cross-utility $\kappa^+ = \kappa^{XY} + \kappa^{YX}$ (while fixing $\kappa^{XY}-\kappa^{YX}$), starting and ending with the same value.
        Fig.~\ref{fig:model_phases}E shows the segregation indices measured in the resulting simulations. 
        Some preferences (x-axis of the plot) result in two different segregation indices, depending on whether this preference evolved from a segregated or integrated state. 
        This phenomenon is known as hysteresis: the state of the system depends on its past~\cite{Keim2019}.
        This form of socioeconomic memory emerges at the community level, despite the fact that the individuals in our model have no memory.
        As a result, two communities with an identical set of preferences can be segregated or integrated depending on their histories, and abrupt changes can occur at the edge of the bistability region in phase space.
        This may shed light on a phenomenon known as  \enquote{neighborhood tipping} in sociological research, where a threshold demographic distribution induces a rapid demographic shift in a neighborhood~\cite{Card2008,Zhang2011,Gualdi2015}.
        We note that the region of phase space exhibiting this hysteresis can qualitatively change if utility functions are non-linear (SI. Fig~12).

\section{Conclusion and outlook}
    To sum up, we introduced a data-driven pipeline to construct hydrodynamic descriptions of socially-driven residential motility.
    We illustrated this framework on the example of racial segregation in human residential dynamics.
    After testing the mathematical assumptions that enter a hydrodynamic theory, we validated our model by comparing its predictions to the evolution of demographic distributions in the US Census.
    We observed that our sociohydrodynamic model is sufficient to capture certain trends of residential dynamics in the US, even though our model neglects many important personal, sociological, and geographical aspects of residential choices.
    We showed that segregation can persist even in the absence of external influence due to an emergent memory in the population of memoryless agents.
    
    More broadly, our work illustrates how to incorporate individual choices in a hydrodynamic theory to provide a precise mathematical mapping from micromotives to macrobehavior~\cite{Schelling1980}.
    Beyond humans, it could be applied to motile micro-organisms~\cite{Kerr2002, Velicer2003, Hibbing2009, Kelsic2015, VanDyken2013, Ratzke2020,Miller2001,Waters2005} evolving~\cite{Mills1967,Smith1985,Chen1991} or adapting~\cite{Balaban2004, Kussell2005, Murugan2021} in spatially extended time-varying environments.
    Further, our method may also be applied to motion through abstract rather than physical spaces, such as the chase and escape dynamics in antigenic space during hosts and pathogens coevolution ~\cite{Marchi2021,Lassig2017} or cell fate decisions during development~\cite{TuringAlan1952,Gilmour2017,Lefebvre2023}.

\section{Methods}
\subsection{US Census Data}

    \begin{figure*}
        \centering
        \includegraphics[width=0.75\textwidth]{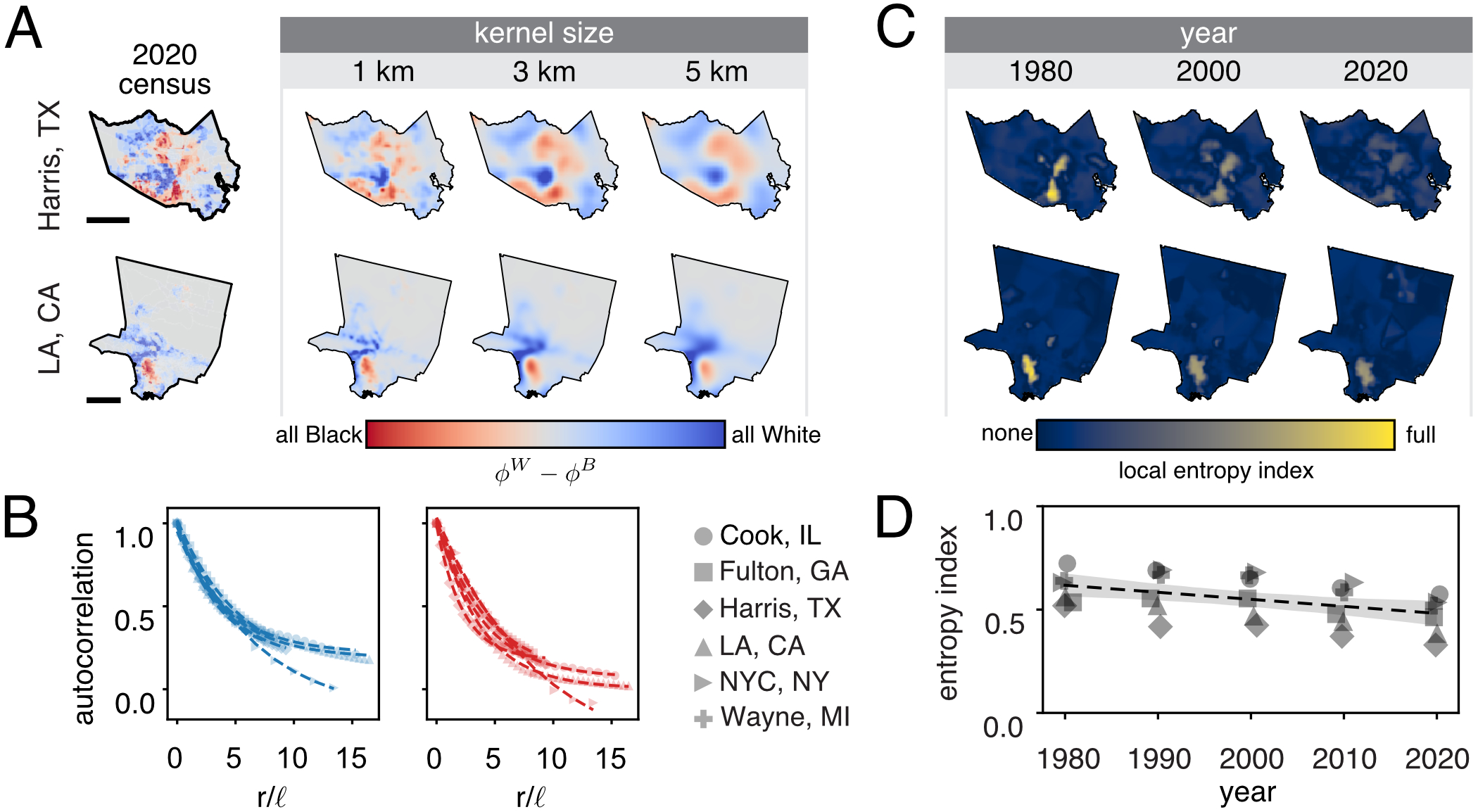}
        \caption{
            \textbf{Population distributions are slow variables across the US.}
            (A) Results of smoothing populations in various regions around the US using increasing kernel sizes. From top to bottom, the regions are: Harris County TX and Los Angeles County CA. Colors represent the proportion of local populations that are non-Hispanic White (blue) and non-Hispanic Black (red). Scale bars are 25 km. A Gaussian filter is used to smooth the populations.
            (B) Autocorrelation functions for White (blue) and Black (red) populations in six US regions (markers). Each set of data is fit to a decaying exponential (dashed line). The x-axis, measuring distance, is normalized by the median linear census tract size, $\ell$ of each region.
            Normalizing the correlation length by the $\ell$ accounts for differing densities, as each census tract is created to contain $\sim~10^3$ residents. The typical correlation lengths for White and Black populations in the regions we analyzed are $\langle \xi_W \rangle = 7.3 \pm 2.5 \ \ell$ and $\langle \xi_B \rangle = 4.3 \pm 1.0 \ \ell$.
            (C) Local segregation, $h_i$ (\eqref{eq:entropy_index_local}) for the two regions in A, shown from Census Data in 1980, 2000, and 2020.
            (D) Entropy index, $\mathcal{S}_\mathrm{H}$ (\eqref{eq:entropy_index}) measured as the weighted sum of the local segregation (Methods). Markers denote the same counties as B, and the dashed line is a linear fit to all data.
        }
        \label{fig:slowVars}
    \end{figure*}
    
    Population data is collected from the decennial US census at the census tract level for decades 1980-2020, aggregated using the IPUMS database \cite{Manson2022}.
    Interpolation from census tracts to a square grid is performed using areal-weighted interpolation on population densities in units of $\#/(10 \mathrm{m})^2$~\cite{Lam1983}.
    Smoothing is then done using a Gaussian filter with width $\sigma$.
    GIS file information are provided in the SI.

\subsection{Hydrodynamic variables}
    We use US Census data at the census tract level to find population densities of each group $a \in \lbrace W, B \rbrace$ at position $\mathbf{r}$ at time $t$, denoted as $\rho^a(\mathbf{r}, t)$.
    We approximate the housing availability as $h(\mathbf{r}) = \underset{t}{\mathrm{max}} \left( \sum_a \rho^a(\mathbf{r}, t) \right)$.
    Finally, we define the fill fraction as $\phi^a(\mathbf{r},t) = \rho^a(\mathbf{r}, t) /  \capacity$, where $\capacity = \underset{\mathbf{r}}{\mathrm{max}} \left( h(\mathbf{r}) \right)$ acts as a carrying capacity.
    One could consider fill fractions that are normalized by a spatiotemporal capacity to capture how the dynamics of housing availability impacts residential dynamics~\cite{Becharat2024}.
    This is beyond the scope of this work.

\subsection{Measuring segregation}
    We define the local demographic distribution of census tract $i$ using the probability vector $\mathbf{p}_i$, whose $m^\mathrm{th}$ element gives the proportion of the population at location $i$ belonging to group $m$. 
    Similarly, we define the demographic distribution of the county using the probability vector $\mathbf{p}$, whose $m^\mathrm{th}$ element gives the proportion of the county population belonging to group $m$. 
    We measure the local segregation using the entropy index~\cite{Reardon2004},defined as
    \begin{equation}
        \label{eq:entropy_index_local}
        h_i = \dfrac{D_\mathrm{KL}\left( \mathbf{p}_i || \mathbf{p} \right)}{H\left( \mathbf{p} \right)} \geq 0
    \end{equation}
    where $D_\mathrm{KL}\left( \mathbf{p}_i || \mathbf{p} \right)$ is the Kullback-Leibler divergence and $H(\mathbf{p})$ is the Shannon entropy. The condition for perfect integration is $h_i=0$, which occurs if a census tract's demographic distribution matches that of the county, $\mathbf{p}_i = \mathbf{p}$, \textit{not} when demographic groups are present in equal numbers.
    In order to quantify the segregation present in both Census data and simulations, we utilize the entropy index, also called a divergence index~\cite{Roberto2015}, which measures how distinct the demographic distribution in a local area (e.g. a census tract) is from the demographic distribution of the entire region (e.g. a county).
    
    Let $i~\in~[1, 2, \ldots, N]$ index local areas and $m~\in~[1, 2, \ldots, M]$ index demographic groups.
    We define $p_i^m$ to be the proportion of the population in the local area $i$ that is composed of individuals from group $m$, and thereby define the probability vector $\mathbf{P}_i~=~\left( p_i^1, p_i^2, \ldots \right)$.
    Similarly, we define $p^m$ to be the proportion of the population in the entire region that is composed of individuals from group $m$, and thereby define the probability vector $\mathbf{P}~=~\left( p^1, p^2, \ldots \right)$.

    \begin{subequations}
    With these two quantities, we construct the local measure of segregation at location $i$, $h_i$, that is shown in Fig.~\ref{fig:slowVars}C,
    \begin{equation}
            h_i = \dfrac{\sum\limits_m p_i^m \ln \left( \dfrac{p_i^m}{p^m} \right)}{-\sum\limits_m p^m \ln p^m} = \dfrac{D_\mathrm{KL}\left( \mathbf{P}_i || \mathbf{P} \right)}{H\left( \mathbf{P} \right)}
    \end{equation}
    where $D_\mathrm{KL}(\mathbf{P}_i||\mathbf{P}) \geq 0$ is the Kullback-Leibler divergence between $\mathbf{P}_i$ and $\mathbf{P}$, and $H(\mathbf{P})$ is the Shannon entropy of $\mathbf{P}$.
    
    We define a single scalar measure of a region's segregation, the entropy index, $\mathcal{S}_\mathrm{H}$, using a weighted mean of $h_i$. Specifically, we weight $h_i$ by the fraction of the total population that lives in local area $i$, $f_i$, giving
    \begin{equation}
            \mathcal{S}_\mathrm{H} = \sum_i f_i h_i.
            \label{eq:entropy_index}
    \end{equation}
    \end{subequations}

\subsection{Sociohydrodynamic diffusion matrix}
    We provide the full derivation of our sociohydrodynamic theory starting from an agent-based model in the SI, including the effects of altruism. Here, we simply report the diffusion matrix in \eqref{eq:sociohydro}. Assuming that the utility does not have an explicit spatial dependence, i.e. $\pi^a = \pi^a(\vphi)$, we have
    \begin{align}
        J^a(x,t) &= -\sum_b D^{ab} \partial_x \phi^b + \phi^a \vac \partial_x \pi^a + \Gamma^{ab} \partial_x^3 \phi^b
        \\
        &= -\sum_b \mathsf{D}^{ab} \partial_x \phi^b + \Gamma^{ab} \partial_x^3 \phi^b
    \end{align}
    where $\phi^0 = 1 - \sum_b \phi^b$ is the vacancy fraction and $\mathsf{D}^{ab}$ are elements of the matrix
    \begin{equation}
    \begin{aligned}
        \mathsf{D}(\vphi) =&
        -T
        \begin{pmatrix}
            1 - \phi^Y & \phi^X \\
            \phi^Y & 1 - \phi^X
        \end{pmatrix}
        +
        \vac
        \begin{pmatrix}
            \phi^X \dfrac{\partial \utility^X}{\partial \phi^X}
            &
            \phi^X  \dfrac{\partial \utility^X}{\partial \phi^Y} 
            \\
            \phi^Y \dfrac{\partial \utility^Y}{\partial \phi^X}
            &
            \phi^Y \dfrac{\partial \utility^Y}{\partial \phi^Y} 
        \end{pmatrix}
    \end{aligned}
    \end{equation}
    The first term arises due to volume exclusion effects with finite carrying capacity of the lattice sites in the agent-based model. The second term arises from the utility functions.
    Similarly, we find
    \begin{equation}
        \Gamma^{ab} = \delta^{ab} \phi^b \vac,
    \end{equation}
    where $\delta^{ab}$ is the Kronecker delta.

\subsection{Numerical methods}
    Parameter inference is done using linear regression on US Census data. 
    We first interpolate the US Census data using a $3^\mathrm{rd}$-order spline to estimate populations between decennial census years.
    We then estimate time derivatives and spatial gradients using finite differences.
    A feature matrix $A$ is constructed using each term in \eqref{eq:sociohydro} as a column, and a target vector $b$ is created using the time derivatives.
    The coefficients in \eqref{eq:sociohydro} are collected in a vector $x$, and we invert the equation $Ax=b$.
    
    For the simulations shown in Fig.~\ref{fig:sociosim}, we solve \eqref{eq:sociohydro} over county boundaries using a finite-volume method.
    We construct a triangular mesh over the county geometry using GMSH, and use the Python package FiPy~\cite{Guyer2009} to solve \eqref{eq:sociohydro}. More details can be found in the SI.
    
    For the simulations shown in Fig.~\ref{fig:model_phases}, we simulate \eqref{eq:sociohydro} in 1 spatial dimension using a finite difference method, with a 4$^\mathrm{th}$ order discretization in space and 1$^\mathrm{st}$ order discretization in time. Unless otherwise stated, we set $T = 0.1$ and $\Gamma=1$, and use a time step of $\Delta t = 0.1$ and $\Delta x = 0.625$.

\subsection{Linear utility functions}
    The linear utility functions used in Fig.~\ref{fig:model_phases} are given by
    \begin{subequations}
    \begin{align}
        \pi^a(\vphi) &= \sum_b \kappa^{ab} \phi^b \\
        \kappa &=
        \begin{pmatrix}
            \kappa^{XX} & \kappa^{XY} \\
            \kappa^{YX} & \kappa^{YY}
        \end{pmatrix}
    \end{align}
    \end{subequations}
    The matrix elements $\kappa^{ab}$ quantify how the utility of $a$ is affected by the presence of $b$. We call $\kappa^{XX}$ and $\kappa^{YY}$ the self-utility coefficients, and $\kappa^{XY}$ and $\kappa^{YX}$ the cross-utility coefficients.
    We find it convenient to define $\kappa^\pm = \kappa^{YX} \pm \kappa^{XY}$. We dub $\kappa^+$ the \enquote{mutual} cross-utility coefficient as it measures the degree of mutual (dis)like between the $X$ and $Y$. Likewise, we call $\kappa^-$ the \enquote{incompatible} cross-utility coefficient as it measures the incompatibility of the two utilities with each other (see Fig.~\ref{fig:model_phases}).
    
\subsection{Assessing locality via neural networks}
    We used a convolutional neural network to predict the dynamics $(\partial_t \phi^W, \partial_t \phi^B)$ from an initial condition $(\phi^W, \phi^B)$. Briefly, it contains two convolutional modules which computes latent features at the scale of the input data and using a downsampled representation in order to aggregate spatial information over short or long distances. 
    The network predicts the discrete time derivative and forecasts the next time step as
    \begin{equation}
        \phi^a(t + \Delta t) = \phi^a(t) + f^a_{\NN}[\vec{\phi}(t); h] \,\Delta t
    \end{equation}
    Here, the population index $a \in (W, B)$ and the capacity $h$ is taken as the maximum occupancy at each coordinate over the observed time range. The full network details are given in the SI.
    We used the trained network to predict population dynamics over the 40-year window spanned by decennial Census data. 
     
    To assess locality in the predicted dynamics, we computed the output saliencies for the trained network. Saliency is a measure of how much a model's predictions depend on its inputs and is here defined as~\cite{Simonyan2014}
    \begin{equation}
     K^{ab}({\mathbf{r}_i, \mathbf{r}_j}) = \frac{\partial f^a_{\NN}[\vec{\phi}(\mathbf{r}_i)]}{\partial \phi^b (\mathbf{r}_j)}
    \end{equation}
    Here $a,b$ are population indices while $i,j$ refer to spatial coordinates within each county. We compute $K^{ab}(\mathbf{r}_i, \mathbf{r}_j)$ from each input-output pair in each county dataset, for 100 randomly sampled output points $\mathbf{r}_i$ and all input points $\mathbf{r}_j$.
    We azimuthally average this into a set of curves $K^{ab}(|\mathbf{r}_a - \mathbf{r}_b|)$ which are plotted in Figure~\ref{fig:nn_results}C.


\bibliography{refs}

\clearpage

\input{supplement.tex}

\end{document}

%% file: preamble.tex
\usepackage{amsmath}
\usepackage{amssymb}
\usepackage{stmaryrd}
\usepackage{fourier-orns}
\usepackage{commath}
\usepackage{mathrsfs}
\usepackage{braket}
\usepackage[svgnames]{xcolor}
\usepackage{bm}
\usepackage{csquotes}
\usepackage{graphicx}
\usepackage{booktabs}
\usepackage{enumitem}
\usepackage{siunitx}
\usepackage{microtype}
\usepackage{tikz}
\usepackage{hyperref}
\usepackage{bookmark}
\hypersetup{
  colorlinks=true,
  linkcolor=DarkBlue,
  urlcolor=DarkBlue,
  citecolor=DarkGreen
}
\usepackage[many]{tcolorbox}

\newcommand{\capacity}{\mathcal{N}}
\newcommand{\utility}{\pi}
\newcommand{\vphi}{\vec{\phi}}
\newcommand{\un}{\underline{n}}
\newcommand{\uphi}{\underline{\phi}}
\newcommand{\vac}{\phi^0}
\newcommand{\NN}{
 {\mathchoice
  {\includegraphics[height=1.0em]{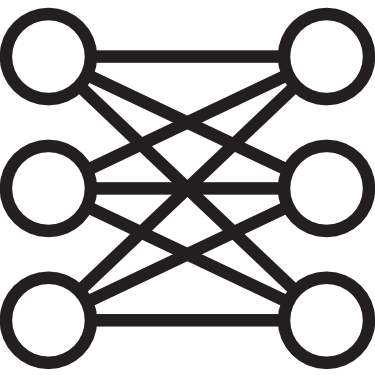}}
  {\includegraphics[height=1.0em]{figures/NN.jpg}}
  {\includegraphics[height=0.8em]{figures/NN.jpg}}
  {\includegraphics[height=0.9em]{figures/NN.jpg}}
 }
}
\newcommand{\mat}[1]{\mathsf{#1}}
\newcommand{\Tr}[1]{\mathrm{Tr} \ #1}

\definecolor{bkrd}{HTML}{ebf8fa}
\definecolor{title}{HTML}{087FC7}
\newtcolorbox{myBox}{
    boxrule = 0pt,
    colback = bkrd 
}

%% file: supplement.tex
\onecolumngrid

\begin{center}
{\Large \textbf{Supplementary Materials}}
\end{center}

\title{Supplement to \\ Sociohydrodynamics: data-driven modelling of social behavior}
\author{Daniel S. Seara}
\affiliation{University of Chicago, James Franck Institute, 929 E 57th Street, Chicago, IL 60637}

\author{Jonathan Colen}
\affiliation{University of Chicago, James Franck Institute, 929 E 57th Street, Chicago, IL 60637}
\affiliation{University of Chicago, Department of Physics, 5720 S Ellis Ave, Chicago, IL 60637}
\affiliation{Old Dominion University, Joint Institute on Advanced Computing for Environmental Studies, 1070 University Blvd, Portsmouth, VA, 23703}

\author{Michel Fruchart}
\affiliation{University of Chicago, James Franck Institute, 929 E 57th Street, Chicago, IL 60637}
\affiliation{University of Chicago, Department of Physics, 5720 S Ellis Ave, Chicago, IL 60637}
\affiliation{ESPCI, Laboratoire Gulliver, 10 rue Vauquelin, 75231 Paris cedex 05}

\author{Yael Avni}
\affiliation{University of Chicago, James Franck Institute, x929 E 57th Street, Chicago, IL 60637}

\author{David Martin}
\affiliation{University of Chicago, Kadanoff Center for Theoretical Physics and Enrico Fermi Institute, 933 E 56th St, Chicago, IL 60637}

\author{Vincenzo Vitelli}
\affiliation{University of Chicago, James Franck Institute, 929 E 57th Street, Chicago, IL 60637}
\affiliation{University of Chicago, Kadanoff Center for Theoretical Physics, 933 E 56th St, Chicago, IL 60637}
    \date{\today}
\maketitle

\renewcommand{\thepage}{S\arabic{page}}
\setcounter{page}{1}
\setcounter{section}{0}
\setcounter{figure}{0}
\setcounter{table}{0}
\setcounter{equation}{0}


\section{US Census Data}

    \begin{figure}[h]
    \centering
    \includegraphics[width=0.9\textwidth]{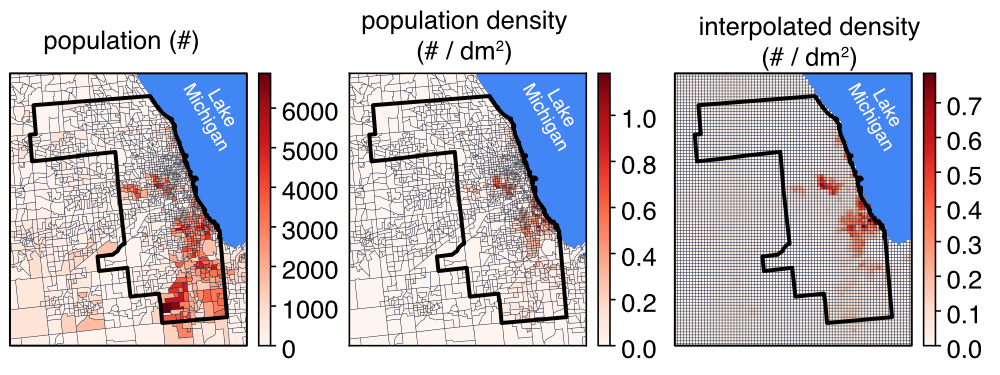}
    \caption{
        \textbf{Census data interpolation.}
        Example of interpolation of Census data to a regular grid, focusing on Black residents of Cook County, Illinois.
        Starting with Census data (left, 2020), we first convert to population densities (middle) in units of $\#/\mathrm{dm}^2$.
        We then use areal-weighted interpolation~\cite{Tobler1979} to interpolate population densities to a regular grid.
        The same step is also done to interpolate to a triangular grid instead when running finite-volume simulations.
        }
        \label{fig:interpolation}
    \end{figure}
    
    The US population data are decennial census data at the census tract level for decades 1980-2010 \cite{Manson2022}.
    Populations labeled as \enquote{White} or \enquote{Black} were taken to be non-Hispanic White and non-Hispanic Black.
    We further used GIS data to give proper physical coordinates to the population data using the 2008 TIGER/Line GIS for years 1980-2000, and the 2020 TIGER/Line GIS for years 2010 and 2020.
    
    To coarse-grain and analyze the population data, we perform an interpolation for each studied region separately.
    After reading the GIS shapefile, we convert the total population numbers to population densities by dividing by the census tract area.\
    This population density is then interpolated to a regular grid using areal-weighted interpolation~\cite{Tobler1979}.
    Areal-weighted interpolation is a method developed by geographers to interpolate data aggregated using one one set of spatial units to a different set of spatial units.
    Consider that we begin with a given set of values $n_i$ distributed over spatial units given by $s_i$, where $i$ indexes each spatial unit (e.g. a single census tract).
    We would like to interpolate the values $n_i$ to a new set of spatial units $\tilde{s}_j$ to give $\tilde{n}_j$.
    Areal-weighted interpolation says to assign
    $$\tilde{n}_j = n_i \dfrac{A_{ij}}{A_i},$$
    where $A_{ij}$ is the area of the intersection of $s_i$ with $s_j$, and $A_i$ is the area of $s_i$.
    If both sets $s_i$ and $\tilde{s}_j$ cover the same area, then $\sum_i n_i = \sum_j \tilde{n}_j$ and the total number is conserved.

    In our case, we are working with population numbers $n_i^a$ of White and Black populations, $a = \lbrace W, B \rbrace$ and where $i$ indexes the spatial location. The population density is given by $\rho^a_i = n^a_i/A_i$.
    In practice, we read census and GIS shapefiles using \href{https://geopandas.org/en/stable/}{\texttt{geopandas}}.
    The areal weighted interpolation is performed using the Python package \href{https://pysal.org/tobler/}{\texttt{tobler}}, treating the population densities as an intensive variable (see SFig.~\ref{fig:interpolation}).
    Once on a regular square grid, we can write $\rho_i^a = \rho^a(\mathbf{r}_i)$, where $\mathbf{r}_i$ denotes the center of grid point $i$.
    We then convert from densities to fill fractions, $\phi^a(\mathbf{r}_i)$ by finding the maximum density within the county over all time, 
    $$\rho^* = \underset{i}{\mathrm{max}} \left(\sum_{a=\lbrace W, B \rbrace} \rho^a(\mathbf{r}_i) \right)$$
    We then divide the density field by this normalizing factor, giving
    $\phi^a(\mathbf{r}_i) = \rho^a(\mathbf{r}_i) / \rho^*$.

    We measure the correlation length of the fluctuations of the two populations using standard Fourier Transform methods. Specifically, for each population we calculate $\delta \phi^a(x, t) = \phi^a(x, t) - \langle \phi^a \rangle$, where the average is taken over space and time. We then calculate the spatial Fourier transform, $\mathcal{F}_q \left[ \delta \phi^a(x, t) \right] = \delta \phi^a(q, t)$, and calculate the autocorrelation function as
    \begin{equation}
        C_{aa}(x) = \mathcal{F}^{-1}_q \left[ \langle |\delta \phi(q, t)|^2 \rangle_t \right].
    \end{equation}
    Across the regions shown in Fig.~\ref{fig:interpolation}, we measure a mean correlation length for White and Black populations to be $\langle \xi_w \rangle = 11.7 \pm 4.2$ km and $\langle \xi_b \rangle = 7.9 \pm 3.9$ km.

    The typical census tract size, $\ell$, for the different areas studied are found using the areas of all census tracts, $\lbrace A_i \rbrace$, and defining $\ell = \mathrm{median}\left( \lbrace \sqrt{A_i} \rbrace \right)$ (Table~\ref{tab:tractSize}).
    
    \begin{table}[h]
        \centering
        \begin{tabular}{c|c}
            \textbf{Region} & $\ell$ (km) \\
            \hline \hline
            Cook County, IL & 1.37 \\
            Fulton County, GA & 2.78 \\
            Harris County, TX & 2.58 \\
            Los Angeles County, CA & 2.31 \\
            New York City, NY & 0.60 \\
            Wayne County, MI & 1.60
        \end{tabular}
        \caption{Median census tract size for regions studied. Median is taken over all census tracts from 1980-2020.}
        \label{tab:tractSize}
    \end{table}

\section{Segregation indices}
    In order to quantify both the measured distribution of human populations as well as our simulations, we utilize segregation indices. Mathematically, a segregation index is measured by a functional $\mathcal{S}$ that takes the distribution of $\geq 2$ populations to generate a scalar measure of how spatially separated the populations \cite{Reardon2004}. We will normalize our indices such that $0$ means the system is perfectly integrated, and $1$ means the system is completely segregated. We stress that \enquote{integration} does not means that a location has $50-50$ representation of each group, but rather that proportion of each group present locally is equal to the proportion of each group present globally. For example, if a region comprises 70\% White residents and 30\% Black residents, the region is integrated if each location within the region also comprises 70\% White residents and 30\% Black residents.

    Our notation is defined in Table~\ref{tab:segregation_notation} and largely follows \cite{Reardon2004}. We call a \enquote{region} the full geographic area we consider (e.g. Cook County, New York City), and a \textit{neighborhood} a small area within the region (e.g. an individual census tract). The entropy index used in the main text, which we will denote here as $\mathcal{S}_\mathrm{H}$, is a weighted mean of an entropy index field $h_i$, defined as
    \begin{subequations}
    \label{eq:segregation_index}
    \begin{align}
        h_i &= \dfrac{\sum_m p_i^m \ln \left( \dfrac{p_i^m}{p^m} \right)}{-\sum_m p_m \ln p_m} \\
        \mathcal{S}_\mathrm{H}[p_i^m] &= \dfrac{1}{T} \sum_i t_i h_i.
    \end{align}
    \end{subequations}
    The spatial maps of segregation in Figs.~1\&3 show $h_i$, which measures the Kullback-Leibler divergence between the local composition distribution $p_i^m$ to the region's composition distribution $p_m$, normalized by the Shannon entropy of the region's composition. The entropy index is the population-weighted mean of $h_i$.

    This choice of index is not unique. Another common index used in the sociology literature is the so-called \enquote{dissimilarity index}, which we will denote as $\mathcal{S}_\mathrm{D}$. Like $\mathcal{S}_\mathrm{H}$, it measures how different local compositions are compared to the global composition. Again, it is given by a weighted mean of a field, $d_i$, defined for binary populations as
    \begin{subequations}
    \begin{align}
        d_i &= \dfrac{1}{2}\dfrac{|p_i^m - p^m|}{p_m(1-p_m)} \\
        \mathcal{S}_\mathrm{D} &= \dfrac{1}{T} \sum_i t_i d_i.
    \end{align}
    \end{subequations}
    We note that the binary population assumption ensures $d_i$ is the same regardless of which subgroup $m$ is chosen.

    \begin{table}[h]
        \centering
        \begin{tabular}{r|l}
            \textbf{Variable} & \textbf{Definition} \\ \hline
            $m$ & index of subgroup \\
            $i$ & index of neighborhood \\
            $t_i^m$ & total number of group $m$ in neighborhood $i$ \\
            $t^m = \sum_i t_i^m$ & total number of $m$ in region \\
            $t_i = \sum_m t_i^m$ & total number of people in neighborhood \\
            $T = \sum_{m, i} t_i^m$ & total number of people in region \\
            $p_i^m = t_i^m / t_i$ & proportion of residents in neighborhood $i$ of type $m$ \\
            $p^m = t^m / T$ & proportion of residents in region of type $m$ \\
        \end{tabular}
        \caption{Notation used for defining and calculating segregation indices.}
        \label{tab:segregation_notation}
    \end{table}

    In Fig.~\ref{fig:indices}, we reproduce the results from the annealing simulations shown in Fig.~6C along-side measurements using the dissimilarity index, illustrating that the two measurements give the same qualitative results.

\section{Assessing locality via neural networks}
\begin{figure}[h]
    \centering
    \includegraphics[width=0.9\linewidth]{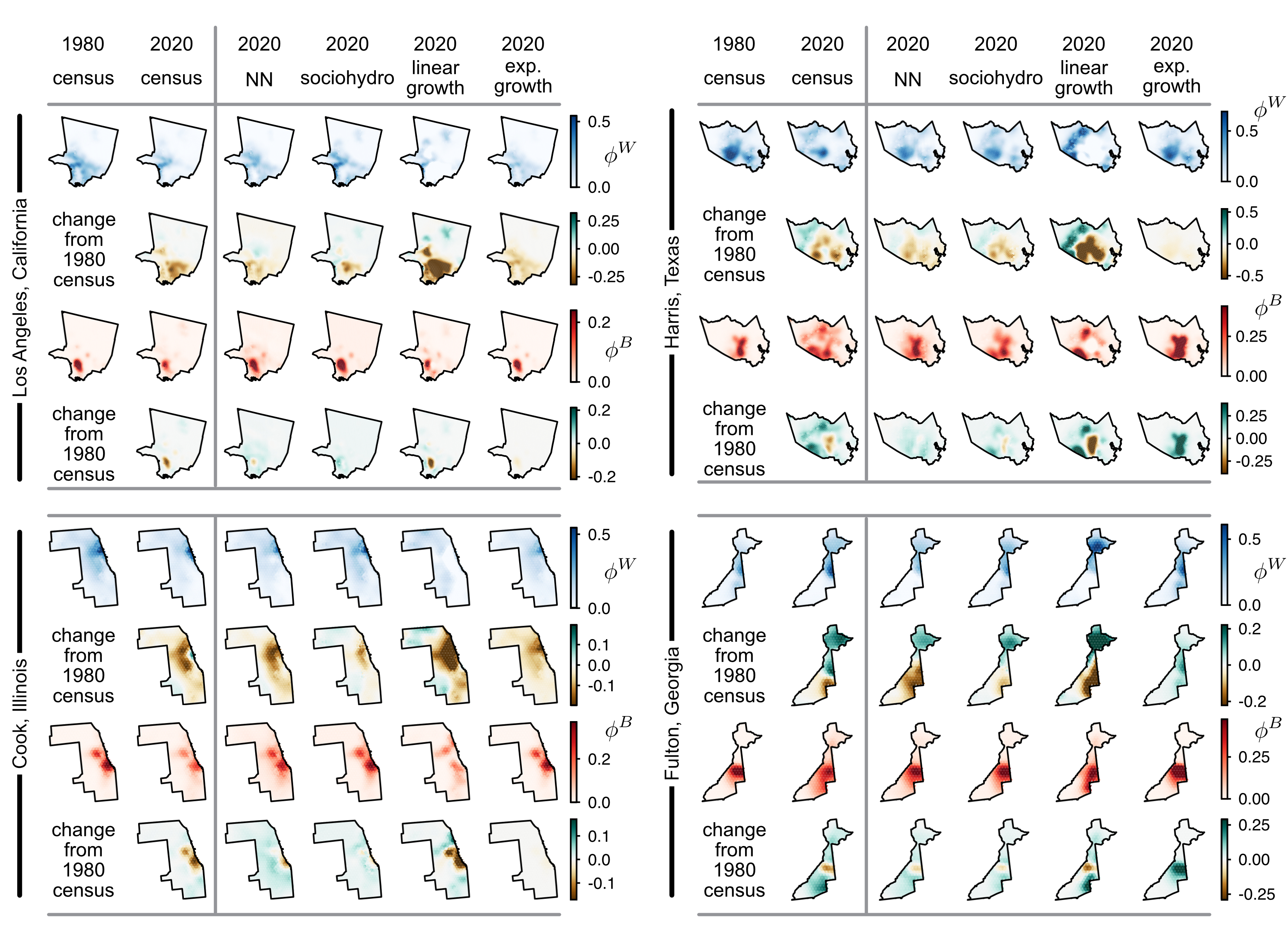}
    \caption{
        \textbf{Detailed maps of various population predictions}.
        Here, we show explicit spatial comparisons between various models to predict demographic distributions for four counties -- Los Angeles, CA (top left), Harris, TX (top right), Cook, IL (bottom left), and Fulton, GA (bottom right).
        Each county has the same layout.
        The first column shows the distribution from the 1980 census, which serves as our initial condition.
        The second column shows the distribution from the 2020 census, which serves as our target.
        The following columns show predictions from the neural network, sociohydrodynamic simulations, linear extrapolation/growth, and exponential growth, respectively.
        The first and third rows show the distributions for the White and Black populations, respectively.
        The second and fourth rows show the difference of the map above it with the 1980 census data, illustrating population changes rather than absolute magnitudes.
    }
    \label{fig:socioSimSI}
\end{figure}

\begin{figure}[h]
    \centering
    \includegraphics[width=0.5\linewidth]{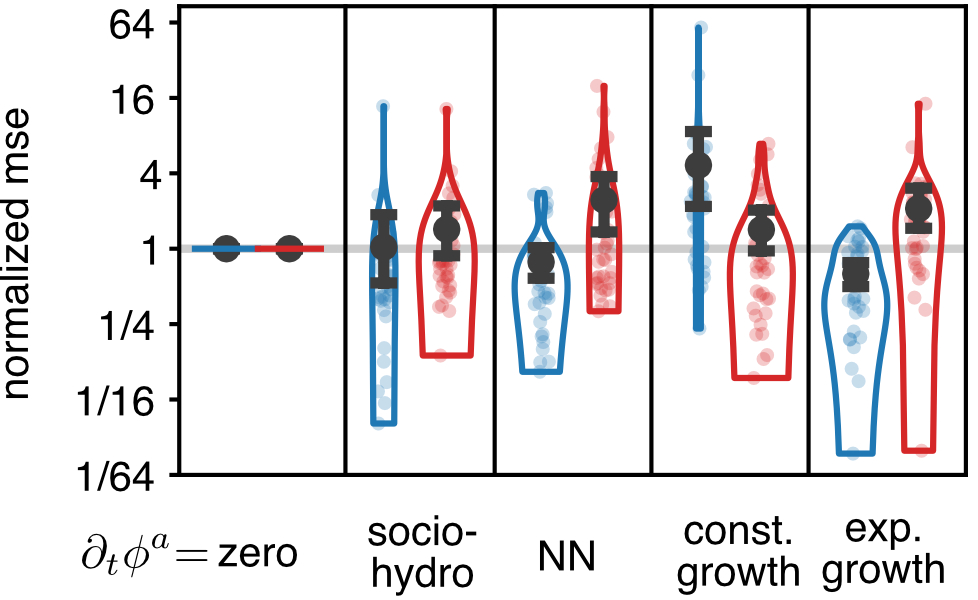}
    \caption{
    \textbf{Mean-square error for all counties.}
        Dots denote the normalized mean-square error for all 35 counties with populations $>10^6$ for White (blue) and Black (red) populations.
        The curves show the density of points for each population.
        The black circles and error bars show the mean $\pm$ standard deviation within each category.
    }
    \label{fig:msesAll}
\end{figure}

    We trained and evaluated neural networks using decennial Census Data (1980-2020) from 34 counties with total populations greater than 1 million.
    For each county, we first downsampled the population data to a scale of 1 pixel = 1 km and performed Gaussian smoothing ($\sigma = 3$ km) to soften sharp gradients in the sampled areas. Next, we converted the raw populations into an occupation fraction $\phi$ by dividing the population at each pixel by the maximum total occupancy within the region over the entire time range. Each training example contained an input-output pair: the input was an image of occupation fractions for the county interpolated to a random time point $t_0 \in [1980, 2020]$, while the output was a second image of occupation fractions at some future time $t_0 + \Delta t$ where $\Delta t \in [0, 1]$. The machine learning model was tasked with predicting this future state $\vec{\phi} (t_0+\Delta t)$ from the initial condition $\vec{\phi}(t_0)$. 

    \subsection{Neural network architecture and training}
    
        We used a convolutional neural network to predict dynamics from the input data. The full architecture is outlined in Table~\ref{tab:nn_arch} and implemented using the Pytorch library~\footnote{Code is available online at \url{https://github.com/jcolen/sociohydro}}. Briefly, it contains two sequential modules of 8 convolutional blocks. The first computes latent features at the scale of the input data, while the second computes additional information using a downsampled representation of these latent features. As this second layer operates at a coarse-grained level, it can aggregate spatial information over a longer range. Throughout the network, we use ConvNext-style blocks shown in Table~\ref{tab:convnext}~\cite{Liu2022}.
        
        \begin{table}[]
            \centering
            \begin{tabular}{c|c|c|c}
                    Module & Size in & Size out & Details \\
                    \hline
                    \hline
                    Read-in & $2\times H\times W$ & $C\times H\times W$ & 1$\times$ConvNext \\
                    CNN1 & $C\times H\times W$ & $C\times H\times W$ & 8$\times$ConvNext \\
                    Downsample & $C\times H\times W$ & $C \times H/4 \times W/4$ & Strided Conv2d \\
                    CNN2 & $C \times H/4 \times W/4$ & $C \times H/4 \times W/4$ & 8$\times$ConvNext \\
                    Upsample & $C \times H/4 \times W/4$ & $C\times H\times W$ & Interpolate \\
                    Skip & & $2C\times H\times W$ & Concat CNN1, \\
                    & & & CNN 2 outputs \\
                    Read-out & $2C\times H\times W$ & $2\times H\times W$ & Conv2d 
            \end{tabular}
            \caption{Neural network architecture. ConvNext layers are summarized in Table~\ref{tab:convnext}. We use a hidden size $C = 64$. The spatial dimensions $H,W$ are determined by the input data. 
            }
            \label{tab:nn_arch}
        \end{table}
        
        \begin{table}
            \centering
            \begin{tabular}{c|c| c c c}
                    Layer & Details & Channels & Kernel & Groups \\
                    \hline
                    \hline
                    Conv2d & Mix space & $N \rightarrow N$ & 7x7 & N \\
                    Conv2d & Mix channels & $N \rightarrow 4N$ & 1x1 & 1 \\
                    Sin & Activation & & & \\
                    Conv2d & Inverse bottleneck & $4N \rightarrow N$ & 1x1 & 1 \\
                    Dropout & $p = 0.1$ & & & 
            \end{tabular}
            \caption{ConvNext-style block architecture}
            \label{tab:convnext}
        \end{table}
        
        This neural network processes a 3-channel image $(\phi^W, \phi^B, h)$ and outputs another 2-channel image representing the predicted time derivative of the input $(\partial_t \phi^W, \partial_t \phi^B)$. 
        Here $h$ was a fixed local capacity defined as the maximum occupancy at each pixel over the 40-year Census period. We evaluated the quality of these predictions by integrating them using a finite-difference discretization over 40 years starting from an initial condition at 1980.
        The network learns to minimize the loss
        \begin{equation}
            \mathcal{L} = \frac{1}{N} \sum_{i=1}^N \bigg \lvert 
                \phi^W_i(t_0 + \Delta t) - \hat{\phi}^W_i (t_0 + \Delta t)
            \bigg \rvert + 
            \bigg \lvert 
                \phi^B_i(t_0 + \Delta t) - \hat{\phi}^B_i (t_0 + \Delta t)
            \bigg \rvert
        \end{equation}
        Here $\hat{\phi}^a_i$ is the predicted occupation fraction within the boundary of the county for sample index $i$.
        We train the model for 100 epochs with batch size 8. We use the Adam optimizer with learning rate $\lambda = 0.0003$ and exponential learning rate decay with $\gamma = 0.98$. The network was trained using all available Census data from 30 of the collected counties and was tested on four counties: Cook IL, Harris TX, Fulton GA, and Los Angeles CA. During training, the network saw samples from the first decade of Census data (1980-1990) for each of these four counties. The results presented in Figure~3 are extrapolations beyond the training regime for the remaining 30 years (1990-2020) in each test county.

        \begin{table}[]
        \label{tab:nn_mse}
        \begin{tabular}{r||c|c|c||c|c|c}
        County          & \multicolumn{3}{c||}{MSE}           & \multicolumn{3}{c}{Rel. MSE} \\ \hline
                        & zero    & linear growth & NN       & zero  & linear growth & NN    \\ \cline{2-7} 
        Fulton, GA      & 0.00114 & 0.00181       & 0.000768 & 1.0   & 1.583         & 0.674 \\
        Cook, IL        & 0.0103  & 0.0202        & 0.0045   & 1.0   & 1.966         & 0.437 \\
        Harris, TX      & 0.00119 & 0.00429       & 0.00119  & 1.0   & 2.202         & 0.613 \\
        Los Angeles, CA & 0.00371 & 0.013         & 0.00186  & 1.0   & 3.495         & 0.500
        \end{tabular}
        \end{table}
        
    \subsection{Post-processing and saliency analysis}
    \label{sec:postprocessing}
    
        After training, we use the network to predict population dynamics over the 40-year window spanned by the decennial Census data. Specifically, we solve the following initial value problem
        \begin{align}
            \partial_t \vec{\phi} (t, \mathbf{r}) &= f_\NN \left[\vec{\phi} (t, \mathbf{r}); h\right] \\
            \vec{\phi}(1980, \mathbf{r}) &= \text{US Census data from 1980}
        \end{align}
        To do this, we discretize the time derivative using finite differences as $\partial_t \vec{\phi} \approx (\vec{\phi}_{t+\Delta t} - \vec{\phi}_t) / \Delta t$ and iteratively compute $\vec{\phi}_{t+\Delta t}$ using a finite element approach. We used a time step $\Delta t = 1$ year and integrated the above equations as a closed loop for 40 years. Beyond the initial condition, the neural network used no Census data to generate these predictions. 
        
        To assess locality in the predictions of the machine learning model, we computed output saliencies for the trained neural networks. Saliency is a measure of how much a model's predictions depend on its inputs and is here defined as~\cite{Simonyan2014}
        \begin{equation}
            K^{ab}({\mathbf{r}_i, \mathbf{r}_j}) = \frac{\partial f_\NN^a (\mathbf{r}_i)}{\partial \phi^b (\mathbf{r}_j)}
        \end{equation}
        Here $a,b$ are population indices while $i,j$ refer to spatial coordinates within each county. We compute $K^{ab}(\mathbf{r}_i, \mathbf{r}_j)$ from each input-output pair in each county dataset, for 100 randomly sampled output points $\mathbf{r}_i$ and all input points $\mathbf{r}_j$.
        By azimuthally averaging, we aggregate this into a set of curves $K^{ab}(|\mathbf{r}_a - \mathbf{r}_b|)$ which are plotted in Figure~3 of the main text.

    \subsection{Benchmarking saliency analysis with Monte Carlo simulations}

    \begin{figure}[h]
        \centering
        \includegraphics[width=0.9\linewidth]{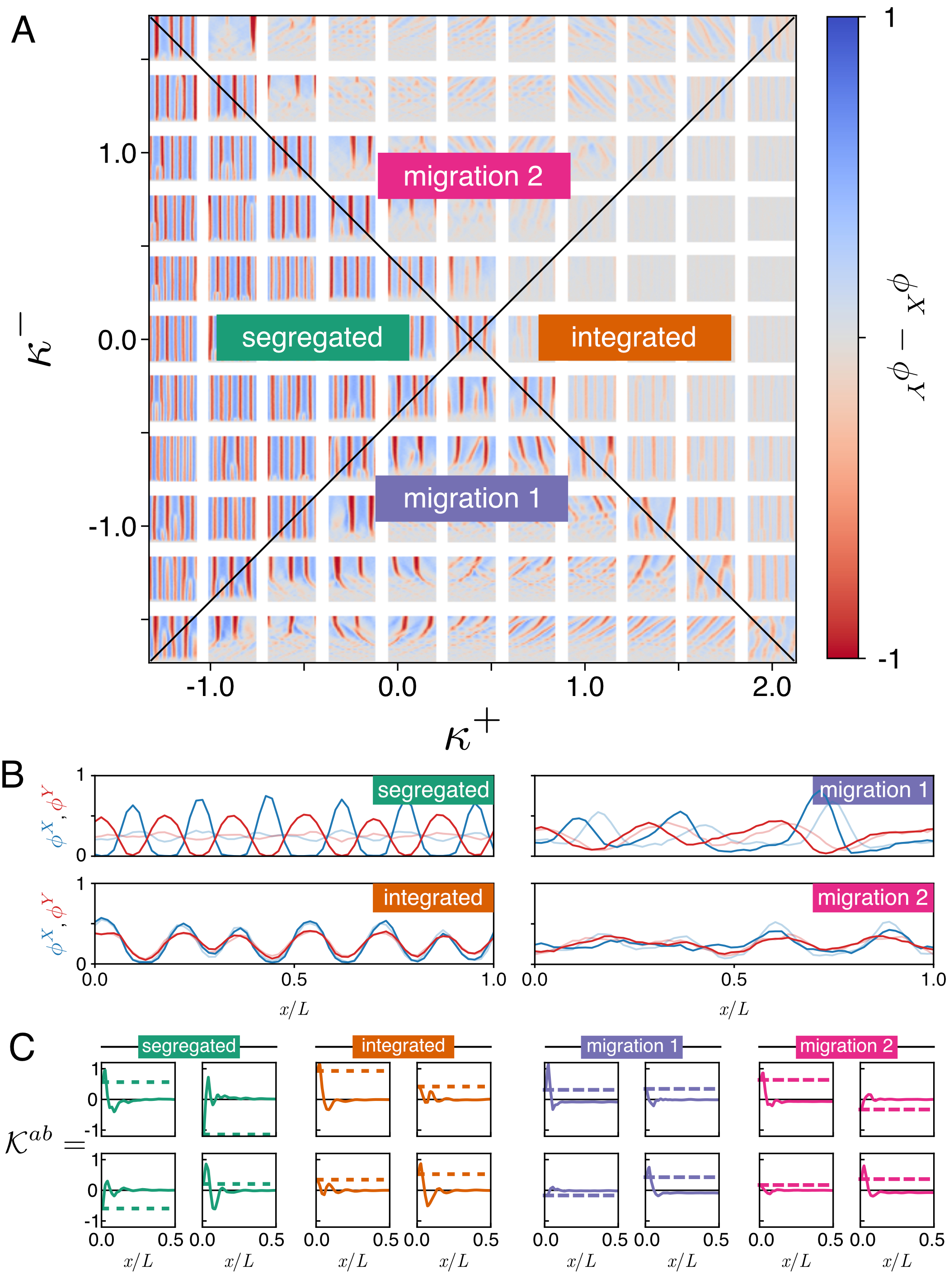}
        \caption{
            \textbf{Saliency for agent-based simulations.}
            (A) Result of Monte Carlo simulations of 1D Schelling model using linear utility functions, $\pi^a = \kappa^{ab} \phi^b$, with $\kappa^{XX} = \kappa^{YY} = 1$ and $\kappa^+ \in [-1.2, 2.0]$, $\kappa^- \in [-1.6, 1.6]$, giving a phase space diagram over $D^\pm \in [-0.1, 0.1]$, with temperature $T = 0.1$ and $\Gamma = 1$.
            Each plot inside the phase space is a kymograph showing the dynamics of $\phi^X - \phi^Y$ over time.
            The phase space is divided into 4 quadrants based on the phenotype of each region -- segregation, integration, migration 1, and migration 2 (blue, orange, green red, respectively).
            (B) Exemplary trajectories in each phase. Light colored lines indicate the fill fractions at an earlier time, and the dark colored lines indicate the fill fractions at later times.
            (C) Saliencies measured in each phase.
            Solid lines show the saliencies extracted from neural networks trained on trajectories in each phase. 
            Solid lines show the saliencies measured from the Monte Carlo simulations. 
            The semi-transparent lines show the value of $\mathcal{K}^{ab}(x = 0)$.
            The black horizontal line indicates $\mathcal{K}^{ab} = 0$.
            Saliencies for each phase are arranged as they would be in a matrix, with $\mathcal{K}^{XX}, \mathcal{K}^{XY}, \mathcal{K}^{YX}, \mathcal{K}^{YY}$ in the top left, top right, bottom left, and bottom right, respectively.
            }
        \label{fig:nnResults_MC}
    \end{figure}

        To benchmark the behavior of the neural network models and saliency analysis pipeline, we applied it to agent-based simulations of a 1D Schelling model using a range of utility functions. We divided teh phase space into 4 quadrants based on the phenotype of each region -- segregation, integration, migration 1, and migration 2. We trained unique neural networks (architecture equivalent to Table~\ref{tab:nn_arch} with Conv2d $\rightarrow$ Conv1d) on sampled trajectories from each phenotype. Once trained, we extracted saliencies $\mathcal{K}^{ab}$ from each neural network and plotted them in Fig.~\ref{fig:nnResults_MC}.

        Across each phase, we found that $\mathcal{K}^{XX}$ and $\mathcal{K}^{YY}$ were unchanged as we did not change $\kappa^{XX}$ and $\kappa^{YY}$ in the phase diagram. Looking at the cross-saliencies, we found $\mathcal{K}^{XY}(x=0)$ ad $\mathcal{K}^{YX}(x=0)$ to have signs that matched our expectations given their phase. Namely, they were both negative (positive) in the segregated (integrated) phase, stemming from the sign of $\kappa^+$. In the two migration phases, they had opposite sign. In particular, $\mathcal{K}^{XY} > 0 (<0)$ and $\mathcal{K}^{YX}<0 (>0)$ when $\kappa^- = \kappa^{YX} - \kappa^{XY} < 0 (>0)$.

    \subsection{Saliency of a non-local Ising model}

        \begin{figure*}
            \centering
            \includegraphics[width=\linewidth]{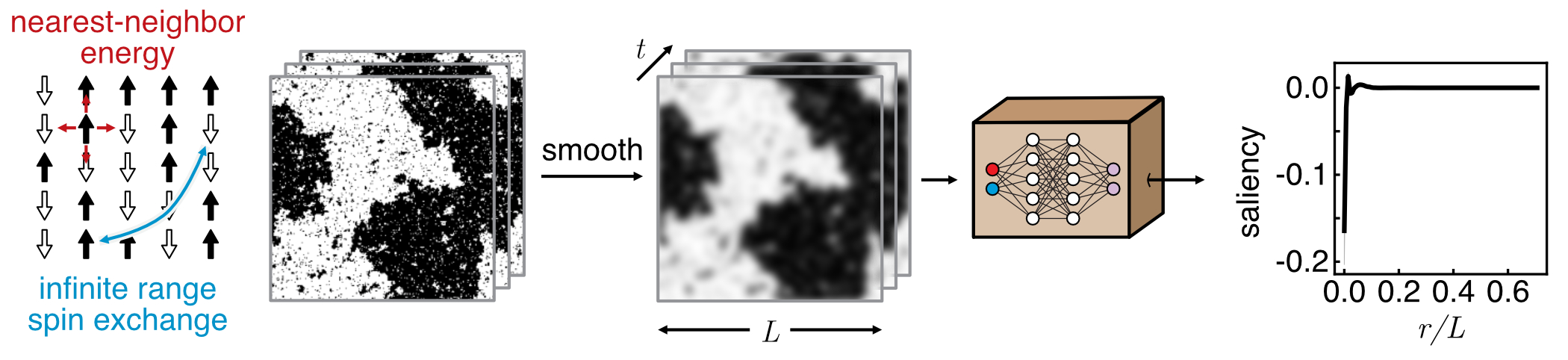}
            \caption{
                \textbf{Coarse-graining non-local motion leads to local dynamics.}
                We implement a 2D Ising model whose dynamics are given by exchanging spins. Any two spins in the lattice can be exchanged, but the energy is computed locally.
                The resulting dynamics are then smoothied, and used inputs into a convolutional neural network to predict the dynamics and compute a saliency.
                The result of the saliency is shown on the right side of the text, and shows that the NN function is highly localized around $r = 0$.
            }
            \label{fig:nonlocalIsing}
        \end{figure*}
    
        What would the saliency give if the underlying dynamics exhibit non-local displacements?
        A simple model for this is a 2D Ising model whose dynamics involve swapping any two arbitrarily chosen spins, i.e. Kawasaki dynamics with infinite range exchanges~\cite{Conti2002}.
        More precisely, consider a set of spins $\mathbf{s}$ on a 2D square lattice. The Hamiltonian of the system is given by
        $$H(\mathbf{s}) = -\sum_{\langle i, j \rangle} J s_i s_j,$$
        where the coupling strength $J > 0$ and $\sum_{\langle i , j \rangle}$ indicates a sum over nearest neighbor pairs of spins.
        The dynamics are generated by swapping any two spins $i$ and $j$. If $\mathbf{s} = (s_1, s_2, \ldots, s_i, \ldots, s_j, \ldots)$, we define the swapped state $\mathbf{s}' = (s_1, s_2, \ldots, s_j, \ldots, s_i, \ldots)$.
        The system is evolved using Glauber dynamics, where the probability of accepting a proposed spin swap is given by
        $$w(\mathbf{s}'|\mathbf{s}) = \dfrac{1}{1 + \exp[\beta ( H(\mathbf{s}') - H(\mathbf{s}) )]},$$
        where $\beta$ is an inverse temperature.
        Thus, this system has local interactions, and long-range displacements (Fig.~\ref{fig:nonlocalIsing})

        Simulating this system gives us a trajectory $\mathbf{s}(t)$.
        We smooth the spin field with a Gaussian kernel and use the smoothed fields as an input into the same neural network architecture used to predict the dynamics of the Census data.
        We then measure the saliency of the trained system, which is shown on the right-hand side of Fig.~\ref{fig:nonlocalIsing}.
        We see that the neural network learns a highly localized dynamical rule, despite the infinite-range motility of the underlying spins.

        These results show that a coarse-grained field, even with long-range mobility, can evolve according to a local set of dynamics.
        This analysis further supports our use of a local PDE when simulations residential dynamics whose microscopic origins are surely non-local.

\section{The sociohydrodynamic equation}
    Our sociohydrodynamic equation is
    \begin{equation}
        \label{eq:sociohydroSI}
        \partial_t \phi^a = \nabla \cdot \left(D^{ab}(\vphi) \nabla \phi^b - \phi^a \vac \nabla\pi^a - \Gamma^{ab}(\vphi) \nabla^3\phi^b \right) + S^a(\vphi)
    \end{equation}
    This is an equation for advection-diffusion plus growth, where the advection velocity is given by $\mathbf{v}^a = \vac \nabla \pi^a$ and we require $\Gamma^{ab} > 0$ for stability.
    In the sections below, we discuss the phenomenological origins of this model, in particular for the forms of the growth $S^a$ and utility functions $\pi^a$ considered in the main text.
    We also discuss the origin of the $\nabla^3$ term as a necessary, stabilizing term.
    Finally, we describe a microscopic, agent-based model that we explicitly coarse-grain to obtain \eqref{eq:sociohydroSI}, including all prefactors of the vacancy $\vac$.

\subsection{Growth laws for US populations}
    \begin{figure}[h]
        \centering
        \includegraphics{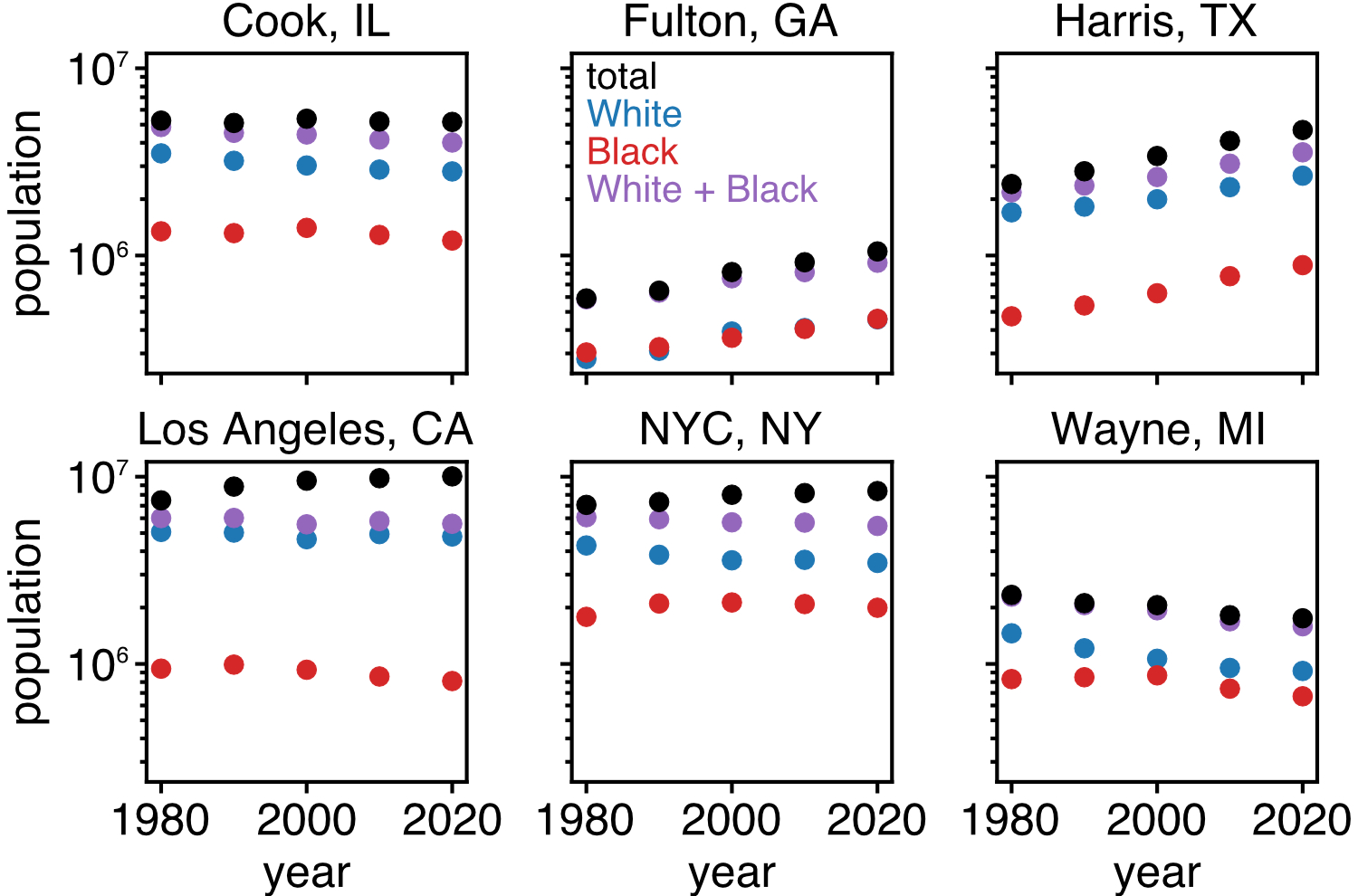}
        \caption{\textbf{Total population numbers in six areas around the US.}
        Population numbers are measured from decennial US Census data from 1980-2020, shown on a semi-log axis.
        Total, White, Black, and White + Black populations in black, blue, red, and purple, respectively.
        The growing gap between the total and White + Black population numbers indicate an increasing proportion of other races in those areas, predominantly Hispanic populations.
        The populations change roughly linearly on the semi-log axis, reflecting a roughly exponential increase in population.}
        \label{fig:total_population}
    \end{figure}

    As discussed in the main text, we do not build a model for the growth of populations in the US counties we consider (for the interested reader, we refer to Refs.~\cite{Verbavatz2020,Reia2022}).
    Instead, we fit the observed growth across counties.
    Figure~\ref{fig:total_population} shows that population growth in various US counties is roughly exponential, $S^a(\vphi) = r^a \phi^a$ with distinct coefficients $r^a$ for each county.
    In the inference and simulations performed throughout the text, we fit the coefficient $r^a$ for each county and model growth as this simple exponential law.

\subsection{Utility functions from survey questions}
\begin{figure}
    \centering
    \includegraphics[width=\textwidth]{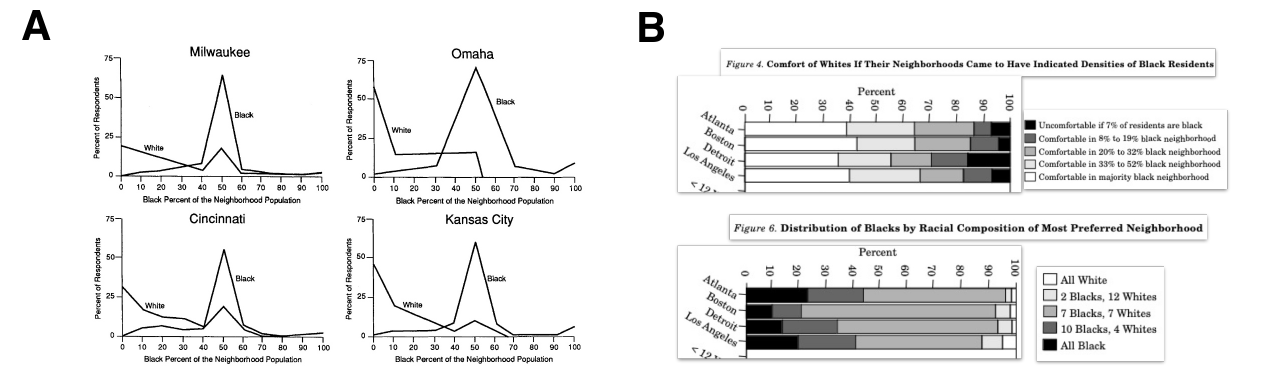}
    \caption{\textbf{Survey results for neighborhood preferences.}
    (A) Reproduction of Fig.4 of Ref.~\cite{Clark1991}. Residents across all cities were asked to \enquote{Now suppose you have been looking for a house or apartment and have found a nice place you can afford. It could be located in neighborhoods with different racial groups. What mixture of people would you prefer? Would you prefer a neighborhood that is... (combinations of 100\% white, 90\% white and 10\% black, and so on through 100\% black were read to respondents)}.
    (B) Reproduces portions of Figs 4 \& 6 from Ref.~\cite{Farley1997}, with the corresponding questions written on top of each bar plot. Both cases show the generic features found in Fig.~1C, which itself reproduced data from Ref.~\cite{Farley1993} -- neighborhood preferences of White respondents is monotonically decreasing with the proportion of Black residents, while Black respondents indicate a preference for mixed neighborhoods.}
    \label{fig:survey_results_supp}
\end{figure}

    The form of the utility functions is, as of yet, arbitrary.
    In order to gain an understanding of their possible form, we turn to sociological survey data that assesses the stated preferences of US citizens for the demographic make-up of their local neighborhood.
    In Fig.~3 of the main text, we show the results of surveys taken in Farley, 1993~\cite{Farley1993} and Clark, 2002~\cite{Clark2002}. Here, we report the precise questions asked in each survey.
    In addition, we show in Fig.~\ref{fig:survey_results_supp} that qualitatively similar results are found in two additional surveys.

    \subsubsection{Farley, 1993 \cite{Farley1993}}
        Surveyed Black respondents were given the following prompt (from Section \enquote{Residential Preferences of Blacks}, bottom of page 20 in \cite{Farley1993}):
        \begin{quote}
            We showed blacks five diagrams of neighborhoods (see figure 7). Each card pictured 15 homes with varying racial compositions ranging from all black to all white. Respondents were asked to imagine that they were searching for a home and found one they could afford. This home was shown as the center of each neighborhood. They were then asked to rank the neighborhoods from most to least attractive. The percent who said a specific neighborhood was their first or second choice is shown in figure 8.
        \end{quote}
        The resulting data are reproduced by the red markers in Fig.~3 (top).

        Surveyed White respondents were given the following prompt (from Section \enquote{Residential Preferences of Whites}, bottom of page 23 in \cite{Farley1993})
        \begin{quote}
            The residential preferences of whites were studied in a manner similar but not exactly identical to that used with blacks. Every white was presented with a series of diagrams showing neighborhoods with 15 homes (see figure 9). We asked them to imagine that they lived in an all-white neighborhood, a realistic assumption for most, and that the center home was theirs. They were then shown the second card, which indicated 1 home occupied by a black family and 14 by whites. Whites were asked how comfortable they would feel in that minimally integrated neighborhood. If they said "comfortable," they were shown cards with higher representations of blacks until they either said \enquote{uncomfortable} or came to the final card, which portrayed a majority-black neighborhood.
        \end{quote}
        The resulting proportion of respondents indicating they were comfortable with each neighborhood are reproduced by the blue markers in Fig.~3 (top).
    
    \subsubsection{Clark, 2002 \cite{Clark2002}}
        Surveyed Black respondents were asked (from page 243 of \cite{Clark2002})
        \begin{quote}
            Now I would like you to imagine that you have been looking for a house and have found a nice house you can afford. This house could be located in several different types of neighborhoods as shown on these cards. (The cards show combinations of 15 own- and other-race houses indicated by stylized drawings.) Would you look through the cards and rearrange them so that the neighborhood that is most attractive to you is on top, the next most attractive second, and so on down the line with the least attractive neighborhood on the bottom.
        \end{quote}
        The results for the proportion of respondents ranking a specific neighborhood as their first choice is reproduced by the red symbols in Fig.~3 (bottom).

        Surveyed White respondents were asked (from page 243 of \cite{Clark2002})
        \begin{quote}
            Now, I’d like you to imagine yourself in a different situation. Suppose you have been looking for a house and have found a nice one you can afford. This house could be located in several different types of neighborhoods, as shown on these cards (similar to those described above). Would you consider moving into any of these neighborhoods?
        \end{quote}
        The results for the proportion of respondents indicating that they would move into a particular neighborhood is reproduced by the blue symbols in Fig.~3 (bottom).
    
    \subsection{Origin of $\nabla^3$ term}

    \begin{figure}[h]
    \centering\includegraphics{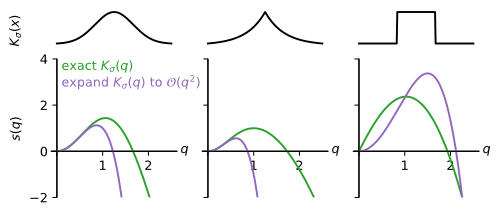}
        \caption{
            \textbf{Exact and expanded non-local interactions.}
            Three examples of the eigenfrequency $s(q)$, depending on the choice of kernel, $K_\sigma(x)$, shown at the top of each column in black. Left column is a Gaussian, $K_\sigma(x) = Z^{-1} e^{-x^2/(2\sigma^2)}$, middle column is a decaying exponential, $K_\sigma(x) = Z^{-1} e^{-|x|/\sigma}$, and right column is a top-hat function, $K_\sigma(x) = Z^{-1} H(\sigma - |x|)$, where $H(x)$ is the Heaviside Function and $Z$ is a normalization constant. 
            The green line plots $s(q)$ using the full spectrum $K_\sigma(q)$, Eq.\eqref{eq:kernel_eigenfrequency}, while the purple line plots $s(q)$ where $K_\sigma(q)$ has been expanded around $q=0$ up to $q^2$, Eq.~\eqref{eq:stability_truncated}.
            In all cases, we take $D=1$, $\phi_0 \vac_0 \pi'_0 = 4$, and $\sigma = 1$.
        }
        \label{fig:kernelStability}
    \end{figure}
    
    Our census data, once smoothed, does not exhibit fast variations at small length-scales.
    Thus, \eqref{eq:sociohydroSI} should be linearly stable at small wavelengths (high wavenumber, $q$).
    Diffusion and (local) advection alone are insufficient to ensure this.
    A natural way to regain stability at high-$q$ is to consider a non-local fitness function, such as those considered in Ref.~\cite{Zakine2024}.
    Below, we will show that this comes at the expense of a more complicated theory.
    The $\nabla^3$ term qualitatively mimics the effect of non-local fitness functions while maintaining analytical simplicity, and is therefore added to the model as a phenomenological term.

    To illustrate, we work with a simplified version of our theory in the case of only one group, denoted by the fill fraction $\phi(x, t) \in [0, 1]$. We introduce the spatially smoothed fill fraction
    \begin{equation}
        \varphi(x, t) = (K_\sigma * \phi)(x, t) = \int dy K_\sigma(x-y)\phi(y, t)
    \end{equation}
    where $K_\sigma(x)$ is a normalized convolution kernel in $L^2$ that is non-increasing in $|x|$ with a characteristic width $\sigma$. We will consider $K_\sigma(x)$ to be an even function, $K_\sigma(-x) = K_\sigma(x)$.
    Examples are Gaussian, decaying exponentials, or top hat functions (see Fig.~\ref{fig:kernelStability}). The model presented in the main text and SM considered the special case $K_\sigma(x) = \delta(x)$.
    
    The non-local interactions manifest by the utility functions becoming functions of $\varphi$ (equivalently, functionals of $\phi$), $\pi = \pi(\varphi(x, t))$. In this case, we can write the sociohydrodynamic equations as
    \begin{subequations}
    \begin{align}
            \partial_t \phi(x, t) &= -\nabla \cdot j \\
            j(x, t) &= -D \nabla \phi + \phi \vac \nabla \pi(\varphi)
    \end{align}
    \end{subequations}
    where the diffusion constant $D>0$ and vacancy fraction $\vac = 1 - \phi$. Linearizing around a uniform solution, $\phi(x) = \phi_0 + \delta \phi(x)$, 
    \begin{equation}
        \partial_t \delta \phi = -\nabla \cdot \left(-D  \nabla \delta \phi + \phi_0 \vac_0 \pi'_0 \nabla \delta \varphi \right) \overset{\mathcal{F}}{\Rightarrow} \partial_t \delta \phi = s(q) \delta \phi,
    \end{equation}
    where $\delta \varphi = K_\sigma * \delta \phi$, $\pi'_0 = (\partial \pi/\partial \varphi)|_{\varphi = \phi_0}$ is the derivative of the utility function with respect to smoothed field $\varphi$ evaluated at $\phi = \phi_0$, and the symbol $\overset{\mathcal{F}}{\Rightarrow}$ indicates moving to Fourier space. The stability of the homogeneous solution is given by the eigenfrequency $s(q)$,
    \begin{equation}
    \label{eq:kernel_eigenfrequency}
        s(q) = -q^2 \left( D - \phi_0 \vac_0 \pi'_0 K_\sigma(q) \right).
    \end{equation}
    An instability arises if $D - \phi_0 \vac_0 \pi'_0 K_\sigma(q) < 0$, or more simply if
    \begin{equation}
        \label{eq:instabilityCondition}
        D - \phi_0 \vac_0 \pi'_0 < 0.
    \end{equation}
    This simplification arises due to the fact that we considered $K_\sigma(x)$ to be normalized and monotonically decreasing and therefore its Fourier transform has a maximum $K_\sigma(q=0) = 1$. Given $K_\sigma(x) \in L^2$, we also have $K_\sigma(q) \in L^2$, meaning $|K_\sigma(q)|$ decays to zero as $q \to \infty$. This means that, even if Eq.~\eqref{eq:instabilityCondition} is satisfied at low $q$, above some $q_c$ it will no longer be satisfied and the system is stable at short wavelengths (Fig.~\ref{fig:kernelStability}).

    If we Taylor expand $K_\sigma(q)$ around $q=0$, we have
    \begin{equation}
        s(q) = -q^2 \left(D - \phi_0 \vac_0 \pi'_0 \sum_{n=0}^\infty c_{2n} q^{2n} \right),
    \end{equation}
    where $c_n = K^{(n)}(0)/n!$, and we only have even powers of $q$ due to $K_\sigma(x)$ being an even function in $x$. If we truncate the Taylor series at $q^2$, we have
    \begin{equation}
    \label{eq:stability_truncated}
        s(q) = -q^2 \left( D - \phi_0 \vac_0 \pi'_0 \right) + q^4 \left( \phi_0 \vac_0 \pi'_0 c_2 \right)
    \end{equation}
    The resulting $q^4$ term can be seen as the result of a $\nabla^3$ term in a flux $j$. This truncation will always stabilize an unstable system at large $q$, as we will now explain.

    If Eq.~\eqref{eq:instabilityCondition} holds, we must have $\pi'_0 > \frac{D}{\phi_0\vac_0} > 0$, and the $q^2$ term in Eq.~\eqref{eq:stability_truncated} becomes unstable (i.e. it has a positive coefficient).
    At large $q$, the $q^4$ term dominates and always has a negative coefficient. As $\phi_0 \vac_0 > 0$ and we are working in a regime where $\pi'_0 > 0$ to satisfy Eq.~\eqref{eq:instabilityCondition}, we are saved by the fact that $c_2 = K''_\sigma(0)/2 < 0$, because $K_\sigma(0)$ is a local maximum.

    There remains one subtlety. Suppose our linearization is stable because $\pi'_0 < 0$, thus violating Eq.~\eqref{eq:instabilityCondition}. In this case, our truncation at $\mathcal{O}(q^2)$ actually \textit{introduces} an instability at large $q$, as $\pi'_0 c_2 > 0$. In this case, we would need to truncate at a higher order of $q$, such as $\mathcal{O}(q^4)$.
    We therefore conclude that there is no consistent way to truncate the non-local interaction model to ensure stability for all kernels and fitness functions. 
    One must keep the fully non-local interactions, making the theory more complicated to study analytically.

    In light of this, we can achieve the same stabilizing effect of non-local interactions by introducing a $\nabla^3$ term in our flux with a fixed sign to ensure stability, in the spirit of Ginzburg-Landau theory.
    While quantitatively different from the non-local interactions (see Fig.~\ref{fig:kernelStability}), the effect is qualitatively similar and does not require the specification of the kernel $K_\sigma(x)$ when attempting to fit to data.
    
    \subsection{The Schelling model}
        Here we describe a set of microscopic dynamics that we will find coarse-grains to our conjectured sociohydrodynamic equations.
        We note that the ability to write the form of \eqref{eq:sociohydroSI}.
        However, having the microscopic model  gives more insight into the origin of otherwise phenomenological coefficients, especially the diffusion matrix $D^{ab}$ and surface tension $\Gamma^{ab}$.
        
        The set up amounts to a bounded-neighborhood variant of the Schelling model \cite{Schelling1971,Grauwin2009,Zhang2011,Grauwin2012,Bertin2021,Zakine2024}.
        We consider a system composed of two types of particles (agents), labeled by $(X, Y)$, distributed over a periodic lattice with $M$ sites and lattice spacing $l$. Each lattice site can contain at most $\capacity$ total particles. At a time $t$, lattice site $i$'s composition is given by the vector $\vec{n}_i(t) = (n^X_i(t), n^Y_i(t))$ giving the number of particles of both types at site $i$. The capacity constraint is $n_i^X(t) + n_i^Y(t) \leq \capacity$ for all $i$ and all $t$. We use $\un(t) \equiv \left( \vec{n}_1(t), \vec{n}_2(t), \ldots, \vec{n}_M(t) \right)$ to indicate the full configuration of the system. Dividing the components of $\un(t)$ by $\capacity$, we further define the (discrete for now) density field configuration $\uphi(t) \equiv \left( \vphi_1(t), \vphi_2(t), \ldots, \vphi_M(t) \right)$ where $\vphi_i(t)=(\phi^X_i(t), \phi^Y_i(t))$ with $\phi_i^m(t) = n_i^m(t) / \capacity$.
    
        \subsubsection{The utility function}
            An agent of type $a$ at lattice site $i$ has a measure of the \textit{utility} of the configuration in which it finds itself, $\utility_i^a(\uphi(t))$. The utility $\utility_i^a$ generically depends on the configuration and on the lattice site $i$. 
            While we do not consider utility functions that explicitly depend on the latter in this work, we note that adding this feature may be necessary for a more realistic description of human residential dynamics.
            
            As noted by Schelling himself \cite{Schelling1971}, any model for residential dynamics that is built only on neighborhood composition preferences ignores at least 2 other factors largely responsible for observed residential patterns -- discrimination (e.g. redlining) and socioeconomic inequality (e.g. income inequality).
            In either case, the end result is that certain areas in space are more or less accessible to certain groups as imposed by factors other than their own personal preference.
            These effects can be at least partially captured by adding an explicit, spatially dependent term to the utility functions which acts like an externally imposed potential function restricting  movement.
            For example, one could consider $\pi^a_i(\uphi) = \alpha i + f(\uphi)$, where $\alpha$ is a constant that determines the utility of a location $i$, in addition to a term $f(\uphi)$ that captures the individuals preferences.
            For the rest of this work, however, we only consider utilities that are functions of the system configuration, $\utility^a(\vphi_i) = \utility^a(\phi^X_i, \phi^Y_i)$.
            Finally, we note that each individual does not count itself as a neighbor in the evaluation of its utility, e.g. an individual of type $X$ calculates its utility as $\utility^X(\vphi_i) = \utility^X(\phi^X_i - \capacity^{-1}, \phi^Y_i)$.
    
        \subsubsection{The transition rates}
            An individual of type $b$ moves by jumping from site $j$ to site $k$. This changes the configuration as
            \begin{equation}
            \label{eq:allowedTransition}
                \uphi \to \uphi' = D^b_{k|j} \uphi \quad \text{such that} \quad (D^b_{k|j} \uphi)^a_i = \phi_i^a + \delta^{ab} \left( \delta_{ik} - \delta_{ij} \right) \delta \phi,
            \end{equation}
            where $\delta \phi = \capacity^{-1}$. We have defined the \textit{diffusion operator} $D^b_{k|j}$ that acts on the configuration $\uphi$ by removing an individual of type $b$ from site $j$ and adding it to site $k$.
            
            An individual makes this jump with a probability depending on a gain function, $g^a(k|j)$, which measures the variation of its utility after moving from site $j$ to site $k$.
            
            The rate we consider is similar to those used in Glauber dynamics of spin lattices \cite{Glauber1963}, as well as in other quantitative socioeconomic models  where it is known as a logit function \cite{Grauwin2009, Bouchaud2013},
            \begin{equation}
                \label{eq:singleParticleRate}
                w^a(k|j) = w^a(g^a(k | j)) = \dfrac{\tau^{-1}}{1 + \exp(-g^a(k|j) / T)},
            \end{equation}
            where $\tau$ gives a time scale for the jumps and $T$ is akin to a temperature that quantifies the uncertainty of a move decreasing the gain function.
            We note that Eq.~\ref{eq:singleParticleRate} is not the only choice possible. Generically, the transition rate should satisfy certain properties, namely $w^a(g \to \infty) \to 1$ and $w^a(g \to -\infty) = 0$. The logit transition rate, in addition of satisfying these properties, also obeys detailed balance when $g^a(k | j) = G(k) - G(j)$, \textit{i.e.} when the gain can be written as the difference of some function $G$ that only depends on the system configuration. If the latter is verified, detailed balance holds and the system will settle into a steady state given by the Boltzmann Distribution of statistical physics, $p(\uphi) \propto e^{G / T}$. This was pointed out for socioeconomic agents in \cite{Grauwin2009}.
            
            In this work, we take the gain to be given by
            \begin{equation}
            \label{eq:gain_micro}
                g^a(k|j) = g^a(\vec{\phi}_k, \vec{\phi}_j) = (1 - \alpha) \left[ \utility^a \left( \vphi_k' \right) - \utility^a(\vphi_j) \right]
                + \alpha \left[ U[\uphi'] - U[\uphi] \right]
                + \Gamma \left[ \Sigma[\uphi'] - \Sigma[\uphi] \right].
            \end{equation}
            The first term in Eq.~\eqref{eq:gain_micro} denotes the change in the moving particle's own utility before and after the move. The second term denotes the change in the global utility given by the functional $U[\uphi]$,
            \begin{equation}
                U[\uphi] = \capacity \sum_{b=X,Y} \sum_{j=1}^M \phi^b_j \pi^b(\vphi_j).
            \end{equation}
            From here, we see that the parameter $\alpha \in [0, 1]$ denotes the relative weight that a particle gives to its own utility with respect to the utility of the entire system. For this reason, we follow \cite{Grauwin2009} and call $\alpha$ an \textit{altruism parameter}. Completely selfish particles ($\alpha = 0$) only consider their own utility, while completely altruistic particles ($\alpha = 1$) only consider the utility of the global configuration. We note that the main text only considered $\alpha = 0$.
        
            The functional $\Sigma[\uphi]$ denotes a contribution to the gain that comes from spatial gradients of the density field,
            \begin{equation}
                \Sigma[\uphi] = \capacity \sum_{b = X, Y} \sum_{j=1}^M \phi_j^b \left( \partial_x^2 \uphi \right)^b_j.
            \end{equation}
            Here, $\partial_x^2 \uphi$ is a discrete Laplacian operator defined as
            \begin{equation}
                (\partial_x^2 \uphi)^b_j = \dfrac{\phi^b_{j+1} - 2\phi^b_j + \phi^b_{j-1}}{\ell^2}.
            \end{equation}
            This function indicates that particles like being at a minimum of the density field, where $\partial_x^2 \uphi > 0$. We will see that this term gives the system a surface tension that promotes spatial uniformity in the density field. Intuitively, this comes from the fact that particles are more likely to move into local minima of the density field ($\partial_x^2 \uphi > 0$) and away from local maxima of the density field ($\partial_x^2 \uphi < 0$), thereby reducing local gradients in the density field. This will prove crucial to obtain a well-behaved hydrodynamic theory.

    \subsection{Coarse-graining the Schelling model}
    Here we coarse-grain our agent-based model and show that it leads to the sociohydrodynamic equations in the main text.
    
        The stochastic evolution of the system is described by the probability distribution $P(\un, t)$, which is assumed to obey the master equation
        \begin{equation}
        \label{eq:masterEq}
            \partial_t P(\un, t) = \sum_{\un'} W(\un | \un') P(\un', t) - W(\un' | \un) P(\un, t),
        \end{equation}
        where $W(\un' | \un)$ is the transition rate from state $\un$ to state $\un'$ and $\sum_{\un'}$ is a sum over all system configurations.
    
        The allowed transitions are given by Eq.~\eqref{eq:allowedTransition}, which reduce the sum over configurations to sums over particle types and pairs of sites $jk$
        \begin{equation}
            \partial_t P(\un, t) = \sum_{b = X,Y} \sum_{jk} W(\un | D^b_{k|j} \un) P(D^b_{k|j} \un, t) - W(D^b_{k|j} \un | \un) P(\un, t).
        \end{equation}
        We can now find the dynamics for the average number of particles of type $a$ at lattice site $i$,
        \begin{subequations}
        \begin{align}
            \partial_t \langle n^a_i \rangle (t) &= \sum_{\un} n_i^a \partial_t P(\un, t)
            \\
            &= \sum_{\un} n^a_i \left( \sum_b\sum_{jk} W(\un | D^b_{k|j} \un) P(D^b_{k|j} \un, t) - W(D^b_{k|j} \un | \un) P(\un, t) \right)
            \\
            &= \sum_{\un} \sum_b \sum_{jk} n^a_i W(\un | D^b_{k|j} \un) P(D^b_{k|j} \un, t) - n^a_i W(D^b_{k|j} \un | \un) P(\un, t) \label{eq:meanDynamics_exact}.
        \end{align}
        \end{subequations}
        We wish to make the right hand side the average over some function of $n^a_i$, i.e. $\partial_t \langle n^a_i \rangle = \langle f(n^a_i) \rangle = \sum_{\un} P(\un, t) f(n^a_i)$ in order to apply the mean-field approximation, \textit{i.e.} $\partial_t \langle n_i \rangle = f(\langle n_i \rangle)$. To this end, we perform a change of variables in the first term of Eq.~\eqref{eq:meanDynamics_exact}, taking
        $$\un \to D^b_{j|k} \un = \left(D^b_{k|j} \right)^{-1} \un$$
        This change of variables leaves the sum over configurations unchanged as both $D^b_{k|j}$ and $D^b_{j|k}$ preserve particle number. This gives
        \begin{subequations}
        \begin{align}
            \partial_t \langle n_i \rangle (t) &= \sum_{\un} \sum_b  \sum_{jk} (D^b_{j|k} \un)^a_i W(D^b_{j|k} \un | \un) P(\un, t) - n_i^a W(D_{k|j}^b \un | \un) P(\un, t)
            \\
            &= \sum_{\un} P(\un, t) \sum_b \sum_{jk} (D^b_{j|k} \un)^a_i W(D^b_{j|k} \un | \un) - n_i^a W(D^b_{k|j} \un | \un)
            \\
            \label{eq:mean_field_D}
            &= \left< \sum_b \sum_{jk} (D^b_{j|k} \un)^a_i W(D^b_{j|k} \un | \un) - n_i^a W(D^b_{k|j} \un | \un) \right>.
        \end{align}
        \end{subequations}
        In the following, we take the mean-field approximation and drop the angle brackets.
    
        As written, $W(D^b_{k|j} \un | \un)$ gives the rate for \textit{some} particle of type $b$ to move from site $j$ to site $k$ in the configuration $\un$. This is not the single particle jump rate given by Eq.~\ref{eq:singleParticleRate}, $w^b(k|j)$. Rather, the two are related by
        \begin{equation}
            W(D^b_{k|j} \un | \un) = n_j^b \left( 1 - \sum_c \phi_k^c \right) w^b(k|j) (\delta_{k, j+1} + \delta_{k, j-1}).
        \end{equation}
        Intuitively, the above formula states that the rate at which a $b$ particle moves from site $j$ to site $k$ is given by the rate at which a single $b$ particle makes that transition ($w^b(k|j)$) scaled by the number of $b$ particles at $j$ ($n_j^b$) and the fraction of vacant sites at $k$ ($1 - \sum_c \phi_k^c)$. This prevents particles from entering full sites. In the following, we denote the total vacancy fraction by
            $$\vac_k = 1 - \sum_c \phi^c_k.$$
        Finally, we have also restricted motion to neighboring lattice sites, specifying that $k = j \pm 1$ with the Kronecker deltas.
    
        Plugging this decomposition of $W(D^b_{k|j} \un | \un)$ into the mean-field equation \eqref{eq:mean_field_D} and writing $(D^b_{j|k} \un)^a_i$ explicitly, we obtain
        \begin{equation}
        \begin{split}
            \partial_t n_i^a = \sum_b \sum_{jk}
            & \left(n_i^a + \delta^{ab}(\delta_{ij} - \delta_{ik}) \right) n_k^b \vac_j w^b(j|k) \left( \delta_{j, k+1} + \delta_{j, k-1} \right) \\
            & - n_i^a n_j^b \vac_k w^b(k|j) \left( \delta_{k, j+1} + \delta_{k, j-1} \right).
        \end{split}
        \end{equation}
        The first and last terms cancel as they are identical up to a relabeling of $j \leftrightarrow k$ and are summed over all $j,k$. After summing over $b$ in the middle term, we have
        \begin{equation}
            \partial_t n_i^a = \sum_{jk}  n_k^a \vac_j w^a(j|k) \left( \delta_{ij} - \delta_{ik} \right) \left( \delta_{j, k+1} + \delta_{j, k-1} \right).
        \end{equation}
        The sum over $jk$ is significantly simplified due to the two sets of $\delta$-functions and gives
        \begin{equation}
            \partial_t n_i^a = n^a_{i-1} \vac_i w^a(i|i-1) + n^a_{i+1} \vac_i w^a(i|i+1) - n_i^a \left[ \vac_{i - 1} w^a(i - 1|i) + \vac_{i + 1} w^a(i + 1|i) \right].
        \end{equation}
    
        This has the form of a conservation law,
        \begin{equation}
        \begin{aligned}
            \partial_t n_i^a &= J_{i-1}^a - J_i^a
            \\
            J_i^a &= n_i^a \vac_{i+1} w^a(i+1|i) - n_{i+1}^a \vac_i w^a(i+1|i)
        \end{aligned}
        \end{equation}
        Moving to continuous space, the lattice site $i$ is at position $x = i \ell$, where $\ell$ is the lattice spacing. Upon taking $\ell \to 0$ and dividing by $\capacity$, we can approximate the above as
        \begin{equation}
        \label{eq:hydro_conservation_j}
            \partial_t \phi^a(x, t) = j^a(x - \ell) - j^a(x) \approx -\ell \partial_x j^a(x)
        \end{equation}
    
        Now we are left to specify the particle current,
        \begin{equation}
            j^a(x) = \phi^a(x) \vac(x + \ell) w^a(x+\ell | x) - \phi^a(x + \ell) \vac(x) w^a(x | x+\ell)
        \end{equation}
        Performing a Taylor expansion for $\ell \to 0$ up to $\mathcal{O}(\ell^2)$ gives
        \begin{equation}
        \begin{aligned}[b]
            j^a(x) =& \phi^a \left(\vac + \ell \partial_x \vac \right) \left( w^a(x|x) + \ell \partial_x \vphi \cdot \nabla_{\vphi}^{(1|0)} w^a(x|x) \right)
            - (\phi^a + \ell \partial_x \phi^a) \vac \left( w^a(x|x) + \ell \partial_x \vphi \cdot \nabla_{\vphi}^{(0|1)} w^a(x|x) \right) + \mathcal{O}(\ell^2)
            \\
            \label{eq:current_first_order}
            =& \ell \left[ \left(\phi^a \partial_x \vac - \vac \partial_x \phi^a \right) w^a(x|x) + \phi^a \vac \partial_x \vphi \cdot \left( \nabla_{\vphi}^{(1|0)} w^a(x|x) - \nabla_{\vphi}^{(0|1)} w^a(x|x) \right) \right] + \mathcal{O}(\ell^2),
        \end{aligned}
        \end{equation}
        where we have canceled all terms that do not depend on $\ell$. Furthermore, we have used the notation $\partial_x \vec{\phi} = \left( \partial_x \phi^X, \partial_x \phi^Y \right)$,
        \begin{equation}
            \nabla_{\vphi}^{(\mu|\nu)} w^a(y|x) =
            \begin{pmatrix}
                \partial_{\phi^X}^{(\mu|\nu)} w^a(y|x) \\
                \partial_{\phi^Y}^{(\mu|\nu)} w^a(y|x)
            \end{pmatrix}
        \end{equation}
        and
        \begin{equation}
            \partial_{\phi^b}^{(\mu|\nu)}w^a(y|x) = \dfrac{\partial^{(\mu + \nu)} w^a(y|x)}{\partial \phi^b(y)^\mu \partial \phi^b(x)^\nu}.
        \end{equation}
        Finally, we take $f(x|x) = \lim_{y, z \to x} f(y|z)$ for any function $f(y|z)$.
    
        We are left to evaluate $w^a(x|x)$ and its derivatives. We begin with the former, which amounts to calculating $g^a(x|x)$. Recalling Eq.~\ref{eq:gain_micro}, we rewrite it in continuous space variables as
        \begin{equation}
        \begin{aligned}
        \label{eq:gain_continuous}
            g^a(y|x) = (1 - \alpha) &\left( \pi^a(\vphi(y) + \delta \phi \hat{e}_a)  - \pi^a(\vphi(x)) \right) 
            \\
            + \alpha &\left( U[D^a(y|x) \vec{\phi}(x')] - U[\vec{\phi}(x')] \right)
            \\
            + \Gamma &\left( \Sigma[ D^a(y|x) \vec{\phi}(x')] - \Sigma[\vec{\phi}(x')] \right),
        \end{aligned}
        \end{equation}
        where we have introduced the notation $\hat{e}_a$ which is a vector with one in the $a$ component, and zero otherwise. This is simply a convenient way to specify that the fraction only changes in the $a$ species when considering $g^a$. Furthermore, there is a dummy index $x'$ in the arguments of $U$ and $\Sigma$, reflecting the fact that these functionals will be integrated over $x'$.
    
        
        The continuous version of the diffusion operator takes the form
        $$D^a(y|x) \vec{\phi}(x') = \phi^b(x') + \delta \phi \delta^{ab} \left( \delta(x'-y) - \delta(x' - x) \right)$$
    
        The second term of the gain is given by
        \begin{equation}
        \begin{aligned}[b]
            U[D^a(y|x) \vec{\phi}(x')] - U[\vec{\phi}(x')] =& \capacity \sum_{b=X,Y} \int dx'  \left[  \left( D^a(y|x) \vec{\phi}(x') \right) \pi^b \left( D^a(y|x) \vec{\phi}(x') \right) - \phi^b(x') \pi^b(\vec{\phi}(x')) \right] \\
            =& \capacity \sum_b \left( \phi^b(y) + \delta \phi \delta^{ab} \right) \pi^b\left( \vec{\phi}(y) + \delta \phi \hat{e}^a \right) - \phi^b(y) \pi^b(\vec{\phi}(y))
            \\
            &+ \left( \phi^b(x) - \delta \phi \delta^{ab} \right) \pi^b\left( \vec{\phi}(x) - \delta \phi \hat{e}^a \right) - \phi^b(x) \pi^b(\vec{\phi}(x)).
        \end{aligned}
        \end{equation}
    
        A similar calculation can be done for the third term proportional to $\Gamma$. With great care taken to keep track of all the correct indices, we only report the result,
        \begin{equation}
        \begin{aligned}
            \Sigma[D^a(y|x) \vec{\phi}(x')] - \Sigma[\vec{\phi}(x)] &= \partial_x^2 \phi^a(y) - \partial_x^2 \phi^a(x) - 3 \delta \phi.
        \end{aligned}
        \end{equation}
    
        Combining the above equations, we arrive at
        \begin{equation}
            g^a(x|x) \approx \delta \phi \left[ (1 - \alpha) \dfrac{\partial \pi^a}{\partial \phi^a} + \alpha \dfrac{\partial^2 u}{(\partial \phi^a)^2} - 3 \right]
        \end{equation}
        where the Laplacian terms in $\Sigma$ cancelled when $y = x$. In the limit where $\delta \phi \to 0$, this term goes to zero, making
        \begin{equation}
        \label{eq:equal_position_jump_rate}
            \lim_{\delta \phi \to 0} w^a(x|x) = w^a(g^a=0) = \dfrac{\tau^{-1}}{2}.
        \end{equation}
    We now turn to computing the derivatives of the rate functions,
        \begin{equation}
        \label{eq:derivative_jump_rate}
            \partial_{\phi^b}^{(1|0)}w^a(y|x) = \dfrac{\partial w^a(y|x)}{\partial \phi^b(y)} = \dfrac{d w^a(g^a)}{d g^a} \dfrac{\partial g^a(\vec{\phi}(y), \vec{\phi}(x))}{\partial \phi^b(y)}.
        \end{equation}
        The first factor in the above equation is given by
        \begin{equation}
            \dfrac{d w^a}{d g^a} = \dfrac{\tau^{-1}}{(1 + e^{-g^a/T})^2} \dfrac{e^{-g^a/T}}{T},
        \end{equation}
        and in the limit of vanishing $\delta \phi$ it reduces to
        \begin{equation}
        \dfrac{d w^a}{d g^a} \underset{\delta \phi \to 0}{\to} \dfrac{\tau^{-1}}{4T}.
        \end{equation}
        To close \eqref{eq:derivative_jump_rate}, we need to compute the derivatives of the gain $g^a$ with respect to the fields using expression~\eqref{eq:gain_continuous}. 
        Substiting the final form of~\eqref{eq:derivative_jump_rate} into \eqref{eq:current_first_order} will then give the expression of the hydrodynamic current.
        We first consider the case $\Gamma=0$ and will discuss how to include the $\Gamma$-dependent contributions to the current later. 
        Retaining only the terms proportional to $\alpha$ and $(1-\alpha)$ in Eq.~\ref{eq:gain_continuous}, we derive them with respect to $\phi^b(y)$ before taking the limit $\delta\phi\to 0$ to obtain
        \begin{equation}
        \label{eq:derivative_phi_y}
            \dfrac{\partial g^a(y|x)}{\partial \phi^b(y)}\Big|_{y = x} \underset{\delta \phi \to 0}{\to} (1-\alpha) \dfrac{\partial \pi^a}{\partial \phi^b} + \alpha \dfrac{\partial^2 u}{\partial \phi^a \partial \phi^b},
        \end{equation}
        where $u(x, t) = \phi^X \pi^X(\vec{\phi}(x))+ \phi^Y \pi^Y(\vec{\phi}(x))$.
        Performing a similar computation, we also obtain
        \begin{equation}
            \lim_{\delta\phi\to 0}\dfrac{\partial g^a(y|x)}{\partial \phi^b(x)} \Big|_{y = x} = -\lim_{\delta\phi\to 0} \dfrac{\partial g^a(y|x)}{\partial \phi^b(y)} \Big|_{y = x}.
        \end{equation}
        Inserting \eqref{eq:derivative_phi_y} into \eqref{eq:derivative_jump_rate}, we obtain the derivative of the jump rate as 
        \begin{equation}
        \label{eq:field_derivatives_rates}
        -\partial_{\phi^b}^{(0|1)}w^a(x|x)=\partial_{\phi^b}^{(1|0)}w^a(x|x) = \dfrac{\tau^{-1}}{4T}\left((1-\alpha) \dfrac{\partial \pi^a}{\partial \phi^b} + \alpha \dfrac{\partial^2 u}{\partial \phi^a \partial \phi^b}\right)\;.
        \end{equation}
        We can now inject \eqref{eq:field_derivatives_rates} and \eqref{eq:equal_position_jump_rate} in expression \eqref{eq:current_first_order} to obtain the particle current $j^a_0$ in the $\Gamma=0$ case as
        \begin{equation}
        \label{eq:current_gamma_0}
        \begin{aligned}[b]
            j^a_0(x) &= \dfrac{\ell \tau^{-1}}{2} \left[ (\phi^a \partial_x \vac - \vac \partial_x \phi^a) + T^{-1} \phi^a \vac  \left( (1-\alpha) \partial_x \pi^a + \alpha \partial_x \dfrac{\partial u}{\partial \phi^a} \right) \right]\;.
        \end{aligned}
        \end{equation}
        
        When $\Gamma \neq 0$, additional contributions have to be taken into account in \eqref{eq:current_gamma_0}.
        Instead of deriving them by using expression \eqref{eq:derivative_jump_rate} together with \eqref{eq:gain_continuous}, which is not a straightforward derivation due to the presence of both functional and spatial gradients, we will use arguments drawn from equilibrium physics. 
        First, we note that the term proportional to $\Gamma$ in \eqref{eq:gain_continuous} is a purely relaxational, equilibrium contribution to the gain. Indeed, it is written as a difference of a single function evaluated before and after the microscopic jump. Having a gain function with this property necessarily gives rise to a Boltzmann-like steady-state distribution with a well-defined potential function $F[\vec{\phi}(x)]$. In the case of the $\Sigma$ functional, which penalizes gradients, this potential function is given by $F[\vec{\phi}(x)]= \int dx \ \Gamma^a \frac{|\partial_x \phi^a|^2}{2}$. Taking into account volume exclusions, the contribution $j^a_{\Gamma}$ to the current due to the gradient-penalizing terms is given by the equilibrium formula $j^a_{\Gamma} = M^{ab}(\vec{\phi}) \partial_x \delta F/\delta \phi^b$, with a fraction-dependent mobility $M^{ab}(\vec{\phi}) = \delta^{ab} \phi^b \vac$. Inserting the expression of $F[\vec{\phi}(x)]$ in the latter formula for $j^a_{\Gamma}$ gives
        \begin{equation}
            j^a_{\Gamma}=\Gamma^a \phi^a \vac \partial_x^3 \phi^a.
        \end{equation}
        Adding the above contribution to \eqref{eq:current_gamma_0} gives us the final form of the current as
        \begin{equation}
        \begin{aligned}[b]
            j^a(x)=j^a_0(x)+j^a_{\Gamma}(x) &= \dfrac{\ell \tau^{-1}}{2} \left[ (\phi^a \partial_x \vac - \vac \partial_x \phi^a) + T^{-1} \phi^a \vac  \left( (1-\alpha) \partial_x \pi^a + \alpha \partial_x \dfrac{\partial u}{\partial \phi^a} + \Gamma^a \partial_x^3 \phi^a \right) \right]\;.
        \end{aligned}
        \end{equation}
    
        We now make an assumption about the rate of jumping $\tau$: we assume that it follows an Eyring equation \cite{Eyring1935} with $\tau^{-1} \propto T$.
        The latter indicates that reactions happen faster as the temperature increases. In particular, we assume $\tau^{-1} = 2 T \tau_0^{-1}$, where $\tau_0$ is a time-scale.
        In this case, the current becomes
        \begin{equation}
        \label{eq:current_final}
            j^a(x) = \ell \tau_0^{-1} \left[T \left( \phi^a \partial_x \vac - \vac \partial_x \phi^a\right) + \phi^a \vac \left( (1- \alpha) \partial_x \pi^a + \alpha \dfrac{\partial u}{\partial \phi^a} + \Gamma^a \partial_x^3 \phi^a \right) \right].
        \end{equation}
    Putting  \eqref{eq:current_final} into \eqref{eq:hydro_conservation_j} and setting $\ell = 1$, and $\tau_0^{-1} = 1$, we arrive at our (non-dimensionalized) sociohydrodynamics equation
        \begin{equation}
        \label{eq:sociohydro}
        \begin{aligned}
            \partial_t \phi^a(x, t) &= T \left(\vac \partial_x^2 \phi^a - \phi^a \partial_x^2 \vac \right) -  \partial_x \left( \phi^a\vac \left[ (1-\alpha) \partial_x \utility^a + \alpha \partial_x \dfrac{\partial u}{\partial \phi^a} + \Gamma^a \partial_x^3 \phi^a \right] \right)
            \\
            &= -\partial_x \left( \sum_b \mat{D}^{ab}(\vphi) \partial_x \phi^b + \Gamma^{ab}(\vphi) \partial_x^3 \phi^a \right)
        \end{aligned}
        \end{equation}
        where $\mat{D}^{ab}(\vphi)$ are the elements of the density dependent diffusion matrix, written for 2 groups explicitly as
        \begin{equation}
            \mat{D}(\vphi) =
            -T
            \begin{pmatrix}
                1 - \phi^Y & \phi^X \\
                \phi^Y & 1 - \phi^X
            \end{pmatrix}
            +
            \vac
            \begin{pmatrix}
                \phi^X \left( (1 - \alpha) \dfrac{\partial \utility^X}{\partial \phi^X} + \alpha \dfrac{\partial^2 u}{(\partial \phi^X)^2} \right)
                &
                \phi^X \left( (1 - \alpha) \dfrac{\partial \utility^X}{\partial \phi^Y} + \alpha \dfrac{\partial^2 u}{\partial \phi^X \partial \phi^Y} \right)
                \\
                \phi^Y \left((1 - \alpha) \dfrac{\partial \utility^Y}{\partial \phi^X} + \alpha \dfrac{\partial^2 u}{\partial \phi^X \partial \phi^Y} \right)
                &
                \phi^Y \left((1 - \alpha) \dfrac{\partial \utility^Y}{\partial \phi^Y} + \alpha \dfrac{\partial^2 u}{(\partial \phi^Y)^2} \right)
            \end{pmatrix}.
        \end{equation}
        The fourth-order derivative has coefficients $\Gamma^{ab} = \sum_b \delta^{ab} \Gamma^b \phi^b \vac$, which are components of a diagonal matrix given our assumptions on the underlying dynamics.
    
    
        Generically, the diffusion matrix can be written as
        \begin{equation}
            \mat{D}(\vphi) =
            \begin{pmatrix}
                D^{XX} & D^{XY} \\
                D^{YX} & D^{YY}
            \end{pmatrix}
            =
            \begin{pmatrix}
                D^{XX} & D_+ - D_- \\
                D_+ + D_- & D^{YY}
            \end{pmatrix}
        \end{equation}
        In terms of $D_\pm$, the off-diagonal terms are given by
        \begin{equation}
        \label{eq:offdiag_diffusion_coefficient}
            D_\pm = \dfrac{D^{YX} \pm D^{XY}}{2} =
            -T \dfrac{\phi^Y \pm \phi^X}{2}
            +
            \dfrac{\vac}{2}
            \left[
                (1 - \alpha)
                \left(
                    \phi^Y \dfrac{\partial \utility^Y}{\partial \phi^X} \pm \phi^X \dfrac{\partial \utility^X}{\partial \phi^Y}
                \right)
                +
                \alpha (\phi^Y \pm \phi^X) \dfrac{\partial^2 u}{\partial \phi^X \partial \phi^Y} 
            \right]
            %
        \end{equation}
        From \eqref{eq:offdiag_diffusion_coefficient}, we see that $D_-$ measures the breaking of the compatibility condition. This becomes especially clear when and $\alpha = 0$, giving
        $$
        D_- = \dfrac{\vac}{2} \left( \dfrac{\partial(\phi^Y\pi^Y)}{\partial \phi^X} - \dfrac{\partial (\phi^X \pi^X)}{\partial \phi^Y} \right)
        $$

\section{Numerical methods}
    \subsection{Linear regression}
        Here, we give details on the regression used to infer the parameters of our equation of motion.
        For convenience, we reproduce our equation of motion here with a generic, quadratic utility function.
        \begin{equation}
        \label{eq:sociohydroRegression}
        \begin{aligned}
            \partial_t \phi^a &= -\nabla\cdot \left(-T^a(\vac \nabla \phi^a - \phi^a \nabla \vac) + \phi^a \vac \left( \nabla \pi^a + \Gamma^a \nabla^3\phi^a\right)\right) + S^a(\vphi) \\
            \pi^a(\vphi) &= \sum_{j,k = \lbrace W, B \rbrace} \kappa^{ab} \phi^b + \nu^{abc}\phi^b \phi^k.
        \end{aligned}
        \end{equation}
        We choose the quadratic utility to reflect the non-linearities found in reported preferences from social surveys.
        With this, we have 7 parameters to fit for the flux of each field $\phi^a$ -- $(T^a, \Gamma^a, \kappa^{aa}, \kappa^{ab}, \nu^{aaa}, \nu^{aab}, \nu^{abb})$. Here, $\nu^{aab}$ is a stand-in for $(\nu^{aab} + \nu^{aba})$ as they both multiply the same factor of $\phi^a \phi^b$ with $a \neq b$.

\begin{figure}[h]
            \centering
            \includegraphics{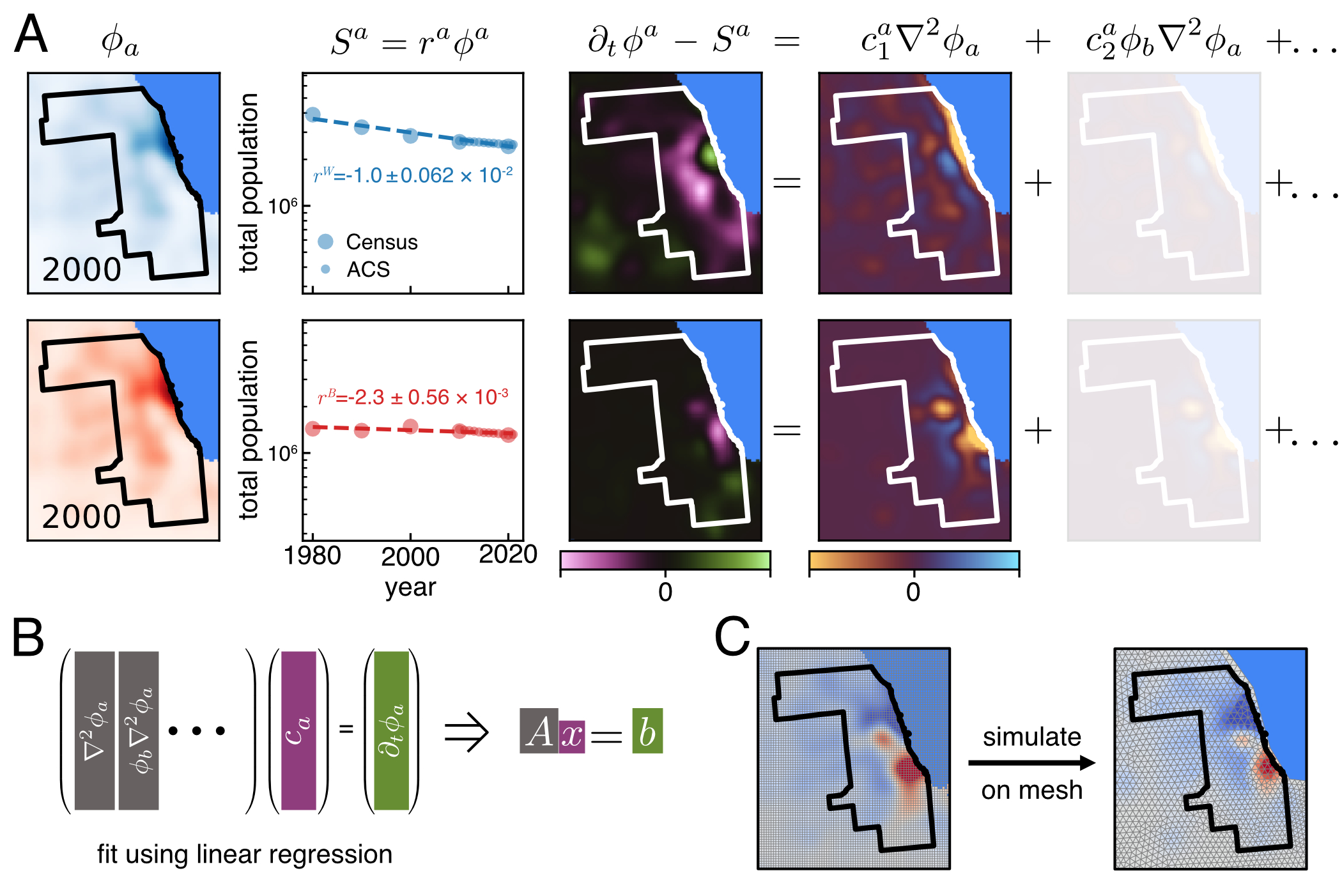}
            \caption{
            \textbf{Inference pipeline.}
            (A) Starting from census data interpolated onto a square mesh, we first fit the total population over time to an exponential function, giving our growth term $S^a$ for each population $a$.
            We then estimate time-derivatives and spatial gradients using a Savitzky-Golay filter.
            We group each of the spatial derivative terms into terms that are multiplied by individual coefficients in the flux of Eq.~\eqref{eq:sociohydroSI}, and equate them to $\partial_t \phi^a - S^a$.
            (B) Each such term is one feature of our linear regression task, i.e. it forms one column of a matrix $A$ that when multiplied by a vector $x$ of coefficients, should give the time derivatives, collected in the vector $b$.
            (C) After linear regression, we interpolate our system onto a triangular mesh and numerically solve Eq.~\eqref{eq:sociohydroSI} using a finite-volume method~\cite{Guyer2009}.
            }
            \label{fig:inferencePipeline}
        \end{figure}

        To accomplish our inference task, we first subtract $S^a$ from both sides of Eq.~\eqref{eq:sociohydroRegression} and then use ordinary least squares regression on $\partial_t \phi^a - S^a$ to estimate the remaining coefficients in the flux of Eq.~\eqref{eq:sociohydroRegression}, see Fig.~\ref{fig:inferencePipeline}.
        In order to ensure stability of the solutions, we restrict the terms $T^a$ and $\Gamma^a$ to be positive.
        After inferring the coefficients, we then solve our PDE on the county geometries starting from the population distributions in 1980.

    \subsection{Solving PDEs on counties}
        Here, we give further details about the methods used to solve our equations over individual counties.   
        We create our mesh and solve our equations of motion in Python using \texttt{FiPy}, which implements a finite-volume method~\cite{Guyer2009}.

        In order to minimize the effects of boundary conditions, we simulate dynamics over a region $1.5\times$ the size of each individual county.
        We simplify the external boundaries by buffering by 1 km and simplifying the boundary to have line segments of a minimum of 1 km using the \texttt{geopandas} functions \texttt{buffer} and \texttt{simplify}, respectively.
        This process is especially helpful to eliminate small-scale details along coasts. 
        We only simulate the largest contiguous geographic area left after this process, mostly to avoid needing to consider the dynamics between islands and the mainland (for example, Catalina Island off the coast of Los Angeles).
        We note that this process will often merge areas separated by rivers (e.g. the island of Manhattan with the boroughs of Queens and Brooklyn in New York).
        This allows us to approximate dynamics between areas that are physically separated but connected by bridges and tunnels.

        We use \texttt{FiPy}'s interface with \texttt{gmsh} to create a triangular mesh within this simplified boundary, with a mean mesh area of 3 km$^2$.
        We then interpolate the population densities over time to the generated mesh using areal-weighted interpolation (see Fig.~\ref{fig:interpolation}) and convert the densities to fill fractions as discussed above.
        Finally, we smooth fill fractions using a Gaussian filter with standard deviation of 3 km.
        We then solve the sociohydrodynamic equations over the mesh while enforcing no-flux boundary conditions.

        While we simulate areas larger than our counties of interest, we evaluate errors only within each county.
        Our errors are measured using the population densities, as
        $$\mathrm{mse} = \left\langle \sum_{a = \lbrace W, B \rbrace} \sum_\mathrm{r} \left( \rho^a _\mathrm{sociohydro} (\mathbf{r}, 2020) - \rho^a_\mathrm{census}(\mathbf{r}, 2020) \right)^2 \right\rangle$$

    \section{Linear Stability Analysis}
        We now make predictions about the onset of patterns and their subsequent dynamics using linear stability analysis. We consider perturbing around a spatially uniform state densities, $\vphi(x, t) = \vphi_0 + \delta \vphi(x, t)$, where the components $\phi_0^a$ satisfy $\phi_0^a > 0$ and $\sum_c \phi_0^c \leq 1$. Writing the perturbation as a plane wave, $\delta \vphi(x, t) = e^{st - iqx} \delta \vphi_0 $, where $s = \sigma + i \omega$ is a complex number describing the growth rate $\sigma$ and oscillation frequency $\omega$ of the perturbation with wavenumber $q$, our equations of motion reduce to the eigenvalue problem
        \begin{equation}
            s \ \delta \phi_0^a = J^{ab}(\vphi_0) \delta \phi_0^b
        \end{equation}
        with the Jacobian matrix with elements
        \begin{equation}
        \label{eq:jacobian}
            J^{ab}(\vphi_0) = q^2 D^{ab}(\vphi_0) - q^4 \Gamma^{ab}(\vphi_0),
        \end{equation}
        Given $\Gamma^{ab}$ is a diagonal matrix with positive elements (which we will always assume to be true), the $q^4$ term acts to stabilize patterns at large wavenumber $q$, or short wavelengths.
        The system's stability is entirely determined by the eigenvalues $s$ of the Jacobian matrix $\mat{J}$. 
        In the case with two groups $X$ and $Y$,
        \begin{equation}
            \mat{J} =
            \begin{pmatrix}
                J^{XX} & J^{XY} \\
                J^{YX} &  J^{YY}\\
            \end{pmatrix}
            =
            \begin{pmatrix}
                J^{XX} & (J_+ - J_-)/2 \\
                (J_+ + J_-)/2 &  J^{YY}\\
            \end{pmatrix}.
        \end{equation}
        The eigenvalues take the usual form for $2 \times 2$ matrices,
        \begin{equation}
            s_\pm = \dfrac{\Tr{\mat{J}}}{2} \pm \sqrt{\left(\dfrac{\Tr{\mat{J}}}{2} \right)^2 - \det{\mat{J}}} = \dfrac{1}{2} \left( J^{XX} + J^{YY} \pm \sqrt{(J^{XX} - J^{YY})^2 + J_+^2 - J_-^2} \right).
        \end{equation}
        From here, we see that $\omega = \mathrm{Im}(s) \neq 0$ only when $J_-^2 > J_+^2 + (J^{XX}-J^{YY})^2$.
        As the $q^4$ term in \eqref{eq:jacobian} is diagonal, this condition is controlled entirely by $\mat{D}$.

\section{Linear utilities}

\begin{figure}[t!]
    \centering
    \includegraphics[width=\textwidth]{figures/model_phases_SI.jpg}
    \caption{\textbf{Amplitude, segregation, and velocity.} 
    (A) Amplitude of patterns measured using Eq.~\ref{eq:pattern_amplitude} in the $(D_+, D_0)$ plane for $D_- = 0.01$.
    (B) Shows the quantification of the segregation using Eq.~\ref{eq:segregation_index}.
    (C) migration speed of the pattern as defined in \eqref{eq:pattern_velocity}. All these cases begin with uniform initial conditions.
    (D-F) Segregation index for forward pass (D), backward pass (E), and their difference (F) using annealing protocol described in Sec.~\ref{sec:linear_utilities}, starting from a maximum $\kappa_+ = 1.52$ and a minimum $\kappa_+ = -0.88$, with a step of $\Delta \kappa_+ = 0.16$.
    (G-I) Wave velocity for forward pass (G), backward pass (H), and their difference (I) using annealing protocol described in Sec.~\ref{sec:linear_utilities}, starting from a maximum $\kappa_+ = 1.52$ and a minimum $\kappa_+ = -0.88$, with a step of $\Delta \kappa_+ = 0.16$.}
    \label{fig:linear_utility_phases}
\end{figure}
\label{sec:linear_utilities}
    Here, we will specialize to the particularly simple class of utilities that are linear in the densities. Specifically, we will consider
    \begin{equation}
    \label{eq:utilities}
        \utility^a(\vphi) = \sum_{b = X, Y} \kappa^{ab} \phi^b 
    \end{equation}
    where $\kappa^{ab}$ are constants. In all of the following, we set $\kappa^{XX} = \kappa^{YY} = 1$, thus measuring the utilities in units of the self-utilities. As described in the methods, we also define $\kappa^\pm = \kappa^{YX} \pm \kappa^{XY}$.
    
    With this choice of utilities, the equations of motion then become
    \begin{equation}
    \label{eq:sociohydro_linearUtility}
    \begin{split}
        \partial_t \phi^X =& -\partial_x \left[ -T \left( (1 - \phi^Y) \partial_x \phi^X + \phi^X \partial_x \phi^Y \right) + \phi^X \vac \partial_x \left( (1 + \alpha) \phi^X + \gamma_- \phi^Y + \Gamma^X \partial_x^2 \phi^X \right) \right]
        \\
        \partial_t \phi^Y =& -\partial_x \left[ -T \left( (1 - \phi^X) \partial_x \phi^Y + \phi^Y \partial_x \phi^X \right) + \phi^Y \vac \partial_x \left( \gamma_+ \phi^X + (1 + \alpha) \phi^Y + \Gamma^Y \partial_x^2 \phi^Y \right) \right],
    \end{split}
    \end{equation}
    where we have defined
    \begin{equation}
        \gamma_\pm = \dfrac{\kappa^+ (1 +\alpha) \pm \kappa^- (1-\alpha)}{2}.
    \end{equation}


    Considering the case $\Gamma^X = \Gamma^Y$, then stability From \eqref{eq:sociohydro_linearUtility}, we extract the diffusion matrix as
    \begin{equation}
        \mat{D}(\vphi) = -T
        \begin{pmatrix}
            1 - \phi^Y & \phi^X \\
            \phi^Y & 1 - \phi^X
        \end{pmatrix}
        + \vac
        \begin{pmatrix}
            \phi^X (1 + \alpha) & \phi^X \gamma_-
            \\
            \phi^Y \gamma_+ & \phi^Y (1 + \alpha)
        \end{pmatrix}.
    \end{equation}
    Considering the case of equal densities, $\phi^X = \phi^Y = \phi$ where $0 \leq \phi \leq 0.5$, the diffusion matrix becomes
    \begin{equation}
    \mat{D}(\vphi) =
        \begin{pmatrix}
            D_0 & D_+ - D_-
            \\
            D_+ + D_- & D_0
        \end{pmatrix},
    \end{equation}
    where $D_0$, $D_{+}$ and $D_{-}$ are given by
    \begin{align}
        D_0 &= -T(1 - \phi) + \phi(1 - 2\phi) (1 + \alpha)
        &
        D_+ &= - \dfrac{\phi \left(2T - \kappa^+ (1 + \alpha) (1 - 2\phi) \right)}{2}
        &
        D_- &= \dfrac{\phi(1 - 2\phi) (1 - \alpha)}{2} \kappa^- \;.
    \end{align}
    We readily compute the eigenvalues of $\mat{D}$ as
    \begin{equation}
        \lambda_D^\pm = D_0 \pm \sqrt{D_+^2 - D_-^2},
    \end{equation}
    as well as its eigenvectors, which are given by
    \begin{equation}
        \vec{v}^\pm_D =
        \begin{pmatrix}
            \dfrac{\pm \sqrt{D_+^2 - D_-^2}}{D_+ + D_-}
            \\
            1
        \end{pmatrix}\;.
    \end{equation}
    $\lambda_D^{\pm}$ become imaginary when
    \begin{equation}
        |D_-| > |D_+| \Rightarrow |\kappa^-| > \pm \dfrac{2T - \kappa^+ (1 + \alpha) (1 - 2\phi)}{(1 - 2\phi)(1 - \alpha)}.
    \end{equation}
    Furthermore, we expect to find pattern formation when $\mathrm{Re}(\lambda_D^\pm) > 0$. This occurs when
    \begin{equation}
        D_0 \geq \pm \mathrm{Re}\left[ D_+^2 - D_-^2 \right].
    \end{equation}

    Fig.~\ref{fig:linear_utility_phases}A shows the amplitude of patterns found, measured using
    \begin{equation}
    \label{eq:pattern_amplitude}
        \mathcal{A}[\phi(x)] = \dfrac{1}{L} \int_0^L dx \ \sqrt{ \left(\phi(x) - \langle \phi \rangle \right)^2}.
    \end{equation}
    We see that the onset of patterns, given by $\mathcal{A} > 0$, agrees well with the linear stability prediction shown with the white dashed line.

    We measure the velocity of the resulting patterns using~\cite{Brauns2023,FrohoffHuelsmann2023b}
    \begin{equation}
    \label{eq:pattern_velocity}
        v = -\dfrac{\int_0^L dx \ \partial_t \phi \partial_x \phi}{\int_0^L dx \ \left( \partial_x \phi \right)^2}
    \end{equation}

    \begin{figure}
        \centering
        \includegraphics{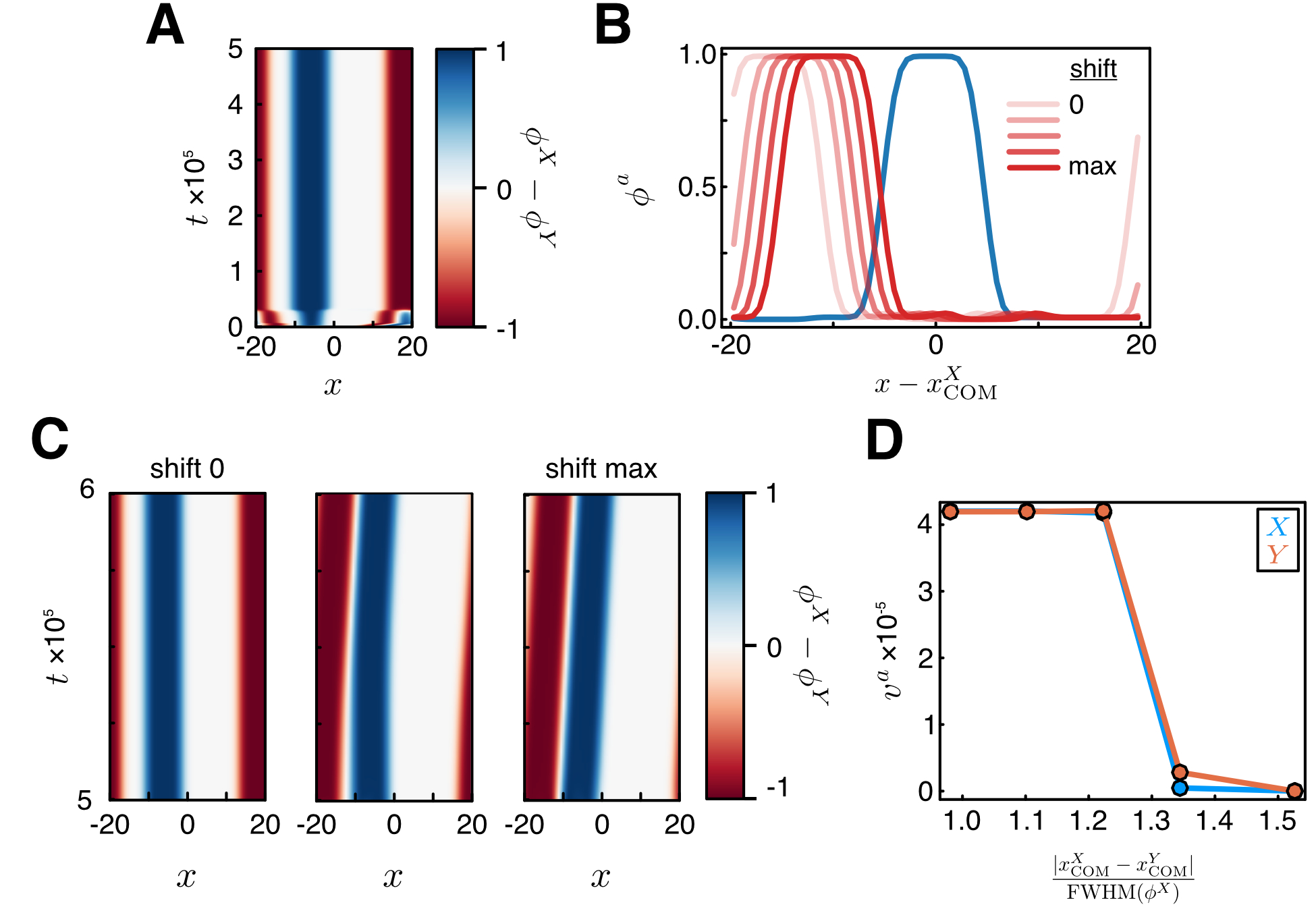}
        \caption{\textbf{Migration speed depends on overlap between populations.}
        (A) Result of simulation using linear utilities, $\kappa_+ = 0.5$, $\kappa_- = 0.2$. The wave velocity after $t = t_f = 5 \times 10^5$ is zero.
        (B) Illustration of \enquote{shifting} the final distribution, $\phi^Y(t_f)$ (red). Shifts are done in units of the lattice spacing, $\Delta x = 0.625$, with a maximum shift of $10 \Delta x$.
        (C) Results of restarting simulations after shifting $\phi^Y$ by different amounts.
        (D) Measurement of wave velocity after $t = 10^5$ of restarted dynamics as a function of the shift, showing that the distributions find a unique velocity once they overlap enough.}
        \label{fig:overlap}
    \end{figure}

    \subsection{Phase space criteria.}
    We define the different phases using the following criteria:
    \begin{itemize}
        \item \textbf{Segregation:} $\mathcal{S}_H > 0.7$
        \item \textbf{Migration:} $v^a \geq 10^{-9}$
        \item \textbf{Integration:} $\mathcal{S}_H < 0.14$
    \end{itemize}
    
    In order to find regions of coexistence, we perform an annealing procedure, where we change the values of parameters cyclically (Fig.~\ref{fig:linear_utility_phases}D-I). We run simulations for a fixed value of $\kappa_-$ and a maximal value of $\kappa_+$. After the system has reaches a steady state, we decrease $\kappa_+$ by a small amount until we reach a minimum $\kappa_+$. This constitutes a \enquote{backward} sweep. We then reverse this process, increasing $\kappa_+$ until we reach its initial value. This constitutes a \enquote{forward} sweep. The coexistence region was found by $\mathcal{S}_H^\mathrm{forward} - \mathcal{S}_H^\mathrm{backward} \geq 0.07$ 

    \begin{figure}
        \centering
        \includegraphics[width=0.75\linewidth]{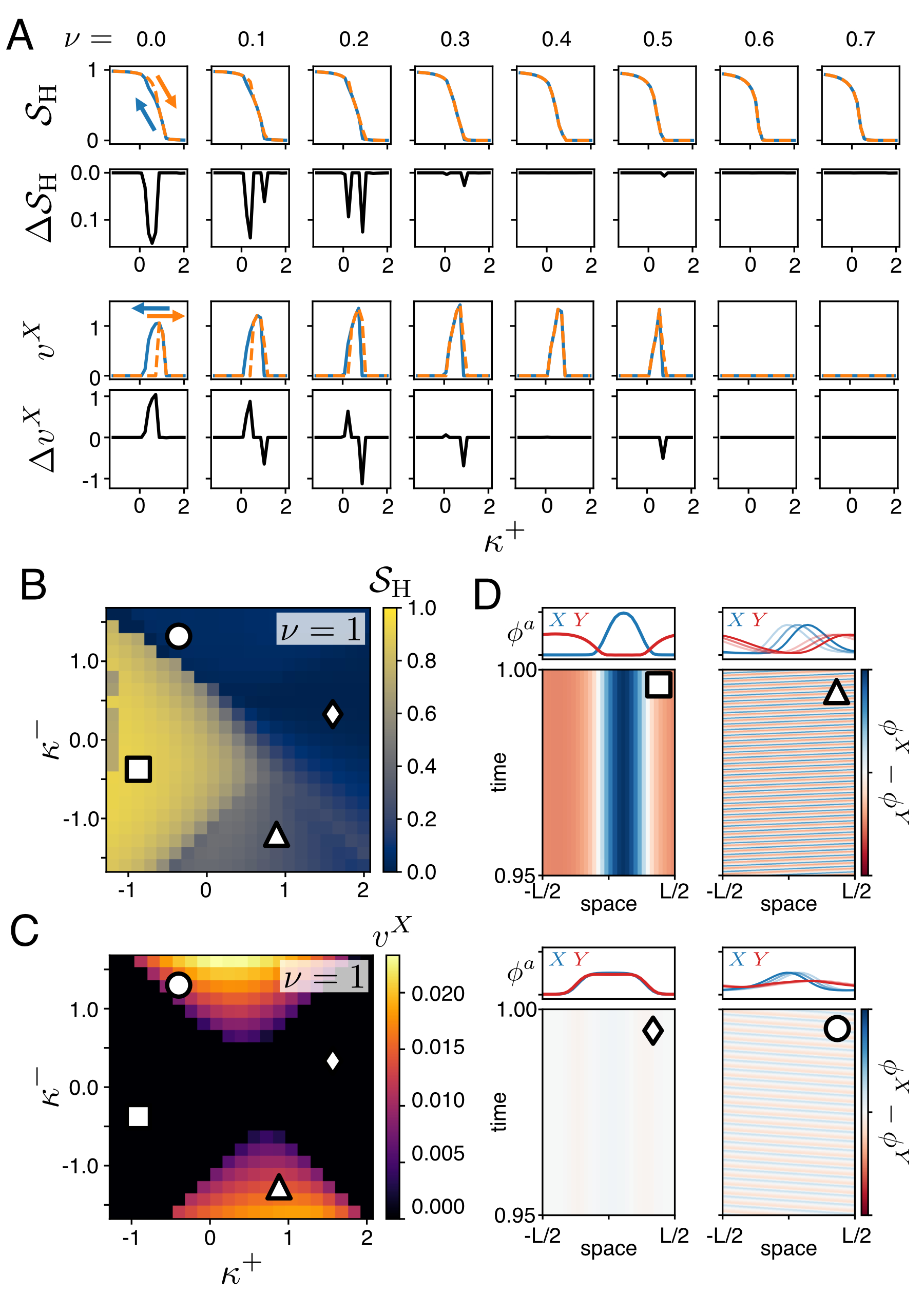}
        \caption{\textbf{Phase space of with non-linear utility function.}
        (A) Results of 1D simulations for the quadratic utility functions $\pi^X = \phi^X + \kappa^{XY} \phi^Y$, $\pi^Y = \kappa^{YX} \phi^X +  \phi^Y(1-\nu \phi^Y)$. Defining $\kappa^\pm = \kappa^{YX} \pm \kappa^{XY}$, parameters are $\kappa^- = 0.3, T = 0.1, \Gamma=1$, simulated on a domain of size $L=20$.
        A 1D simulation is initialized at $\kappa^+ = 2.0$, and $\kappa^+$ is slowly decreased to a minimal value (blue curve) before its trajectory is reversed (orange curve).
        The hysteresis is seen to disappear as the non-linear parameter, $\nu$, is increased.
        (B) Segregation index, $\mathcal{S}_\mathrm{H}$, in a phase space spanned by $\kappa^\pm$ with utilities given by Eq.~\eqref{eq:nonlinearUtility} with $\nu = 1$.
        Other parameters are the same as $A$.
        (C) Similar to (B), but measuring the velocity of traveling waves.
        (D) Kymographs showing the dynamics of select values of $(\kappa^+, \kappa^-)$.
        In particular, we show $(\kappa^+, \kappa^-)_\square = (0.88, -0.48)$, $(\kappa^+, \kappa^-)_\triangle = (0.88, -1.28)$, $(\kappa^+, \kappa^-)_\lozenge = (1.52, 0.48)$, and $(\kappa^+, \kappa^-)_\circ = (-0.56, 1.28)$}
        \label{fig:nonlinearHysteresis}
    \end{figure}


\section{The potential function}
    In this section, we make connections with the potential function for the Schelling model discovered in Ref.~\cite{Grauwin2009,Grauwin2012}.
    In particular, we rederive the \enquote{link function}, the non-trivial addition to a potential function that shows that the equilibrium state of bounded-neighborhood Schelling models need not coincide with the state that maximizes the utility function of all agents.

    The advantage of using Eq.~\eqref{eq:singleParticleRate} as the transition rate is its connection to detailed balance. Namely, if one can show that
    $$g^a(k|j) = F[\uphi'] - F[\uphi],$$
    then the system obeys detailed balance and will settle into a steady state given by the Boltzmann distribution, $P(\uphi) \propto e^{F[\uphi] / T}$ \cite{Evans2005}. To find this potential function, $F[\uphi]$, we will proceed term by term in Eq.~\eqref{eq:gain_micro} and show that they can each be written as the difference of some function depending on $\uphi$. The full potential function will then be a sum of these individual contributions.

    This condition is manifestly satisfied by the second and third terms in Eq.~\eqref{eq:gain_micro}, which are already written in the requisited form of a difference of a function of $\uphi$. However, it is not immediately obvious that the first term can be written as a function of the full configuration, as it explicitly depends on the type of particle moving (i.e. has a dependence on the index $a$) and on the specific lattice sites in consideration (i.e. has dependence on the indices $j,k$). Luckily, \cite{Grauwin2009} showed that it is indeed possible if the utilities obey a specific restriction. Specifically, the function is given by
    \begin{equation}
        \Lambda(\uphi) = \dfrac{1}{2} \sum_i  \left[\sum_{n=0}^{n^X_i - 1} \left( \utility^X \left( \dfrac{n}{\capacity}, \dfrac{n^Y_i - 1}{\capacity} \right) + \utility^X \left(\dfrac{n}{\capacity}, 0 \right) \right)
        + \sum_{m=0}^{n^Y_i - 1} \left( \utility^Y \left(\dfrac{n^X_i}{\capacity}, \dfrac{m}{\capacity} \right) + \utility^Y \left(0, \dfrac{m}{\capacity} \right) \right) \right],
    \end{equation}
    and the restriction on the utilities is
    \begin{equation}
    \label{eq:discreteCompatibility}
        \utility^X \left( \dfrac{n^X}{\capacity}, \dfrac{n^Y + 1}{\capacity}\right) - \utility^X \left( \dfrac{n^X}{\capacity}, \dfrac{n^Y}{\capacity}\right) = \utility^Y \left( \dfrac{n^X+1}{\capacity}, \dfrac{n^Y}{\capacity} \right) - \utility^Y \left( \dfrac{n^X}{\capacity}, \dfrac{n^Y}{\capacity} \right).
    \end{equation}
    This is the discrete version of the compatibility condition introduced in the main text. In the next section, we explicitly show that $\Lambda(\uphi)$, which we call a \enquote{link} function, is indeed a potential function for the first term of Eq.~\eqref{eq:gain_micro}.

    We are now in a position to give the form of our potential function, $F[\uphi]$, where we follow $\S~4.1.1$ in \cite{Bertin2021}. As stated above, given the satisfaction of Eq.~\eqref{eq:discreteCompatibility}, our system will reach a steady state given by
    $$P[\uphi] = \dfrac{1}{Z}\exp \left( \dfrac{\capacity}{T} \tilde{F}[\uphi] \right)$$
    where $Z$ is a normalization factor. The potential function can be written as a sum over lattice sites,
    $$\tilde{F}[\uphi] = \sum_i \tilde{f}(\vphi_i),$$
    with the density
    \begin{subequations}
    \begin{align}
    \tilde{f}(\vec{\phi}_i) &= T \tilde{S}(\vphi_i) + (1 - \alpha)\tilde{\lambda}(\vphi_i) + \alpha \tilde{u}(\vphi_i) + \Gamma \tilde{\sigma}(\vphi_i)
    \\
    \tilde{S}(\vphi_i) &= \dfrac{1}{\capacity} \ln \left( \dfrac{\capacity!}{n_i^X! n_i^Y! \left( \capacity - n_i^X - n_i^Y \right)!} \right)
    \\
    \tilde{\lambda}(\vphi_i) &= \dfrac{1}{2 \capacity} \left[ \sum_{n=0}^{n^X_i - 1} \utility^X \left( \dfrac{n}{\capacity}, \dfrac{n^Y_i - 1}{\capacity} \right) + \utility^X \left(\dfrac{n}{\capacity}, 0 \right) + \sum_{m=0}^{n^Y_i - 1}  \utility^Y \left(\dfrac{n^X_i}{\capacity}, \dfrac{m}{\capacity} \right) + \utility^Y \left(0, \dfrac{m}{\capacity} \right) \right]
    \\
    \tilde{u}(\vphi_i) &= \sum_{b = X,Y} \phi^b \utility^b(\vphi_i)
    \\
    \tilde{\sigma}(\vphi_i) &= \sum_{b = X,Y} \dfrac{\phi^b_i ( \partial_x^2 \uphi)^b_i}{2}.
    \end{align}
    \end{subequations}

    The term $S(\vphi_i)$ comes from counting the number of ways of arranging $n^X$ and $n^Y$ particles in a lattice site with $\capacity$ total sites. Taking the limit $\lbrace \capacity, n_i^a \rbrace \to \infty$ while keeping $\phi^a_i = n_i^a / \capacity$ constant, we can expand the factorials in $\tilde{S}$ using Stirling's approximation, $\ln n! \approx n \ln n - n$, to get
    \begin{equation}
        \tilde{S}(\vphi_i) \to S(\vphi_i) =  -\phi^X_i \ln \phi^X_i - \phi^Y_i \ln \phi^Y_i - \vac_i \ln \vac_i,
    \end{equation}
    where $\vac_i = 1 - \phi^X_i - \phi^Y_i$ is the number of vacancies at site $i$. This is the entropy of mixing.

    Furthermore, we can convert the sums in $\lambda(\vphi_i)$ to integrals, giving
    \begin{equation}
        \tilde{\lambda}(\vphi_i) \to \lambda(\vphi_i) =  \dfrac{1}{2} \left[ \int_0^{\phi^X_i} d\psi \  \utility^X \left( \psi, \phi^Y_i \right) + \utility^X \left(\psi, 0 \right) + \int_0^{\phi^Y_i} d\psi' \ \utility^Y \left( \phi^X_i, \psi' \right) + \utility^Y \left(0, \psi' \right) \right].
    \end{equation}
    Finally, we also take the continuum limit, $\ell \to 0$, while keeping the total density, $\sum_i \phi^a_i \to \ell^{-1} \int \phi^a(x) dx $ constant. For all the local terms, this amounts to converting discrete $i$ indices to a continuous $x$ coordinate. In addition, the discrete Laplacian operator in $\sigma(\vphi_i)$ becomes the usual differential Laplacian operator.

    We now finally arrive at our potential function,
    $$F[\vphi(x)] = \int f(\vphi(x)) dx$$
    with the potential density
    \begin{subequations}
    \begin{align}
        f(\vphi, \partial_x \vphi) =& T S(\vphi) + (1 - \alpha) \lambda(\vphi) +  \alpha u(\vphi) - \Gamma \sigma(\vphi) \label{eq:freeEnergyDensity}
        \\
        S(\vphi) =& -\phi^X \ln (\phi^X) - \phi^Y \ln (\phi^Y) - \vac \ln\vac \label{eq:entropyMix}
        \\
        \lambda(\vphi) =& \dfrac{1}{2} \Bigg[ \int_0^{\phi^X} d\phi' \left( \utility^X(\phi', \phi^Y) + \utility^X(\phi', 0) \right) + \int_0^{\phi^Y} d\phi'' \left( \utility^Y(\phi^X, \phi'') + \utility^Y(0, \phi'') \right) \Bigg]
        \label{eq:selfishFitness}
        \\
        u(\vphi) =& \sum_{a=X,Y} \phi^a \utility^a(\vphi) \label{eq:globalFitness}
        \\
        \sigma(\vphi) =& \sum_{a=X,Y} \dfrac{| \partial_x \phi^a (x)|^2}{2}
    \end{align}
    \end{subequations}
    Note that we have used integration by parts on $\int \sigma(\vphi(x)) dx$ to obtain the minus sign in front of $\Gamma$. In the large $\capacity$ limit, the compatibility condition Eq.~\ref{eq:discreteCompatibility} is now given by
    \begin{equation}
    \label{eq:continuousCompatibility}
        \dfrac{\partial \utility^X(\phi^X, \phi^Y)}{\partial \phi^Y} = \dfrac{\partial \utility^Y(\phi^X, \phi^Y)}{\partial \phi^X}
    \end{equation}

    \subsection{Demonstration of $\Lambda(\uphi)$ as a potential function}
        Here, we will explicitly show that the difference of the utility function of an individual agent is given by the difference of the link function defined over the entire population, i.e.
        $$\Lambda(\uphi') - \Lambda(\uphi) = \utility^a(\vec{\phi}'_k) - \utility^a(\vec{\phi}_j) = \utility^a(\vec{\phi}_k) - \utility^a(\vec{\phi}_j - \delta \phi \hat{e}_a),$$
        where $\uphi' = D^a_{k|j} \uphi$. For concreteness, let us set $a = X$. The calculation for $a = Y$ proceeds identically. In this case,
        \begin{equation}
            \Lambda(D^X_{k|j} \uphi) = \dfrac{1}{2} \sum_i
            \left[
            \sum_{n'=0}^{n^X_i - 1 + \delta_{ik} + \delta_{ij}}
            \utility^X \left( \dfrac{n'}{\capacity}, \dfrac{n^Y_i}{\capacity} \right)
            +
            \utility^X \left(\dfrac{n'}{\capacity}, 0 \right)
            +
            \sum_{m'=0}^{n^Y_i - 1}
            \utility^Y \left(\dfrac{n^X_i - 1 + \delta_{ik} - \delta_{ij}}{\capacity}, \dfrac{m'}{\capacity} \right)
            +
            \utility^Y \left(0, \dfrac{m'}{\capacity} \right) \right].
        \end{equation}
        Each term in the sum over $i$ in $\Lambda(\uphi') - \Lambda(\uphi)$ is zero unless $i = \lbrace j, k \rbrace$. We are left with
        \begin{equation}
        \begin{split}
            2(\Lambda(D^X_{k|j} \uphi) - \Lambda(\uphi)) =
            & \textcolor{DarkRed}{\sum_{n_k'=0}^{n_k^X}
            \utility^X \left( \dfrac{n_k'}{\capacity}, \dfrac{n_k^Y}{\capacity} \right)
            +
            \utility^X \left( \dfrac{n_k'}{\capacity}, 0 \right)}
                + \textcolor{DarkBlue}{\sum_{m_k'=0}^{n_k^Y-1}
            \utility^Y \left( \dfrac{n_k^X + 1}{\capacity}, \dfrac{m_k'}{\capacity} \right)}
            +
            \utility^Y \left( 0, \dfrac{m_k'}{\capacity} \right)
            \\
            - & \textcolor{DarkRed}{\sum_{n_k=0}^{n_k^X - 1}
            \utility^X \left( \dfrac{n_k}{\capacity}, \dfrac{n_k^Y}{\capacity} \right)
            +
            \utility^X \left( \dfrac{n_k}{\capacity}, 0 \right)}
            - \textcolor{DarkBlue}{\sum_{m_k=0}^{n_k^Y - 1}
            \utility^Y \left( \dfrac{n_k^X}{\capacity}, \dfrac{m_k}{\capacity} \right)}
            +
            \utility^Y \left( 0, \dfrac{m_k}{\capacity} \right)
            \\
            + & \textcolor{DarkGreen}{\sum_{n_j'=0}^{n_j^X - 2}
            \utility^X \left( \dfrac{n_j'}{\capacity}, \dfrac{n_j^Y}{\capacity} \right)
            +
            \utility^X \left( \dfrac{n_j'}{\capacity}, 0 \right)}
            + \textcolor{DarkMagenta}{\sum_{m_j'=0}^{n_j^Y - 1}
            \utility^Y \left( \dfrac{n_j^X - 1}{\capacity}, \dfrac{m_j'}{\capacity} \right)}
            +
            \utility^Y \left( 0, \dfrac{m_j'}{\capacity} \right)
            \\
            - & \textcolor{DarkGreen}{\sum_{n_j=0}^{n_j^X - 1}
            \utility^X \left( \dfrac{n_j'}{\capacity}, \dfrac{n_j^Y}{\capacity} \right)
            +
            \utility^X \left( \dfrac{n_j'}{\capacity}, 0 \right)}
            - \textcolor{DarkMagenta}{\sum_{m_j=0}^{n_j^Y-1}
            \utility^Y \left( \dfrac{n_j^X}{\capacity}, \dfrac{m_j}{\capacity} \right)}
            +
            \utility^Y \left( 0, \dfrac{m_j}{\capacity} \right).
        \end{split}
        \end{equation}
        The first two lines come from the $i = k$ term of $\Lambda(\uphi') - \Lambda(\uphi)$, while the final two lines from the $i = j$ term of $\Lambda(\uphi') - \Lambda(\uphi)$, respectively. We have color-coded the different terms for clarity in what follows.

        To proceed, we notice that the very last terms in the first two lines cancel, as do the very last terms in the final two lines. Combining the remaining terms with the same colors, we have
        \begin{equation}
        \begin{split}
            2(\Lambda(D^X_{k|j} \uphi) - \Lambda(\uphi)) =
            & \textcolor{DarkRed}{\utility^X \left( \dfrac{n_k^X}{\capacity}, \dfrac{n_k^Y}{\capacity} \right) + \utility^X \left( \dfrac{n_k^X}{\capacity}, 0 \right)}
            -
            \textcolor{DarkGreen}{\utility^X \left( \dfrac{n_j^X - 1}{\capacity}, \dfrac{n_j^Y}{\capacity} \right) - \utility^X \left( \dfrac{n_j^X - 1}{\capacity}, 0 \right)}
            \\
            +
            & \textcolor{DarkBlue}{\sum_{m_k=0}^{n_k^Y - 1} \utility^Y \left( \dfrac{n_k^X + 1}{\capacity}, \dfrac{m_k}{\capacity} \right) - \utility^Y \left( \dfrac{n_k^X}{\capacity}, \dfrac{m_k}{\capacity} \right)}
            +
            \textcolor{DarkMagenta}{\sum_{m_j=0}^{n_j^Y - 1} \utility^Y \left( \dfrac{n_j^X - 1}{\capacity}, \dfrac{m_j}{\capacity} \right) - \utility^Y \left( \dfrac{n_j^X}{\capacity}, \dfrac{m_j}{\capacity} \right)}.
        \end{split}
        \end{equation}
        Here, we are forced to use Eq.~\eqref{eq:discreteCompatibility} to change the last two summands as
        \begin{equation*}
        \begin{split}
            \textcolor{DarkBlue}{
            \utility^Y \left( \dfrac{n_k^X+ 1}{\capacity}, \dfrac{m_k}{\capacity} \right) - \utility^Y \left( \dfrac{n_k^X}{\capacity}, \dfrac{m_k}{\capacity} \right)}
            &=
            \textcolor{DarkBlue}{
            \utility^X \left( \dfrac{n_k^X}{\capacity}, \dfrac{m_k + 1}{\capacity} \right) - \utility^X \left( \dfrac{n_k^X}{\capacity}, \dfrac{m_k}{\capacity} \right)}
            \\
            \textcolor{DarkMagenta}{
            \utility^Y \left( \dfrac{n_j^X - 1}{\capacity}, \dfrac{m_j}{\capacity} \right) - \utility^Y \left( \dfrac{n_j^X}{\capacity}, \dfrac{m_j}{\capacity} \right)}
            &=
            \textcolor{DarkMagenta}{
            \utility^X \left( \dfrac{n_j^X - 1}{\capacity}, \dfrac{m_j}{\capacity} \right) - \utility^X \left( \dfrac{n_j^X - 1}{\capacity}, \dfrac{m_j + 1}{\capacity} \right)}.
        \end{split}
        \end{equation*}
        Evaluating the sums, we find
        \begin{equation}
        \begin{split}
            2(\Lambda(D^X_{k|j} \uphi) - \Lambda(\uphi)) =
            & \textcolor{DarkRed}{\utility^X \left( \dfrac{n_k^X}{\capacity}, \dfrac{n_k^Y}{\capacity} \right) + \utility^X \left( \dfrac{n_k^X}{\capacity}, 0 \right)}
            -
            \textcolor{DarkGreen}{\utility^X \left( \dfrac{n_j^X - 1}{\capacity}, \dfrac{n_j^Y}{\capacity} \right) - \utility^X \left( \dfrac{n_j^X - 1}{\capacity}, 0 \right)}
            \\
            +
            & \textcolor{DarkBlue}{\utility^X \left( \dfrac{n_k^X}{\capacity}, \dfrac{n_k^Y}{\capacity} \right) - \utility^X \left( \dfrac{n_k^X}{\capacity}, 0 \right)}
            +
            \textcolor{DarkMagenta}{\utility^X \left( \dfrac{n_j^X - 1}{\capacity}, 0 \right) - \utility^X \left( \dfrac{n_j^X - 1}{\capacity}, \dfrac{n_j^Y}{\capacity} \right)}.
        \end{split}
        \end{equation}
        Finally, combining terms and canceling others, we arrive at our desired result
        \begin{equation}
            \Lambda(D^X_{k|j}\uphi) - \Lambda(\uphi) = \utility^X(\vphi_k) - \utility^X(\vphi_j - \delta \phi \hat{e}^X).
        \end{equation}

    \subsection{Relaxational dynamics}
        Given the existence of the utility potential function $F[\vphi(x)]$, one can define a dynamics for $\phi^a(x, t)$ such that $F[\vphi(x)]$ plays the role of a Lyapunov function ensuring the relaxation of the system toward a stationary state, as done for physical systems \cite{Hohenberg1977}. One important difference is that the utility potential function will be \textit{maximized}, as contrasted with a physical free energy which is minimized.
        We note that this is not the general case for sociohydrodyanmics. It is a special case where the incompatibility condition is satisfied.

        As the microscopic dynamics consist solely of particles moving, the macroscopic dynamics must conserve the total particle number,
        \begin{equation}
            \partial_t \phi^a(x, t) = - \partial_x J^a(x, t).
        \end{equation}
        where $\partial_x = \partial/\partial x$. Analogously to physical system, we take the particle current $J^a(x, t)$ to obey the constitutive relation
        \begin{equation}
        \label{eq:fluxConstitutive}
            J^a(x, t) = + M^{ab}(\vphi) \partial_x \dfrac{\delta F}{\delta \phi^b}.
        \end{equation}
        $M^{ab}(\vphi)$ is a density dependent mobility matrix, which we take to be given by
        \begin{equation}
            M^{ab}(\vphi) = \delta^{ab} \phi^b \vac.
        \end{equation}
        Recalling $\vac = 1 - \sum_c \phi^c$, this form enforces that both empty ($\phi^b = 0$) and full ($\vac = 0$) locations give zero-flux.

        $\delta F / \delta \phi^b$ is the functional derivative of the potential $F[\vphi(x)]$ that plays the role of a chemical potential \cite{Lemoy2011}. Eq.~\eqref{eq:fluxConstitutive} says that the particle current flows \textit{up} spatial gradients of the chemical potential. We now evaluate the chemical potential as
        \begin{equation}
            \dfrac{\delta F}{\delta \phi^b} = -T \left(\ln \phi^b - \ln \vac \right) + (1-\alpha) \dfrac{\partial \lambda}{\partial \phi^b} + \alpha \dfrac{\partial u}{\partial \phi^b} + \Gamma \partial_x^2 \phi^b.
        \end{equation}
        Taking the gradient, we obtain
        \begin{equation}
            \partial_x \dfrac{\delta F}{\delta \phi^b} = -T \dfrac{\vac \partial_x \phi^b + \phi^b \partial_x \vac}{\phi^b \vac} + (1 - \alpha) \partial_x \dfrac{\partial \lambda}{\partial \phi^b} + \alpha \partial_x \dfrac{\partial u}{\partial \phi^b} + \Gamma \partial_x \partial_x^2 \phi^b.
        \end{equation}
        Finally multiplying by the mobility matrix $M^{ab} = \delta^{ab} \phi^b \vac$, we get our general form for relaxational sociohydrodynamics
        \begin{equation}
        \label{eq:relaxationalDynamics}
            \partial_t \phi^a(x, t) = T \left((1-\phi^b) \partial_x^2 \phi^a + \phi^a \partial_x^2 \phi^b \right) -  \partial_x \left( \phi^a \vac \ \partial_x \left[(1-\alpha)\dfrac{\partial \lambda}{\partial \phi^a} + \alpha \dfrac{\partial u}{\partial \phi^a} + \Gamma^a \partial_x^2 \phi^a \right] \right).
        \end{equation}

        In the special case where the compatibility condition Eq.~\ref{eq:continuousCompatibility} holds, the derivative of the link function $\lambda(\vphi(x))$ is given by (here, we take $b = X$. The calculation for $b = Y$ proceeds analogously)
        \begin{equation}
        \begin{aligned}[b]
            \dfrac{\partial \lambda}{\partial \phi^X} &= \dfrac{1}{2} \left( \utility^X(\phi^X, \phi^Y) + \utility^X(\phi^X, 0) + \int_0^{\phi^Y} d\phi' \dfrac{\partial \utility^Y(\phi^X, \phi')}{\partial \phi^X} \right)
            \\
            &= \dfrac{1}{2} \left( \utility^X(\phi^X, \phi^Y) + \utility^X(\phi^X, 0) + \int_0^{\phi^Y} d\phi' \dfrac{\partial \utility^X(\phi^X, \phi')}{\partial \phi'} \right)
            \\
            &= \dfrac{1}{2} \left( \utility^X(\phi^X, \phi^Y) + \utility^X(\phi^X, 0) + \utility^X(\phi^X, \phi^Y) - \utility^X(\phi^X, 0) \right)
            \\
            \dfrac{\partial \lambda}{\partial \phi^X} &= \utility^X(\phi^X, \phi^Y).
        \end{aligned}
        \end{equation}
        where we used the compatibility condition Eq.~\eqref{eq:continuousCompatibility} to go from the first to the second line. In this case, Eq.~\eqref{eq:relaxationalDynamics} becomes
        \begin{equation}
        \label{eq:relaxationalDynamics_compat}
            \partial_t \phi^a(x, t) = T \left((1-\phi^b) \partial_x^2 \phi^a + \phi^a \partial_x^2 \phi^b \right) -  \partial_x \left( \phi^a \vac \left[(1-\alpha)\partial_x \utility^a + \partial_x \dfrac{\partial u}{\partial \phi^a} + \Gamma^a \partial_x^3 \phi^a \right] \right).
        \end{equation}
        where $b \neq a$. This is the same equation as Eq.~\ref{eq:sociohydro}, showing that our coarse-grained equations are equivalent to the relaxational dynamics when the compatibility condition holds.

We now consider a microscopic model for growth that is built in the same spirit as agent-based model above. This would lead to a form for $S^a$ which we set to zero in the main text.

We consider individuals within a lattice site can spontaneously die, or reproduce if there is space available in the lattice site, with a rate $k$ dependent on the fitness $f^a$, $k = k(f^a)$. We denote these processes by the chemical reactions
\begin{subequations}
\begin{align}
    a &\overset{k(-f^a)}{\to} \emptyset \\
    a + \emptyset &\overset{k(f^a)}{\to} a + a.
\end{align}
\end{subequations}
The reaction rates are given by
\begin{equation}
    k(f^a) = \dfrac{2r}{1+e^{-\gamma f^a}},
\end{equation}
where $r$ is a rate, $\gamma$ measures the effects of stochasticity in the death/reproduction processes, and the factor of 2 is included for simplicity. Here, we have assumed that both processes have the same $r$ and $\gamma$. The form of $k(f^a)$ assures that reproduction (death) occurs more (less) rapidly with increasing fitness.

Assuming mass action kinetics, the density of population $a$ will obey
\begin{equation}
    \partial_t \phi^a = \phi^a \left( \phi^\emptyset k(f^a) - k(-f^a) \right),
\end{equation}
where $\phi^\emptyset = 1 - \sum_b \phi^b$ measures the density of empty sites available to be occupied by reproduction.
In the limit of weak selection, $\gamma \to 0$, we can expand $k(f^a)$ to find
\begin{equation}
    \partial_t \phi^a = r \phi^a \left(\gamma f^a - \left(\sum_b \phi^b \right) \left(1 + \dfrac{\gamma f^a}{2} \right) \right) = S^a.
\end{equation}


%% file: main.bbl
\begin{thebibliography}{139}%
\makeatletter
\providecommand \@ifxundefined [1]{%
 \@ifx{#1\undefined}
}%
\providecommand \@ifnum [1]{%
 \ifnum #1\expandafter \@firstoftwo
 \else \expandafter \@secondoftwo
 \fi
}%
\providecommand \@ifx [1]{%
 \ifx #1\expandafter \@firstoftwo
 \else \expandafter \@secondoftwo
 \fi
}%
\providecommand \natexlab [1]{#1}%
\providecommand \enquote  [1]{``#1''}%
\providecommand \bibnamefont  [1]{#1}%
\providecommand \bibfnamefont [1]{#1}%
\providecommand \citenamefont [1]{#1}%
\providecommand \href@noop [0]{\@secondoftwo}%
\providecommand \href [0]{\begingroup \@sanitize@url \@href}%
\providecommand \@href[1]{\@@startlink{#1}\@@href}%
\providecommand \@@href[1]{\endgroup#1\@@endlink}%
\providecommand \@sanitize@url [0]{\catcode `\\12\catcode `\$12\catcode
  `\&12\catcode `\#12\catcode `\^12\catcode `\_12\catcode `\%12\relax}%
\providecommand \@@startlink[1]{}%
\providecommand \@@endlink[0]{}%
\providecommand \url  [0]{\begingroup\@sanitize@url \@url }%
\providecommand \@url [1]{\endgroup\@href {#1}{\urlprefix }}%
\providecommand \urlprefix  [0]{URL }%
\providecommand \Eprint [0]{\href }%
\providecommand \doibase [0]{https://doi.org/}%
\providecommand \selectlanguage [0]{\@gobble}%
\providecommand \bibinfo  [0]{\@secondoftwo}%
\providecommand \bibfield  [0]{\@secondoftwo}%
\providecommand \translation [1]{[#1]}%
\providecommand \BibitemOpen [0]{}%
\providecommand \bibitemStop [0]{}%
\providecommand \bibitemNoStop [0]{.\EOS\space}%
\providecommand \EOS [0]{\spacefactor3000\relax}%
\providecommand \BibitemShut  [1]{\csname bibitem#1\endcsname}%
\let\auto@bib@innerbib\@empty
\bibitem [{\citenamefont {Reichenbach}\ \emph {et~al.}(2007)\citenamefont
  {Reichenbach}, \citenamefont {Mobilia},\ and\ \citenamefont
  {Frey}}]{Reichenbach2007}%
  \BibitemOpen
  \bibfield  {author} {\bibinfo {author} {\bibfnamefont {T.}~\bibnamefont
  {Reichenbach}}, \bibinfo {author} {\bibfnamefont {M.}~\bibnamefont
  {Mobilia}},\ and\ \bibinfo {author} {\bibfnamefont {E.}~\bibnamefont
  {Frey}},\ }\bibfield  {title} {\bibinfo {title} {Mobility promotes and
  jeopardizes biodiversity in rock{\textendash}paper{\textendash}scissors
  games},\ }\href {https://doi.org/10.1038/nature06095} {\bibfield  {journal}
  {\bibinfo  {journal} {Nature}\ }\textbf {\bibinfo {volume} {448}},\ \bibinfo
  {pages} {1046} (\bibinfo {year} {2007})}\BibitemShut {NoStop}%
\bibitem [{\citenamefont {Couzin}(2009)}]{Couzin2009}%
  \BibitemOpen
  \bibfield  {author} {\bibinfo {author} {\bibfnamefont {I.~D.}\ \bibnamefont
  {Couzin}},\ }\bibfield  {title} {\bibinfo {title} {Collective cognition in
  animal groups},\ }\href {https://doi.org/10.1016/j.tics.2008.10.002}
  {\bibfield  {journal} {\bibinfo  {journal} {Trends in Cognitive Sciences}\
  }\textbf {\bibinfo {volume} {13}},\ \bibinfo {pages} {36} (\bibinfo {year}
  {2009})}\BibitemShut {NoStop}%
\bibitem [{\citenamefont {Ouellette}(2022)}]{Ouellette2022}%
  \BibitemOpen
  \bibfield  {author} {\bibinfo {author} {\bibfnamefont {N.~T.}\ \bibnamefont
  {Ouellette}},\ }\bibfield  {title} {\bibinfo {title} {A physics perspective
  on collective animal behavior},\ }\href
  {https://doi.org/10.1088/1478-3975/ac4bef} {\bibfield  {journal} {\bibinfo
  {journal} {Physical Biology}\ }\textbf {\bibinfo {volume} {19}},\ \bibinfo
  {pages} {021004} (\bibinfo {year} {2022})}\BibitemShut {NoStop}%
\bibitem [{\citenamefont {Vicsek}\ and\ \citenamefont
  {Zafeiris}(2012)}]{Vicsek2012}%
  \BibitemOpen
  \bibfield  {author} {\bibinfo {author} {\bibfnamefont {T.}~\bibnamefont
  {Vicsek}}\ and\ \bibinfo {author} {\bibfnamefont {A.}~\bibnamefont
  {Zafeiris}},\ }\bibfield  {title} {\bibinfo {title} {Collective motion},\
  }\href {https://doi.org/10.1016/j.physrep.2012.03.004} {\bibfield  {journal}
  {\bibinfo  {journal} {Physics Reports}\ }\textbf {\bibinfo {volume} {517}},\
  \bibinfo {pages} {71} (\bibinfo {year} {2012})}\BibitemShut {NoStop}%
\bibitem [{\citenamefont {Bettencourt}(2013)}]{Bettencourt2013}%
  \BibitemOpen
  \bibfield  {author} {\bibinfo {author} {\bibfnamefont {L.~M.~A.}\
  \bibnamefont {Bettencourt}},\ }\bibfield  {title} {\bibinfo {title} {The
  origins of scaling in cities},\ }\href
  {https://doi.org/10.1126/science.1235823} {\bibfield  {journal} {\bibinfo
  {journal} {Science}\ }\textbf {\bibinfo {volume} {340}},\ \bibinfo {pages}
  {1438} (\bibinfo {year} {2013})}\BibitemShut {NoStop}%
\bibitem [{\citenamefont {Kadanoff}\ and\ \citenamefont
  {Martin}(1963)}]{Kadanoff1963}%
  \BibitemOpen
  \bibfield  {author} {\bibinfo {author} {\bibfnamefont {L.~P.}\ \bibnamefont
  {Kadanoff}}\ and\ \bibinfo {author} {\bibfnamefont {P.~C.}\ \bibnamefont
  {Martin}},\ }\bibfield  {title} {\bibinfo {title} {Hydrodynamic equations and
  correlation functions},\ }\href
  {https://doi.org/10.1016/0003-4916(63)90078-2} {\bibfield  {journal}
  {\bibinfo  {journal} {Annals of Physics}\ }\textbf {\bibinfo {volume} {24}},\
  \bibinfo {pages} {419} (\bibinfo {year} {1963})}\BibitemShut {NoStop}%
\bibitem [{\citenamefont {Landau}\ and\ \citenamefont
  {Lifshitz}(1987)}]{Landau1987}%
  \BibitemOpen
  \bibfield  {author} {\bibinfo {author} {\bibfnamefont {L.~D.}\ \bibnamefont
  {Landau}}\ and\ \bibinfo {author} {\bibfnamefont {E.~M.}\ \bibnamefont
  {Lifshitz}},\ }\href@noop {} {\emph {\bibinfo {title} {Fluid Mechanics :
  Landau and Lifshitz}}}\ (\bibinfo  {publisher} {Elsevier Science \&
  Technology Books},\ \bibinfo {year} {1987})\ p.\ \bibinfo {pages}
  {539}\BibitemShut {NoStop}%
\bibitem [{\citenamefont {Anderson}(1972)}]{Anderson1972}%
  \BibitemOpen
  \bibfield  {author} {\bibinfo {author} {\bibfnamefont {P.~W.}\ \bibnamefont
  {Anderson}},\ }\bibfield  {title} {\bibinfo {title} {More is different},\
  }\href {https://doi.org/10.1126/science.177.4047.393} {\bibfield  {journal}
  {\bibinfo  {journal} {Science}\ }\textbf {\bibinfo {volume} {177}},\ \bibinfo
  {pages} {393} (\bibinfo {year} {1972})}\BibitemShut {NoStop}%
\bibitem [{\citenamefont {Van~Saarloos}\ \emph {et~al.}(2023)\citenamefont
  {Van~Saarloos}, \citenamefont {Vitelli},\ and\ \citenamefont
  {Zeravcic}}]{Saarloos2023}%
  \BibitemOpen
  \bibfield  {author} {\bibinfo {author} {\bibfnamefont {W.}~\bibnamefont
  {Van~Saarloos}}, \bibinfo {author} {\bibfnamefont {V.}~\bibnamefont
  {Vitelli}},\ and\ \bibinfo {author} {\bibfnamefont {Z.}~\bibnamefont
  {Zeravcic}},\ }\href@noop {} {\emph {\bibinfo {title} {Soft Matter: Concepts,
  Phenomena and Applications.}}}\ (\bibinfo  {publisher} {Princeton University
  Press},\ \bibinfo {year} {2023})\BibitemShut {NoStop}%
\bibitem [{\citenamefont {Sokolov}\ \emph {et~al.}(2007)\citenamefont
  {Sokolov}, \citenamefont {Aranson}, \citenamefont {Kessler},\ and\
  \citenamefont {Goldstein}}]{Sokolov2007}%
  \BibitemOpen
  \bibfield  {author} {\bibinfo {author} {\bibfnamefont {A.}~\bibnamefont
  {Sokolov}}, \bibinfo {author} {\bibfnamefont {I.~S.}\ \bibnamefont
  {Aranson}}, \bibinfo {author} {\bibfnamefont {J.~O.}\ \bibnamefont
  {Kessler}},\ and\ \bibinfo {author} {\bibfnamefont {R.~E.}\ \bibnamefont
  {Goldstein}},\ }\bibfield  {title} {\bibinfo {title} {{Concentration
  Dependence of the Collective Dynamics of Swimming Bacteria}},\ }\href
  {https://doi.org/10.1103/PhysRevLett.98.158102} {\bibfield  {journal}
  {\bibinfo  {journal} {Physical Review Letters}\ }\textbf {\bibinfo {volume}
  {98}},\ \bibinfo {pages} {158102} (\bibinfo {year} {2007})}\BibitemShut
  {NoStop}%
\bibitem [{\citenamefont {Wensink}\ \emph {et~al.}(2012)\citenamefont
  {Wensink}, \citenamefont {Dunkel}, \citenamefont {Heidenreich}, \citenamefont
  {Drescher}, \citenamefont {Goldstein}, \citenamefont {Löwen},\ and\
  \citenamefont {Yeomans}}]{Wensink2012}%
  \BibitemOpen
  \bibfield  {author} {\bibinfo {author} {\bibfnamefont {H.~H.}\ \bibnamefont
  {Wensink}}, \bibinfo {author} {\bibfnamefont {J.}~\bibnamefont {Dunkel}},
  \bibinfo {author} {\bibfnamefont {S.}~\bibnamefont {Heidenreich}}, \bibinfo
  {author} {\bibfnamefont {K.}~\bibnamefont {Drescher}}, \bibinfo {author}
  {\bibfnamefont {R.~E.}\ \bibnamefont {Goldstein}}, \bibinfo {author}
  {\bibfnamefont {H.}~\bibnamefont {Löwen}},\ and\ \bibinfo {author}
  {\bibfnamefont {J.~M.}\ \bibnamefont {Yeomans}},\ }\bibfield  {title}
  {\bibinfo {title} {Meso-scale turbulence in living fluids},\ }\href
  {https://doi.org/10.1073/pnas.1202032109} {\bibfield  {journal} {\bibinfo
  {journal} {Proceedings of the National Academy of Sciences}\ }\textbf
  {\bibinfo {volume} {109}},\ \bibinfo {pages} {14308} (\bibinfo {year}
  {2012})}\BibitemShut {NoStop}%
\bibitem [{\citenamefont {Wioland}\ \emph {et~al.}(2013)\citenamefont
  {Wioland}, \citenamefont {Woodhouse}, \citenamefont {Dunkel}, \citenamefont
  {Kessler},\ and\ \citenamefont {Goldstein}}]{Wioland2013}%
  \BibitemOpen
  \bibfield  {author} {\bibinfo {author} {\bibfnamefont {H.}~\bibnamefont
  {Wioland}}, \bibinfo {author} {\bibfnamefont {F.~G.}\ \bibnamefont
  {Woodhouse}}, \bibinfo {author} {\bibfnamefont {J.}~\bibnamefont {Dunkel}},
  \bibinfo {author} {\bibfnamefont {J.~O.}\ \bibnamefont {Kessler}},\ and\
  \bibinfo {author} {\bibfnamefont {R.~E.}\ \bibnamefont {Goldstein}},\
  }\bibfield  {title} {\bibinfo {title} {Confinement stabilizes a bacterial
  suspension into a spiral vortex},\ }\href
  {https://doi.org/10.1103/physrevlett.110.268102} {\bibfield  {journal}
  {\bibinfo  {journal} {Physical Review Letters}\ }\textbf {\bibinfo {volume}
  {110}},\ \bibinfo {pages} {268102} (\bibinfo {year} {2013})}\BibitemShut
  {NoStop}%
\bibitem [{\citenamefont {Li}\ \emph {et~al.}(2018)\citenamefont {Li},
  \citenamefont {Shi}, \citenamefont {Huang}, \citenamefont {Chen},
  \citenamefont {Xiao}, \citenamefont {Liu}, \citenamefont {Chaté},\ and\
  \citenamefont {Zhang}}]{Li2018}%
  \BibitemOpen
  \bibfield  {author} {\bibinfo {author} {\bibfnamefont {H.}~\bibnamefont
  {Li}}, \bibinfo {author} {\bibfnamefont {X.-q.}\ \bibnamefont {Shi}},
  \bibinfo {author} {\bibfnamefont {M.}~\bibnamefont {Huang}}, \bibinfo
  {author} {\bibfnamefont {X.}~\bibnamefont {Chen}}, \bibinfo {author}
  {\bibfnamefont {M.}~\bibnamefont {Xiao}}, \bibinfo {author} {\bibfnamefont
  {C.}~\bibnamefont {Liu}}, \bibinfo {author} {\bibfnamefont {H.}~\bibnamefont
  {Chaté}},\ and\ \bibinfo {author} {\bibfnamefont {H.~P.}\ \bibnamefont
  {Zhang}},\ }\bibfield  {title} {\bibinfo {title} {Data-driven quantitative
  modeling of bacterial active nematics},\ }\href
  {https://doi.org/10.1073/pnas.1812570116} {\bibfield  {journal} {\bibinfo
  {journal} {Proceedings of the National Academy of Sciences}\ }\textbf
  {\bibinfo {volume} {116}},\ \bibinfo {pages} {777} (\bibinfo {year}
  {2018})}\BibitemShut {NoStop}%
\bibitem [{\citenamefont {Copenhagen}\ \emph {et~al.}(2020)\citenamefont
  {Copenhagen}, \citenamefont {Alert}, \citenamefont {Wingreen},\ and\
  \citenamefont {Shaevitz}}]{Copenhagen2020}%
  \BibitemOpen
  \bibfield  {author} {\bibinfo {author} {\bibfnamefont {K.}~\bibnamefont
  {Copenhagen}}, \bibinfo {author} {\bibfnamefont {R.}~\bibnamefont {Alert}},
  \bibinfo {author} {\bibfnamefont {N.~S.}\ \bibnamefont {Wingreen}},\ and\
  \bibinfo {author} {\bibfnamefont {J.~W.}\ \bibnamefont {Shaevitz}},\
  }\bibfield  {title} {\bibinfo {title} {Topological defects promote layer
  formation in myxococcus xanthus colonies},\ }\href
  {https://doi.org/10.1038/s41567-020-01056-4} {\bibfield  {journal} {\bibinfo
  {journal} {Nature Physics}\ }\textbf {\bibinfo {volume} {17}},\ \bibinfo
  {pages} {211} (\bibinfo {year} {2020})}\BibitemShut {NoStop}%
\bibitem [{\citenamefont {Curatolo}\ \emph {et~al.}(2020)\citenamefont
  {Curatolo}, \citenamefont {Zhou}, \citenamefont {Zhao}, \citenamefont {Liu},
  \citenamefont {Daerr}, \citenamefont {Tailleur},\ and\ \citenamefont
  {Huang}}]{Curatolo2020}%
  \BibitemOpen
  \bibfield  {author} {\bibinfo {author} {\bibfnamefont {A.~I.}\ \bibnamefont
  {Curatolo}}, \bibinfo {author} {\bibfnamefont {N.}~\bibnamefont {Zhou}},
  \bibinfo {author} {\bibfnamefont {Y.}~\bibnamefont {Zhao}}, \bibinfo {author}
  {\bibfnamefont {C.}~\bibnamefont {Liu}}, \bibinfo {author} {\bibfnamefont
  {A.}~\bibnamefont {Daerr}}, \bibinfo {author} {\bibfnamefont
  {J.}~\bibnamefont {Tailleur}},\ and\ \bibinfo {author} {\bibfnamefont
  {J.}~\bibnamefont {Huang}},\ }\bibfield  {title} {\bibinfo {title}
  {Cooperative pattern formation in multi-component bacterial systems through
  reciprocal motility regulation},\ }\href
  {https://doi.org/10.1038/s41567-020-0964-z} {\bibfield  {journal} {\bibinfo
  {journal} {Nature Physics}\ }\textbf {\bibinfo {volume} {16}},\ \bibinfo
  {pages} {1152} (\bibinfo {year} {2020})}\BibitemShut {NoStop}%
\bibitem [{\citenamefont {Mertz}\ \emph {et~al.}(2012)\citenamefont {Mertz},
  \citenamefont {Banerjee}, \citenamefont {Che}, \citenamefont {German},
  \citenamefont {Xu}, \citenamefont {Hyland}, \citenamefont {Marchetti},
  \citenamefont {Horsley},\ and\ \citenamefont {Dufresne}}]{Mertz2012}%
  \BibitemOpen
  \bibfield  {author} {\bibinfo {author} {\bibfnamefont {A.~F.}\ \bibnamefont
  {Mertz}}, \bibinfo {author} {\bibfnamefont {S.}~\bibnamefont {Banerjee}},
  \bibinfo {author} {\bibfnamefont {Y.}~\bibnamefont {Che}}, \bibinfo {author}
  {\bibfnamefont {G.~K.}\ \bibnamefont {German}}, \bibinfo {author}
  {\bibfnamefont {Y.}~\bibnamefont {Xu}}, \bibinfo {author} {\bibfnamefont
  {C.}~\bibnamefont {Hyland}}, \bibinfo {author} {\bibfnamefont {M.~C.}\
  \bibnamefont {Marchetti}}, \bibinfo {author} {\bibfnamefont {V.}~\bibnamefont
  {Horsley}},\ and\ \bibinfo {author} {\bibfnamefont {E.~R.}\ \bibnamefont
  {Dufresne}},\ }\bibfield  {title} {\bibinfo {title} {Scaling of traction
  forces with the size of cohesive cell colonies},\ }\href
  {https://doi.org/10.1103/physrevlett.108.198101} {\bibfield  {journal}
  {\bibinfo  {journal} {Physical Review Letters}\ }\textbf {\bibinfo {volume}
  {108}},\ \bibinfo {pages} {198101} (\bibinfo {year} {2012})}\BibitemShut
  {NoStop}%
\bibitem [{\citenamefont {Saw}\ \emph {et~al.}(2017)\citenamefont {Saw},
  \citenamefont {Doostmohammadi}, \citenamefont {Nier}, \citenamefont
  {Kocgozlu}, \citenamefont {Thampi}, \citenamefont {Toyama}, \citenamefont
  {Marcq}, \citenamefont {Lim}, \citenamefont {Yeomans},\ and\ \citenamefont
  {Ladoux}}]{Saw2017}%
  \BibitemOpen
  \bibfield  {author} {\bibinfo {author} {\bibfnamefont {T.~B.}\ \bibnamefont
  {Saw}}, \bibinfo {author} {\bibfnamefont {A.}~\bibnamefont {Doostmohammadi}},
  \bibinfo {author} {\bibfnamefont {V.}~\bibnamefont {Nier}}, \bibinfo {author}
  {\bibfnamefont {L.}~\bibnamefont {Kocgozlu}}, \bibinfo {author}
  {\bibfnamefont {S.}~\bibnamefont {Thampi}}, \bibinfo {author} {\bibfnamefont
  {Y.}~\bibnamefont {Toyama}}, \bibinfo {author} {\bibfnamefont
  {P.}~\bibnamefont {Marcq}}, \bibinfo {author} {\bibfnamefont {C.~T.}\
  \bibnamefont {Lim}}, \bibinfo {author} {\bibfnamefont {J.~M.}\ \bibnamefont
  {Yeomans}},\ and\ \bibinfo {author} {\bibfnamefont {B.}~\bibnamefont
  {Ladoux}},\ }\bibfield  {title} {\bibinfo {title} {Topological defects in
  epithelia govern cell death and extrusion},\ }\href
  {https://doi.org/10.1038/nature21718} {\bibfield  {journal} {\bibinfo
  {journal} {Nature}\ }\textbf {\bibinfo {volume} {544}},\ \bibinfo {pages}
  {212} (\bibinfo {year} {2017})}\BibitemShut {NoStop}%
\bibitem [{\citenamefont {Pérez-González}\ \emph {et~al.}(2018)\citenamefont
  {Pérez-González}, \citenamefont {Alert}, \citenamefont {Blanch-Mercader},
  \citenamefont {Gómez-González}, \citenamefont {Kolodziej}, \citenamefont
  {Bazellieres}, \citenamefont {Casademunt},\ and\ \citenamefont
  {Trepat}}]{PerezGonzalez2018}%
  \BibitemOpen
  \bibfield  {author} {\bibinfo {author} {\bibfnamefont {C.}~\bibnamefont
  {Pérez-González}}, \bibinfo {author} {\bibfnamefont {R.}~\bibnamefont
  {Alert}}, \bibinfo {author} {\bibfnamefont {C.}~\bibnamefont
  {Blanch-Mercader}}, \bibinfo {author} {\bibfnamefont {M.}~\bibnamefont
  {Gómez-González}}, \bibinfo {author} {\bibfnamefont {T.}~\bibnamefont
  {Kolodziej}}, \bibinfo {author} {\bibfnamefont {E.}~\bibnamefont
  {Bazellieres}}, \bibinfo {author} {\bibfnamefont {J.}~\bibnamefont
  {Casademunt}},\ and\ \bibinfo {author} {\bibfnamefont {X.}~\bibnamefont
  {Trepat}},\ }\bibfield  {title} {\bibinfo {title} {Active wetting of
  epithelial tissues},\ }\href {https://doi.org/10.1038/s41567-018-0279-5}
  {\bibfield  {journal} {\bibinfo  {journal} {Nature Physics}\ }\textbf
  {\bibinfo {volume} {15}},\ \bibinfo {pages} {79} (\bibinfo {year}
  {2018})}\BibitemShut {NoStop}%
\bibitem [{\citenamefont {Streichan}\ \emph {et~al.}(2018)\citenamefont
  {Streichan}, \citenamefont {Lefebvre}, \citenamefont {Noll}, \citenamefont
  {Wieschaus},\ and\ \citenamefont {Shraiman}}]{Streichan2018}%
  \BibitemOpen
  \bibfield  {author} {\bibinfo {author} {\bibfnamefont {S.~J.}\ \bibnamefont
  {Streichan}}, \bibinfo {author} {\bibfnamefont {M.~F.}\ \bibnamefont
  {Lefebvre}}, \bibinfo {author} {\bibfnamefont {N.}~\bibnamefont {Noll}},
  \bibinfo {author} {\bibfnamefont {E.~F.}\ \bibnamefont {Wieschaus}},\ and\
  \bibinfo {author} {\bibfnamefont {B.~I.}\ \bibnamefont {Shraiman}},\
  }\bibfield  {title} {\bibinfo {title} {Global morphogenetic flow is
  accurately predicted by the spatial distribution of myosin motors},\
  }\bibfield  {journal} {\bibinfo  {journal} {eLife}\ }\textbf {\bibinfo
  {volume} {7}},\ \href {https://doi.org/10.7554/elife.27454}
  {10.7554/elife.27454} (\bibinfo {year} {2018})\BibitemShut {NoStop}%
\bibitem [{\citenamefont {Alert}\ \emph {et~al.}(2019)\citenamefont {Alert},
  \citenamefont {Blanch-Mercader},\ and\ \citenamefont
  {Casademunt}}]{Alert2019}%
  \BibitemOpen
  \bibfield  {author} {\bibinfo {author} {\bibfnamefont {R.}~\bibnamefont
  {Alert}}, \bibinfo {author} {\bibfnamefont {C.}~\bibnamefont
  {Blanch-Mercader}},\ and\ \bibinfo {author} {\bibfnamefont {J.}~\bibnamefont
  {Casademunt}},\ }\bibfield  {title} {\bibinfo {title} {Active fingering
  instability in tissue spreading},\ }\href
  {https://doi.org/10.1103/physrevlett.122.088104} {\bibfield  {journal}
  {\bibinfo  {journal} {Physical Review Letters}\ }\textbf {\bibinfo {volume}
  {122}},\ \bibinfo {pages} {088104} (\bibinfo {year} {2019})}\BibitemShut
  {NoStop}%
\bibitem [{\citenamefont {Boocock}\ \emph {et~al.}(2020)\citenamefont
  {Boocock}, \citenamefont {Hino}, \citenamefont {Ruzickova}, \citenamefont
  {Hirashima},\ and\ \citenamefont {Hannezo}}]{Boocock2020}%
  \BibitemOpen
  \bibfield  {author} {\bibinfo {author} {\bibfnamefont {D.}~\bibnamefont
  {Boocock}}, \bibinfo {author} {\bibfnamefont {N.}~\bibnamefont {Hino}},
  \bibinfo {author} {\bibfnamefont {N.}~\bibnamefont {Ruzickova}}, \bibinfo
  {author} {\bibfnamefont {T.}~\bibnamefont {Hirashima}},\ and\ \bibinfo
  {author} {\bibfnamefont {E.}~\bibnamefont {Hannezo}},\ }\bibfield  {title}
  {\bibinfo {title} {Theory of mechanochemical patterning and optimal migration
  in cell monolayers},\ }\href {https://doi.org/10.1038/s41567-020-01037-7}
  {\bibfield  {journal} {\bibinfo  {journal} {Nature Physics}\ }\textbf
  {\bibinfo {volume} {17}},\ \bibinfo {pages} {267} (\bibinfo {year}
  {2020})}\BibitemShut {NoStop}%
\bibitem [{\citenamefont {Yousafzai}\ \emph {et~al.}(2022)\citenamefont
  {Yousafzai}, \citenamefont {Yadav}, \citenamefont {Amiri}, \citenamefont
  {Staddon}, \citenamefont {Errami}, \citenamefont {Jaspard}, \citenamefont
  {Banerjee},\ and\ \citenamefont {Murrell}}]{Yousafzai2022}%
  \BibitemOpen
  \bibfield  {author} {\bibinfo {author} {\bibfnamefont {M.~S.}\ \bibnamefont
  {Yousafzai}}, \bibinfo {author} {\bibfnamefont {V.}~\bibnamefont {Yadav}},
  \bibinfo {author} {\bibfnamefont {S.}~\bibnamefont {Amiri}}, \bibinfo
  {author} {\bibfnamefont {M.~F.}\ \bibnamefont {Staddon}}, \bibinfo {author}
  {\bibfnamefont {Y.}~\bibnamefont {Errami}}, \bibinfo {author} {\bibfnamefont
  {G.}~\bibnamefont {Jaspard}}, \bibinfo {author} {\bibfnamefont
  {S.}~\bibnamefont {Banerjee}},\ and\ \bibinfo {author} {\bibfnamefont
  {M.}~\bibnamefont {Murrell}},\ }\bibfield  {title} {\bibinfo {title}
  {Cell-matrix elastocapillary interactions drive pressure-based wetting of
  cell aggregates},\ }\href {https://doi.org/10.1103/physrevx.12.031027}
  {\bibfield  {journal} {\bibinfo  {journal} {Physical Review X}\ }\textbf
  {\bibinfo {volume} {12}},\ \bibinfo {pages} {031027} (\bibinfo {year}
  {2022})}\BibitemShut {NoStop}%
\bibitem [{\citenamefont {Armengol-Collado}\ \emph {et~al.}(2023)\citenamefont
  {Armengol-Collado}, \citenamefont {Carenza}, \citenamefont {Eckert},
  \citenamefont {Krommydas},\ and\ \citenamefont
  {Giomi}}]{ArmengolCollado2023}%
  \BibitemOpen
  \bibfield  {author} {\bibinfo {author} {\bibfnamefont {J.-M.}\ \bibnamefont
  {Armengol-Collado}}, \bibinfo {author} {\bibfnamefont {L.~N.}\ \bibnamefont
  {Carenza}}, \bibinfo {author} {\bibfnamefont {J.}~\bibnamefont {Eckert}},
  \bibinfo {author} {\bibfnamefont {D.}~\bibnamefont {Krommydas}},\ and\
  \bibinfo {author} {\bibfnamefont {L.}~\bibnamefont {Giomi}},\ }\bibfield
  {title} {\bibinfo {title} {Epithelia are multiscale active liquid crystals},\
  }\href {https://doi.org/10.1038/s41567-023-02179-0} {\bibfield  {journal}
  {\bibinfo  {journal} {Nature Physics}\ }\textbf {\bibinfo {volume} {19}},\
  \bibinfo {pages} {1773} (\bibinfo {year} {2023})}\BibitemShut {NoStop}%
\bibitem [{\citenamefont {Cavagna}\ \emph {et~al.}(2023)\citenamefont
  {Cavagna}, \citenamefont {Di~Carlo}, \citenamefont {Giardina}, \citenamefont
  {Grigera}, \citenamefont {Melillo}, \citenamefont {Parisi}, \citenamefont
  {Pisegna},\ and\ \citenamefont {Scandolo}}]{Cavagna2023}%
  \BibitemOpen
  \bibfield  {author} {\bibinfo {author} {\bibfnamefont {A.}~\bibnamefont
  {Cavagna}}, \bibinfo {author} {\bibfnamefont {L.}~\bibnamefont {Di~Carlo}},
  \bibinfo {author} {\bibfnamefont {I.}~\bibnamefont {Giardina}}, \bibinfo
  {author} {\bibfnamefont {T.~S.}\ \bibnamefont {Grigera}}, \bibinfo {author}
  {\bibfnamefont {S.}~\bibnamefont {Melillo}}, \bibinfo {author} {\bibfnamefont
  {L.}~\bibnamefont {Parisi}}, \bibinfo {author} {\bibfnamefont
  {G.}~\bibnamefont {Pisegna}},\ and\ \bibinfo {author} {\bibfnamefont
  {M.}~\bibnamefont {Scandolo}},\ }\bibfield  {title} {\bibinfo {title}
  {Natural swarms in 3.99 dimensions},\ }\href
  {https://doi.org/10.1038/s41567-023-02028-0} {\bibfield  {journal} {\bibinfo
  {journal} {Nature Physics}\ }\textbf {\bibinfo {volume} {19}},\ \bibinfo
  {pages} {1043} (\bibinfo {year} {2023})}\BibitemShut {NoStop}%
\bibitem [{\citenamefont {Gorbonos}\ \emph {et~al.}(2024)\citenamefont
  {Gorbonos}, \citenamefont {Oberhauser}, \citenamefont {Costello},
  \citenamefont {Günzel}, \citenamefont {Couzin-Fuchs}, \citenamefont
  {Koger},\ and\ \citenamefont {Couzin}}]{Gorbonos2024}%
  \BibitemOpen
  \bibfield  {author} {\bibinfo {author} {\bibfnamefont {D.}~\bibnamefont
  {Gorbonos}}, \bibinfo {author} {\bibfnamefont {F.~B.}\ \bibnamefont
  {Oberhauser}}, \bibinfo {author} {\bibfnamefont {L.~L.}\ \bibnamefont
  {Costello}}, \bibinfo {author} {\bibfnamefont {Y.}~\bibnamefont {Günzel}},
  \bibinfo {author} {\bibfnamefont {E.}~\bibnamefont {Couzin-Fuchs}}, \bibinfo
  {author} {\bibfnamefont {B.}~\bibnamefont {Koger}},\ and\ \bibinfo {author}
  {\bibfnamefont {I.~D.}\ \bibnamefont {Couzin}},\ }\bibfield  {title}
  {\bibinfo {title} {An effective hydrodynamic description of marching
  locusts},\ }\href {https://doi.org/10.1088/1478-3975/ad2219} {\bibfield
  {journal} {\bibinfo  {journal} {Physical Biology}\ }\textbf {\bibinfo
  {volume} {21}},\ \bibinfo {pages} {026004} (\bibinfo {year}
  {2024})}\BibitemShut {NoStop}%
\bibitem [{\citenamefont {Hughes}(2003)}]{Hughes2003}%
  \BibitemOpen
  \bibfield  {author} {\bibinfo {author} {\bibfnamefont {R.~L.}\ \bibnamefont
  {Hughes}},\ }\bibfield  {title} {\bibinfo {title} {The flow of human
  crowds},\ }\href {https://doi.org/10.1146/annurev.fluid.35.101101.161136}
  {\bibfield  {journal} {\bibinfo  {journal} {Annual Review of Fluid
  Mechanics}\ }\textbf {\bibinfo {volume} {35}},\ \bibinfo {pages} {169}
  (\bibinfo {year} {2003})}\BibitemShut {NoStop}%
\bibitem [{\citenamefont {Bain}\ and\ \citenamefont
  {Bartolo}(2019)}]{Bain2019}%
  \BibitemOpen
  \bibfield  {author} {\bibinfo {author} {\bibfnamefont {N.}~\bibnamefont
  {Bain}}\ and\ \bibinfo {author} {\bibfnamefont {D.}~\bibnamefont {Bartolo}},\
  }\bibfield  {title} {\bibinfo {title} {Dynamic response and hydrodynamics of
  polarized crowds},\ }\href {https://doi.org/10.1126/science.aat9891}
  {\bibfield  {journal} {\bibinfo  {journal} {Science}\ }\textbf {\bibinfo
  {volume} {363}},\ \bibinfo {pages} {46} (\bibinfo {year} {2019})}\BibitemShut
  {NoStop}%
\bibitem [{\citenamefont {Gu}\ \emph {et~al.}(2025)\citenamefont {Gu},
  \citenamefont {Guiselin}, \citenamefont {Bain}, \citenamefont {Zuriguel},\
  and\ \citenamefont {Bartolo}}]{Gu2025}%
  \BibitemOpen
  \bibfield  {author} {\bibinfo {author} {\bibfnamefont {F.}~\bibnamefont
  {Gu}}, \bibinfo {author} {\bibfnamefont {B.}~\bibnamefont {Guiselin}},
  \bibinfo {author} {\bibfnamefont {N.}~\bibnamefont {Bain}}, \bibinfo {author}
  {\bibfnamefont {I.}~\bibnamefont {Zuriguel}},\ and\ \bibinfo {author}
  {\bibfnamefont {D.}~\bibnamefont {Bartolo}},\ }\bibfield  {title} {\bibinfo
  {title} {Emergence of collective oscillations in massive human crowds},\
  }\href {https://doi.org/10.1038/s41586-024-08514-6} {\bibfield  {journal}
  {\bibinfo  {journal} {Nature}\ }\textbf {\bibinfo {volume} {638}},\ \bibinfo
  {pages} {112} (\bibinfo {year} {2025})}\BibitemShut {NoStop}%
\bibitem [{\citenamefont {Marchetti}\ \emph {et~al.}(2013)\citenamefont
  {Marchetti}, \citenamefont {Joanny}, \citenamefont {Ramaswamy}, \citenamefont
  {Liverpool}, \citenamefont {Prost}, \citenamefont {Rao},\ and\ \citenamefont
  {Simha}}]{Marchetti2013}%
  \BibitemOpen
  \bibfield  {author} {\bibinfo {author} {\bibfnamefont {M.~C.}\ \bibnamefont
  {Marchetti}}, \bibinfo {author} {\bibfnamefont {J.-F.}\ \bibnamefont
  {Joanny}}, \bibinfo {author} {\bibfnamefont {S.}~\bibnamefont {Ramaswamy}},
  \bibinfo {author} {\bibfnamefont {T.~B.}\ \bibnamefont {Liverpool}}, \bibinfo
  {author} {\bibfnamefont {J.}~\bibnamefont {Prost}}, \bibinfo {author}
  {\bibfnamefont {M.}~\bibnamefont {Rao}},\ and\ \bibinfo {author}
  {\bibfnamefont {R.~A.}\ \bibnamefont {Simha}},\ }\bibfield  {title} {\bibinfo
  {title} {{Hydrodynamics of soft active matter}},\ }\href
  {https://doi.org/10.1103/RevModPhys.85.1143} {\bibfield  {journal} {\bibinfo
  {journal} {Reviews of Modern Physics}\ }\textbf {\bibinfo {volume} {85}},\
  \bibinfo {pages} {1143} (\bibinfo {year} {2013})}\BibitemShut {NoStop}%
\bibitem [{\citenamefont {Vrugt}\ and\ \citenamefont
  {Wittkowski}(2024)}]{Vrugt2024}%
  \BibitemOpen
  \bibfield  {author} {\bibinfo {author} {\bibfnamefont {M.~t.}\ \bibnamefont
  {Vrugt}}\ and\ \bibinfo {author} {\bibfnamefont {R.}~\bibnamefont
  {Wittkowski}},\ }\bibfield  {title} {\bibinfo {title} {A review of active
  matter reviews},\ }\bibfield  {journal} {\bibinfo  {journal} {arXiv}\ }\href
  {https://doi.org/10.48550/ARXIV.2405.15751} {10.48550/ARXIV.2405.15751}
  (\bibinfo {year} {2024}),\ \Eprint {https://arxiv.org/abs/2405.15751}
  {arXiv:2405.15751} \BibitemShut {NoStop}%
\bibitem [{\citenamefont {Helbing}\ and\ \citenamefont
  {Moln{\'{a}}r}(1995)}]{Helbing1995}%
  \BibitemOpen
  \bibfield  {author} {\bibinfo {author} {\bibfnamefont {D.}~\bibnamefont
  {Helbing}}\ and\ \bibinfo {author} {\bibfnamefont {P.}~\bibnamefont
  {Moln{\'{a}}r}},\ }\bibfield  {title} {\bibinfo {title} {Social force model
  for pedestrian dynamics},\ }\href {https://doi.org/10.1103/physreve.51.4282}
  {\bibfield  {journal} {\bibinfo  {journal} {Physical Review E}\ }\textbf
  {\bibinfo {volume} {51}},\ \bibinfo {pages} {4282} (\bibinfo {year}
  {1995})}\BibitemShut {NoStop}%
\bibitem [{\citenamefont {Ballerini}\ \emph {et~al.}(2008)\citenamefont
  {Ballerini}, \citenamefont {Cabibbo}, \citenamefont {Candelier},
  \citenamefont {Cavagna}, \citenamefont {Cisbani}, \citenamefont {Giardina},
  \citenamefont {Lecomte}, \citenamefont {Orlandi}, \citenamefont {Parisi},
  \citenamefont {Procaccini}, \citenamefont {Viale},\ and\ \citenamefont
  {Zdravkovic}}]{Ballerini2008}%
  \BibitemOpen
  \bibfield  {author} {\bibinfo {author} {\bibfnamefont {M.}~\bibnamefont
  {Ballerini}}, \bibinfo {author} {\bibfnamefont {N.}~\bibnamefont {Cabibbo}},
  \bibinfo {author} {\bibfnamefont {R.}~\bibnamefont {Candelier}}, \bibinfo
  {author} {\bibfnamefont {A.}~\bibnamefont {Cavagna}}, \bibinfo {author}
  {\bibfnamefont {E.}~\bibnamefont {Cisbani}}, \bibinfo {author} {\bibfnamefont
  {I.}~\bibnamefont {Giardina}}, \bibinfo {author} {\bibfnamefont
  {V.}~\bibnamefont {Lecomte}}, \bibinfo {author} {\bibfnamefont
  {A.}~\bibnamefont {Orlandi}}, \bibinfo {author} {\bibfnamefont
  {G.}~\bibnamefont {Parisi}}, \bibinfo {author} {\bibfnamefont
  {A.}~\bibnamefont {Procaccini}}, \bibinfo {author} {\bibfnamefont
  {M.}~\bibnamefont {Viale}},\ and\ \bibinfo {author} {\bibfnamefont
  {V.}~\bibnamefont {Zdravkovic}},\ }\bibfield  {title} {\bibinfo {title}
  {Interaction ruling animal collective behavior depends on topological rather
  than metric distance: Evidence from a field study},\ }\href
  {https://doi.org/10.1073/pnas.0711437105} {\bibfield  {journal} {\bibinfo
  {journal} {Proceedings of the National Academy of Sciences}\ }\textbf
  {\bibinfo {volume} {105}},\ \bibinfo {pages} {1232} (\bibinfo {year}
  {2008})}\BibitemShut {NoStop}%
\bibitem [{\citenamefont {Corbetta}\ and\ \citenamefont
  {Toschi}(2023)}]{Corbetta2023}%
  \BibitemOpen
  \bibfield  {author} {\bibinfo {author} {\bibfnamefont {A.}~\bibnamefont
  {Corbetta}}\ and\ \bibinfo {author} {\bibfnamefont {F.}~\bibnamefont
  {Toschi}},\ }\bibfield  {title} {\bibinfo {title} {Physics of human crowds},\
  }\href {https://doi.org/10.1146/annurev-conmatphys-031620-100450} {\bibfield
  {journal} {\bibinfo  {journal} {Annual Review of Condensed Matter Physics}\
  }\textbf {\bibinfo {volume} {14}},\ \bibinfo {pages} {311} (\bibinfo {year}
  {2023})}\BibitemShut {NoStop}%
\bibitem [{\citenamefont {Neumann}\ and\ \citenamefont
  {Morgenstern}(2007)}]{Neumann2007}%
  \BibitemOpen
  \bibfield  {author} {\bibinfo {author} {\bibfnamefont {J.~V.}\ \bibnamefont
  {Neumann}}\ and\ \bibinfo {author} {\bibfnamefont {O.}~\bibnamefont
  {Morgenstern}},\ }\href@noop {} {\emph {\bibinfo {title} {Theory of Games and
  Economic Behavior}}}\ (\bibinfo  {publisher} {Princeton University Press},\
  \bibinfo {year} {2007})\ p.\ \bibinfo {pages} {776}\BibitemShut {NoStop}%
\bibitem [{\citenamefont {Osborne}\ and\ \citenamefont
  {Rubinstein}(2006)}]{Osborne2006}%
  \BibitemOpen
  \bibfield  {author} {\bibinfo {author} {\bibfnamefont {M.~J.}\ \bibnamefont
  {Osborne}}\ and\ \bibinfo {author} {\bibfnamefont {A.}~\bibnamefont
  {Rubinstein}},\ }\href@noop {} {\emph {\bibinfo {title} {A course in game
  theory}}},\ \bibinfo {edition} {12th}\ ed.\ (\bibinfo  {publisher} {MIT
  Press},\ \bibinfo {address} {Cambridge, Mass. [u.a.]},\ \bibinfo {year}
  {2006})\ \bibinfo {note} {literaturverz. S. [321] - 339}\BibitemShut
  {NoStop}%
\bibitem [{\citenamefont {Williams}\ and\ \citenamefont
  {Collins}(2001)}]{Williams2001}%
  \BibitemOpen
  \bibfield  {author} {\bibinfo {author} {\bibfnamefont {D.~R.}\ \bibnamefont
  {Williams}}\ and\ \bibinfo {author} {\bibfnamefont {C.}~\bibnamefont
  {Collins}},\ }\bibfield  {title} {\bibinfo {title} {Racial residential
  segregation: A fundamental cause of racial disparities in health},\ }\href
  {https://doi.org/10.1016/s0033-3549(04)50068-7} {\bibfield  {journal}
  {\bibinfo  {journal} {Public Health Reports}\ }\textbf {\bibinfo {volume}
  {116}},\ \bibinfo {pages} {404} (\bibinfo {year} {2001})}\BibitemShut
  {NoStop}%
\bibitem [{\citenamefont {Pager}\ and\ \citenamefont
  {Shepherd}(2008)}]{Pager2008}%
  \BibitemOpen
  \bibfield  {author} {\bibinfo {author} {\bibfnamefont {D.}~\bibnamefont
  {Pager}}\ and\ \bibinfo {author} {\bibfnamefont {H.}~\bibnamefont
  {Shepherd}},\ }\bibfield  {title} {\bibinfo {title} {The sociology of
  discrimination: Racial discrimination in employment, housing, credit, and
  consumer markets},\ }\href
  {https://doi.org/10.1146/annurev.soc.33.040406.131740} {\bibfield  {journal}
  {\bibinfo  {journal} {Annual Review of Sociology}\ }\textbf {\bibinfo
  {volume} {34}},\ \bibinfo {pages} {181} (\bibinfo {year} {2008})}\BibitemShut
  {NoStop}%
\bibitem [{\citenamefont {Reardon}\ and\ \citenamefont
  {Owens}(2014)}]{Reardon2014}%
  \BibitemOpen
  \bibfield  {author} {\bibinfo {author} {\bibfnamefont {S.~F.}\ \bibnamefont
  {Reardon}}\ and\ \bibinfo {author} {\bibfnamefont {A.}~\bibnamefont
  {Owens}},\ }\bibfield  {title} {\bibinfo {title} {60 years after brown:
  Trends and consequences of school segregation},\ }\href
  {https://doi.org/10.1146/annurev-soc-071913-043152} {\bibfield  {journal}
  {\bibinfo  {journal} {Annual Review of Sociology}\ }\textbf {\bibinfo
  {volume} {40}},\ \bibinfo {pages} {199} (\bibinfo {year} {2014})}\BibitemShut
  {NoStop}%
\bibitem [{\citenamefont {Alexander}\ and\ \citenamefont
  {Currie}(2017)}]{Alexander2017}%
  \BibitemOpen
  \bibfield  {author} {\bibinfo {author} {\bibfnamefont {D.}~\bibnamefont
  {Alexander}}\ and\ \bibinfo {author} {\bibfnamefont {J.}~\bibnamefont
  {Currie}},\ }\bibfield  {title} {\bibinfo {title} {Is it who you are or where
  you live? residential segregation and racial gaps in childhood asthma},\
  }\href {https://doi.org/10.1016/j.jhealeco.2017.07.003} {\bibfield  {journal}
  {\bibinfo  {journal} {Journal of Health Economics}\ }\textbf {\bibinfo
  {volume} {55}},\ \bibinfo {pages} {186} (\bibinfo {year} {2017})}\BibitemShut
  {NoStop}%
\bibitem [{\citenamefont {Charles}(2003)}]{Charles2003}%
  \BibitemOpen
  \bibfield  {author} {\bibinfo {author} {\bibfnamefont {C.~Z.}\ \bibnamefont
  {Charles}},\ }\bibfield  {title} {\bibinfo {title} {The dynamics of racial
  residential segregation},\ }\href
  {https://doi.org/10.1146/annurev.soc.29.010202.100002} {\bibfield  {journal}
  {\bibinfo  {journal} {Annual Review of Sociology}\ }\textbf {\bibinfo
  {volume} {29}},\ \bibinfo {pages} {167} (\bibinfo {year} {2003})}\BibitemShut
  {NoStop}%
\bibitem [{\citenamefont {Hwang}\ and\ \citenamefont
  {McDaniel}(2022)}]{Hwang2022}%
  \BibitemOpen
  \bibfield  {author} {\bibinfo {author} {\bibfnamefont {J.}~\bibnamefont
  {Hwang}}\ and\ \bibinfo {author} {\bibfnamefont {T.~W.}\ \bibnamefont
  {McDaniel}},\ }\bibfield  {title} {\bibinfo {title} {Racialized reshuffling:
  Urban change and the persistence of segregation in the twenty-first
  century},\ }\href {https://doi.org/10.1146/annurev-soc-030420-014126}
  {\bibfield  {journal} {\bibinfo  {journal} {Annual Review of Sociology}\
  }\textbf {\bibinfo {volume} {48}},\ \bibinfo {pages} {397} (\bibinfo {year}
  {2022})}\BibitemShut {NoStop}%
\bibitem [{\citenamefont {Du~Bois}(2018)}]{DuBois2018}%
  \BibitemOpen
  \bibfield  {author} {\bibinfo {author} {\bibfnamefont {W.~E.~B.}\
  \bibnamefont {Du~Bois}},\ }\href@noop {} {\emph {\bibinfo {title} {The souls
  of black folk}}},\ Penguin classics\ (\bibinfo  {publisher} {Penguin Books},\
  \bibinfo {address} {New York},\ \bibinfo {year} {2018})\BibitemShut {NoStop}%
\bibitem [{\citenamefont {Massey}(1990)}]{Massey1990}%
  \BibitemOpen
  \bibfield  {author} {\bibinfo {author} {\bibfnamefont {D.~S.}\ \bibnamefont
  {Massey}},\ }\bibfield  {title} {\bibinfo {title} {American apartheid:
  Segregation and the making of the underclass},\ }\href
  {https://doi.org/10.1086/229532} {\bibfield  {journal} {\bibinfo  {journal}
  {American Journal of Sociology}\ }\textbf {\bibinfo {volume} {96}},\ \bibinfo
  {pages} {329} (\bibinfo {year} {1990})}\BibitemShut {NoStop}%
\bibitem [{\citenamefont {Jargowsky}(1998)}]{Jargowsky1998}%
  \BibitemOpen
  \bibfield  {author} {\bibinfo {author} {\bibfnamefont {P.~A.}\ \bibnamefont
  {Jargowsky}},\ }\href@noop {} {\emph {\bibinfo {title} {Poverty and Place}}}\
  (\bibinfo  {publisher} {Russell Sage Foundation},\ \bibinfo {address} {New
  York},\ \bibinfo {year} {1998})\ \bibinfo {note} {description based upon
  print version of record}\BibitemShut {NoStop}%
\bibitem [{\citenamefont {Taeuber}\ and\ \citenamefont
  {Taeuber}(2008)}]{Taeuber2008}%
  \BibitemOpen
  \bibfield  {author} {\bibinfo {author} {\bibfnamefont {K.~E.}\ \bibnamefont
  {Taeuber}}\ and\ \bibinfo {author} {\bibfnamefont {A.~F.}\ \bibnamefont
  {Taeuber}},\ }\href@noop {} {\emph {\bibinfo {title} {Residential segregation
  and neighborhood change}}}\ (\bibinfo  {publisher} {Transaction Publishers},\
  \bibinfo {year} {2008})\BibitemShut {NoStop}%
\bibitem [{\citenamefont {Manson}\ \emph {et~al.}(2022)\citenamefont {Manson},
  \citenamefont {Schroeder}, \citenamefont {Van~Riper}, \citenamefont
  {Kugler},\ and\ \citenamefont {Ruggles}}]{Manson2022}%
  \BibitemOpen
  \bibfield  {author} {\bibinfo {author} {\bibfnamefont {S.}~\bibnamefont
  {Manson}}, \bibinfo {author} {\bibfnamefont {J.}~\bibnamefont {Schroeder}},
  \bibinfo {author} {\bibfnamefont {D.}~\bibnamefont {Van~Riper}}, \bibinfo
  {author} {\bibfnamefont {T.}~\bibnamefont {Kugler}},\ and\ \bibinfo {author}
  {\bibfnamefont {S.}~\bibnamefont {Ruggles}},\ }\href
  {https://doi.org/10.18128/D050.V17.0} {\bibinfo {title} {National historical
  geographic information system: Version 17.0}} (\bibinfo {year}
  {2022})\BibitemShut {NoStop}%
\bibitem [{\citenamefont {Schelling}(1971)}]{Schelling1971}%
  \BibitemOpen
  \bibfield  {author} {\bibinfo {author} {\bibfnamefont {T.~C.}\ \bibnamefont
  {Schelling}},\ }\bibfield  {title} {\bibinfo {title} {Dynamic models of
  segregation},\ }\href {https://doi.org/10.1080/0022250x.1971.9989794}
  {\bibfield  {journal} {\bibinfo  {journal} {The Journal of Mathematical
  Sociology}\ }\textbf {\bibinfo {volume} {1}},\ \bibinfo {pages} {143}
  (\bibinfo {year} {1971})}\BibitemShut {NoStop}%
\bibitem [{\citenamefont {Schelling}(1980)}]{Schelling1980}%
  \BibitemOpen
  \bibfield  {author} {\bibinfo {author} {\bibfnamefont {T.~C.}\ \bibnamefont
  {Schelling}},\ }\href@noop {} {\emph {\bibinfo {title} {Micromotives and
  macrobehaviour.}}}\ (\bibinfo  {publisher} {W.W.Norton},\ \bibinfo {year}
  {1980})\ p.\ \bibinfo {pages} {256}\BibitemShut {NoStop}%
\bibitem [{\citenamefont {Vinkovi{\'{c}}}\ and\ \citenamefont
  {Kirman}(2006)}]{Vinkovic2006}%
  \BibitemOpen
  \bibfield  {author} {\bibinfo {author} {\bibfnamefont {D.}~\bibnamefont
  {Vinkovi{\'{c}}}}\ and\ \bibinfo {author} {\bibfnamefont {A.}~\bibnamefont
  {Kirman}},\ }\bibfield  {title} {\bibinfo {title} {A physical analogue of the
  schelling model},\ }\href {https://doi.org/10.1073/pnas.0609371103}
  {\bibfield  {journal} {\bibinfo  {journal} {Proceedings of the National
  Academy of Sciences}\ }\textbf {\bibinfo {volume} {103}},\ \bibinfo {pages}
  {19261} (\bibinfo {year} {2006})}\BibitemShut {NoStop}%
\bibitem [{\citenamefont {Grauwin}\ \emph {et~al.}(2009)\citenamefont
  {Grauwin}, \citenamefont {Bertin}, \citenamefont {Lemoy},\ and\ \citenamefont
  {Jensen}}]{Grauwin2009}%
  \BibitemOpen
  \bibfield  {author} {\bibinfo {author} {\bibfnamefont {S.}~\bibnamefont
  {Grauwin}}, \bibinfo {author} {\bibfnamefont {E.}~\bibnamefont {Bertin}},
  \bibinfo {author} {\bibfnamefont {R.}~\bibnamefont {Lemoy}},\ and\ \bibinfo
  {author} {\bibfnamefont {P.}~\bibnamefont {Jensen}},\ }\bibfield  {title}
  {\bibinfo {title} {Competition between collective and individual dynamics},\
  }\href {https://doi.org/10.1073/pnas.0906263106} {\bibfield  {journal}
  {\bibinfo  {journal} {Proceedings of the National Academy of Sciences}\
  }\textbf {\bibinfo {volume} {106}},\ \bibinfo {pages} {20622} (\bibinfo
  {year} {2009})}\BibitemShut {NoStop}%
\bibitem [{\citenamefont {Grauwin}\ \emph {et~al.}(2012)\citenamefont
  {Grauwin}, \citenamefont {Goffette-Nagot},\ and\ \citenamefont
  {Jensen}}]{Grauwin2012}%
  \BibitemOpen
  \bibfield  {author} {\bibinfo {author} {\bibfnamefont {S.}~\bibnamefont
  {Grauwin}}, \bibinfo {author} {\bibfnamefont {F.}~\bibnamefont
  {Goffette-Nagot}},\ and\ \bibinfo {author} {\bibfnamefont {P.}~\bibnamefont
  {Jensen}},\ }\bibfield  {title} {\bibinfo {title} {Dynamic models of
  residential segregation: An analytical solution},\ }\href
  {https://doi.org/10.1016/j.jpubeco.2011.08.011} {\bibfield  {journal}
  {\bibinfo  {journal} {Journal of Public Economics}\ }\textbf {\bibinfo
  {volume} {96}},\ \bibinfo {pages} {124} (\bibinfo {year} {2012})}\BibitemShut
  {NoStop}%
\bibitem [{\citenamefont {Clark}\ and\ \citenamefont
  {Fossett}(2008)}]{Clark2008}%
  \BibitemOpen
  \bibfield  {author} {\bibinfo {author} {\bibfnamefont {W.~A.~V.}\
  \bibnamefont {Clark}}\ and\ \bibinfo {author} {\bibfnamefont
  {M.}~\bibnamefont {Fossett}},\ }\bibfield  {title} {\bibinfo {title}
  {Understanding the social context of the schelling segregation model},\
  }\href {https://doi.org/10.1073/pnas.0708155105} {\bibfield  {journal}
  {\bibinfo  {journal} {Proceedings of the National Academy of Sciences}\
  }\textbf {\bibinfo {volume} {105}},\ \bibinfo {pages} {4109} (\bibinfo {year}
  {2008})}\BibitemShut {NoStop}%
\bibitem [{\citenamefont {Fossett}(2006)}]{Fossett2006}%
  \BibitemOpen
  \bibfield  {author} {\bibinfo {author} {\bibfnamefont {M.}~\bibnamefont
  {Fossett}},\ }\bibfield  {title} {\bibinfo {title} {Ethnic preferences,
  social distance dynamics, and residential segregation: Theoretical
  explorations using simulation analysis$\ast$},\ }\href
  {https://doi.org/10.1080/00222500500544052} {\bibfield  {journal} {\bibinfo
  {journal} {The Journal of Mathematical Sociology}\ }\textbf {\bibinfo
  {volume} {30}},\ \bibinfo {pages} {185} (\bibinfo {year} {2006})}\BibitemShut
  {NoStop}%
\bibitem [{\citenamefont {Gauvin}\ \emph {et~al.}(2009)\citenamefont {Gauvin},
  \citenamefont {Vannimenus},\ and\ \citenamefont {Nadal}}]{Gauvin2009}%
  \BibitemOpen
  \bibfield  {author} {\bibinfo {author} {\bibfnamefont {L.}~\bibnamefont
  {Gauvin}}, \bibinfo {author} {\bibfnamefont {J.}~\bibnamefont {Vannimenus}},\
  and\ \bibinfo {author} {\bibfnamefont {J.-P.}\ \bibnamefont {Nadal}},\
  }\bibfield  {title} {\bibinfo {title} {Phase diagram of a schelling
  segregation model},\ }\href {https://doi.org/10.1140/epjb/e2009-00234-0}
  {\bibfield  {journal} {\bibinfo  {journal} {The European Physical Journal B}\
  }\textbf {\bibinfo {volume} {70}},\ \bibinfo {pages} {293} (\bibinfo {year}
  {2009})}\BibitemShut {NoStop}%
\bibitem [{\citenamefont {Zakine}\ \emph {et~al.}(2024)\citenamefont {Zakine},
  \citenamefont {Garnier-Brun}, \citenamefont {Becharat},\ and\ \citenamefont
  {Benzaquen}}]{Zakine2024}%
  \BibitemOpen
  \bibfield  {author} {\bibinfo {author} {\bibfnamefont {R.}~\bibnamefont
  {Zakine}}, \bibinfo {author} {\bibfnamefont {J.}~\bibnamefont
  {Garnier-Brun}}, \bibinfo {author} {\bibfnamefont {A.-C.}\ \bibnamefont
  {Becharat}},\ and\ \bibinfo {author} {\bibfnamefont {M.}~\bibnamefont
  {Benzaquen}},\ }\bibfield  {title} {\bibinfo {title} {Socioeconomic agents as
  active matter in nonequilibrium sakoda-schelling models},\ }\href
  {https://doi.org/10.1103/PhysRevE.109.044310} {\bibfield  {journal} {\bibinfo
   {journal} {Physical Review E}\ }\textbf {\bibinfo {volume} {109}},\ \bibinfo
  {pages} {044310} (\bibinfo {year} {2024})},\ \Eprint
  {https://arxiv.org/abs/2307.14270} {2307.14270} \BibitemShut {NoStop}%
\bibitem [{\citenamefont {Garnier-Brun}\ \emph {et~al.}(2024)\citenamefont
  {Garnier-Brun}, \citenamefont {Zakine},\ and\ \citenamefont
  {Benzaquen}}]{GarnierBrun2024a}%
  \BibitemOpen
  \bibfield  {author} {\bibinfo {author} {\bibfnamefont {J.}~\bibnamefont
  {Garnier-Brun}}, \bibinfo {author} {\bibfnamefont {R.}~\bibnamefont
  {Zakine}},\ and\ \bibinfo {author} {\bibfnamefont {M.}~\bibnamefont
  {Benzaquen}},\ }\bibfield  {title} {\bibinfo {title} {From nonequilibrium to
  equilibrium: Insights from a two-population occupation model},\ }\bibfield
  {journal} {\bibinfo  {journal} {arXiv}\ }\href
  {https://doi.org/10.48550/ARXIV.2412.14996} {10.48550/ARXIV.2412.14996}
  (\bibinfo {year} {2024}),\ \Eprint {https://arxiv.org/abs/2412.14996}
  {arXiv:2412.14996} \BibitemShut {NoStop}%
\bibitem [{\citenamefont {Becharat}\ \emph {et~al.}(2024)\citenamefont
  {Becharat}, \citenamefont {Benzaquen},\ and\ \citenamefont
  {Bouchaud}}]{Becharat2024}%
  \BibitemOpen
  \bibfield  {author} {\bibinfo {author} {\bibfnamefont {A.-C.}\ \bibnamefont
  {Becharat}}, \bibinfo {author} {\bibfnamefont {M.}~\bibnamefont
  {Benzaquen}},\ and\ \bibinfo {author} {\bibfnamefont {J.-P.}\ \bibnamefont
  {Bouchaud}},\ }\bibfield  {title} {\bibinfo {title} {The diffusive nature of
  housing prices},\ }\bibfield  {journal} {\bibinfo  {journal} {arXiv}\ }\href
  {https://doi.org/10.48550/ARXIV.2412.14624} {10.48550/ARXIV.2412.14624}
  (\bibinfo {year} {2024}),\ \Eprint {https://arxiv.org/abs/2412.14624}
  {arXiv:2412.14624} \BibitemShut {NoStop}%
\bibitem [{\citenamefont {Chen}\ \emph {et~al.}(2020)\citenamefont {Chen},
  \citenamefont {Kinkhabwala}, \citenamefont {Barron}, \citenamefont {Hall},
  \citenamefont {Arias},\ and\ \citenamefont {Cohen}}]{Chen2020}%
  \BibitemOpen
  \bibfield  {author} {\bibinfo {author} {\bibfnamefont {Y.}~\bibnamefont
  {Chen}}, \bibinfo {author} {\bibfnamefont {Y.~A.}\ \bibnamefont
  {Kinkhabwala}}, \bibinfo {author} {\bibfnamefont {B.}~\bibnamefont {Barron}},
  \bibinfo {author} {\bibfnamefont {M.}~\bibnamefont {Hall}}, \bibinfo {author}
  {\bibfnamefont {T.~A.}\ \bibnamefont {Arias}},\ and\ \bibinfo {author}
  {\bibfnamefont {I.}~\bibnamefont {Cohen}},\ }\bibfield  {title} {\bibinfo
  {title} {Small-area population forecast in a segregated city using
  density-functional fluctuation theory},\ }\bibfield  {journal} {\bibinfo
  {journal} {arXiv}\ }\href {https://doi.org/10.48550/ARXIV.2008.09663}
  {10.48550/ARXIV.2008.09663} (\bibinfo {year} {2020}),\ \Eprint
  {https://arxiv.org/abs/2008.09663} {arXiv:2008.09663} \BibitemShut {NoStop}%
\bibitem [{\citenamefont {Kinkhabwala}\ \emph {et~al.}(2021)\citenamefont
  {Kinkhabwala}, \citenamefont {Barron}, \citenamefont {Hall}, \citenamefont
  {Arias},\ and\ \citenamefont {Cohen}}]{Kinkhabwala2021}%
  \BibitemOpen
  \bibfield  {author} {\bibinfo {author} {\bibfnamefont {Y.~A.}\ \bibnamefont
  {Kinkhabwala}}, \bibinfo {author} {\bibfnamefont {B.}~\bibnamefont {Barron}},
  \bibinfo {author} {\bibfnamefont {M.}~\bibnamefont {Hall}}, \bibinfo {author}
  {\bibfnamefont {T.~A.}\ \bibnamefont {Arias}},\ and\ \bibinfo {author}
  {\bibfnamefont {I.}~\bibnamefont {Cohen}},\ }\bibfield  {title} {\bibinfo
  {title} {Forecasting racial dynamics at the neighborhood scale using
  density-functional fluctuation theory},\ }\href@noop {} {\bibfield  {journal}
  {\bibinfo  {journal} {arXiv}\ } (\bibinfo {year} {2021})},\ \Eprint
  {https://arxiv.org/abs/2108.04084} {arXiv:2108.04084} \BibitemShut {NoStop}%
\bibitem [{\citenamefont {Barron}\ \emph {et~al.}(2022)\citenamefont {Barron},
  \citenamefont {Kinkhabwala}, \citenamefont {Hess}, \citenamefont {Hall},
  \citenamefont {Cohen},\ and\ \citenamefont {Arias}}]{Barron2022}%
  \BibitemOpen
  \bibfield  {author} {\bibinfo {author} {\bibfnamefont {B.}~\bibnamefont
  {Barron}}, \bibinfo {author} {\bibfnamefont {Y.~A.}\ \bibnamefont
  {Kinkhabwala}}, \bibinfo {author} {\bibfnamefont {C.}~\bibnamefont {Hess}},
  \bibinfo {author} {\bibfnamefont {M.}~\bibnamefont {Hall}}, \bibinfo {author}
  {\bibfnamefont {I.}~\bibnamefont {Cohen}},\ and\ \bibinfo {author}
  {\bibfnamefont {T.~A.}\ \bibnamefont {Arias}},\ }\bibfield  {title} {\bibinfo
  {title} {Extending the use of information theory in segregation analyses to
  construct comprehensive models of segregation},\ }\href@noop {} {\bibfield
  {journal} {\bibinfo  {journal} {arXiv}\ } (\bibinfo {year} {2022})},\ \Eprint
  {https://arxiv.org/abs/2212.06980} {arXiv:2212.06980} \BibitemShut {NoStop}%
\bibitem [{\citenamefont {Card}\ \emph {et~al.}(2008)\citenamefont {Card},
  \citenamefont {Mas},\ and\ \citenamefont {Rothstein}}]{Card2008}%
  \BibitemOpen
  \bibfield  {author} {\bibinfo {author} {\bibfnamefont {D.}~\bibnamefont
  {Card}}, \bibinfo {author} {\bibfnamefont {A.}~\bibnamefont {Mas}},\ and\
  \bibinfo {author} {\bibfnamefont {J.}~\bibnamefont {Rothstein}},\ }\bibfield
  {title} {\bibinfo {title} {Tipping and the dynamics of segregation},\ }\href
  {https://doi.org/10.1162/qjec.2008.123.1.177} {\bibfield  {journal} {\bibinfo
   {journal} {Quarterly Journal of Economics}\ }\textbf {\bibinfo {volume}
  {123}},\ \bibinfo {pages} {177} (\bibinfo {year} {2008})}\BibitemShut
  {NoStop}%
\bibitem [{\citenamefont {Zhang}(2011)}]{Zhang2011}%
  \BibitemOpen
  \bibfield  {author} {\bibinfo {author} {\bibfnamefont {J.}~\bibnamefont
  {Zhang}},\ }\bibfield  {title} {\bibinfo {title} {Tipping and residental
  segregation: A unified schelling model},\ }\href
  {https://doi.org/10.1111/j.1467-9787.2010.00671.x} {\bibfield  {journal}
  {\bibinfo  {journal} {Journal of Regional Science}\ }\textbf {\bibinfo
  {volume} {51}},\ \bibinfo {pages} {167} (\bibinfo {year} {2011})}\BibitemShut
  {NoStop}%
\bibitem [{\citenamefont {Gualdi}\ \emph {et~al.}(2015)\citenamefont {Gualdi},
  \citenamefont {Tarzia}, \citenamefont {Zamponi},\ and\ \citenamefont
  {Bouchaud}}]{Gualdi2015}%
  \BibitemOpen
  \bibfield  {author} {\bibinfo {author} {\bibfnamefont {S.}~\bibnamefont
  {Gualdi}}, \bibinfo {author} {\bibfnamefont {M.}~\bibnamefont {Tarzia}},
  \bibinfo {author} {\bibfnamefont {F.}~\bibnamefont {Zamponi}},\ and\ \bibinfo
  {author} {\bibfnamefont {J.-P.}\ \bibnamefont {Bouchaud}},\ }\bibfield
  {title} {\bibinfo {title} {Tipping points in macroeconomic agent-based
  models},\ }\href {https://doi.org/10.1016/j.jedc.2014.08.003} {\bibfield
  {journal} {\bibinfo  {journal} {Journal of Economic Dynamics and Control}\
  }\textbf {\bibinfo {volume} {50}},\ \bibinfo {pages} {29} (\bibinfo {year}
  {2015})}\BibitemShut {NoStop}%
\bibitem [{\citenamefont {Schmitt}\ \emph
  {et~al.}(2023{\natexlab{a}})\citenamefont {Schmitt}, \citenamefont {Colen},
  \citenamefont {Sala}, \citenamefont {Devany}, \citenamefont {Seetharaman},
  \citenamefont {Gardel}, \citenamefont {Oakes},\ and\ \citenamefont
  {Vitelli}}]{Schmitt2023}%
  \BibitemOpen
  \bibfield  {author} {\bibinfo {author} {\bibfnamefont {M.~S.}\ \bibnamefont
  {Schmitt}}, \bibinfo {author} {\bibfnamefont {J.}~\bibnamefont {Colen}},
  \bibinfo {author} {\bibfnamefont {S.}~\bibnamefont {Sala}}, \bibinfo {author}
  {\bibfnamefont {J.}~\bibnamefont {Devany}}, \bibinfo {author} {\bibfnamefont
  {S.}~\bibnamefont {Seetharaman}}, \bibinfo {author} {\bibfnamefont {M.~L.}\
  \bibnamefont {Gardel}}, \bibinfo {author} {\bibfnamefont {P.~W.}\
  \bibnamefont {Oakes}},\ and\ \bibinfo {author} {\bibfnamefont
  {V.}~\bibnamefont {Vitelli}},\ }\href@noop {} {\bibinfo {title} {Zyxin is all
  you need: Machine learning adherent cell mechanics}} (\bibinfo {year}
  {2023}{\natexlab{a}}),\ \Eprint {https://arxiv.org/abs/2303.00176}
  {arxiv:2303.00176 [cond-mat, physics:physics]} \BibitemShut {NoStop}%
\bibitem [{\citenamefont {Schmitt}\ \emph
  {et~al.}(2023{\natexlab{b}})\citenamefont {Schmitt}, \citenamefont
  {Koch-Janusz}, \citenamefont {Fruchart}, \citenamefont {Seara}, \citenamefont
  {Rust},\ and\ \citenamefont {Vitelli}}]{Schmitt2023b}%
  \BibitemOpen
  \bibfield  {author} {\bibinfo {author} {\bibfnamefont {M.~S.}\ \bibnamefont
  {Schmitt}}, \bibinfo {author} {\bibfnamefont {M.}~\bibnamefont
  {Koch-Janusz}}, \bibinfo {author} {\bibfnamefont {M.}~\bibnamefont
  {Fruchart}}, \bibinfo {author} {\bibfnamefont {D.~S.}\ \bibnamefont {Seara}},
  \bibinfo {author} {\bibfnamefont {M.}~\bibnamefont {Rust}},\ and\ \bibinfo
  {author} {\bibfnamefont {V.}~\bibnamefont {Vitelli}},\ }\bibfield  {title}
  {\bibinfo {title} {Information theory for data-driven model reduction in
  physics and biology},\ }\bibfield  {journal} {\bibinfo  {journal} {arXiv}\
  }\href {https://doi.org/10.48550/ARXIV.2312.06608}
  {10.48550/ARXIV.2312.06608} (\bibinfo {year} {2023}{\natexlab{b}}),\ \Eprint
  {https://arxiv.org/abs/2312.06608} {2312.06608 [cond-mat.stat-mech]}
  \BibitemShut {NoStop}%
\bibitem [{\citenamefont {Lefebvre}\ \emph {et~al.}(2023)\citenamefont
  {Lefebvre}, \citenamefont {Colen}, \citenamefont {Claussen}, \citenamefont
  {Brauns}, \citenamefont {Raich}, \citenamefont {Mitchell}, \citenamefont
  {Fruchart}, \citenamefont {Vitelli},\ and\ \citenamefont
  {Streichan}}]{Lefebvre2023}%
  \BibitemOpen
  \bibfield  {author} {\bibinfo {author} {\bibfnamefont {M.}~\bibnamefont
  {Lefebvre}}, \bibinfo {author} {\bibfnamefont {J.}~\bibnamefont {Colen}},
  \bibinfo {author} {\bibfnamefont {N.}~\bibnamefont {Claussen}}, \bibinfo
  {author} {\bibfnamefont {F.}~\bibnamefont {Brauns}}, \bibinfo {author}
  {\bibfnamefont {M.}~\bibnamefont {Raich}}, \bibinfo {author} {\bibfnamefont
  {N.}~\bibnamefont {Mitchell}}, \bibinfo {author} {\bibfnamefont
  {M.}~\bibnamefont {Fruchart}}, \bibinfo {author} {\bibfnamefont
  {V.}~\bibnamefont {Vitelli}},\ and\ \bibinfo {author} {\bibfnamefont {S.~J.}\
  \bibnamefont {Streichan}},\ }\bibfield  {title} {\bibinfo {title} {Learning a
  conserved mechanism for early neuroectoderm morphogenesis},\ }\bibfield
  {journal} {\bibinfo  {journal} {bioRxiv}\ }\href
  {https://doi.org/10.1101/2023.12.22.573058} {10.1101/2023.12.22.573058}
  (\bibinfo {year} {2023})\BibitemShut {NoStop}%
\bibitem [{\citenamefont {Duncan}\ and\ \citenamefont
  {Duncan}(1955)}]{Duncan1955}%
  \BibitemOpen
  \bibfield  {author} {\bibinfo {author} {\bibfnamefont {O.~D.}\ \bibnamefont
  {Duncan}}\ and\ \bibinfo {author} {\bibfnamefont {B.}~\bibnamefont
  {Duncan}},\ }\bibfield  {title} {\bibinfo {title} {A methodological analysis
  of segregation indexes},\ }\href {https://doi.org/10.2307/2088328} {\bibfield
   {journal} {\bibinfo  {journal} {American Sociological Review}\ }\textbf
  {\bibinfo {volume} {20}},\ \bibinfo {pages} {210} (\bibinfo {year}
  {1955})}\BibitemShut {NoStop}%
\bibitem [{\citenamefont {Reardon}\ and\ \citenamefont
  {O'Sullivan}(2004)}]{Reardon2004}%
  \BibitemOpen
  \bibfield  {author} {\bibinfo {author} {\bibfnamefont {S.~F.}\ \bibnamefont
  {Reardon}}\ and\ \bibinfo {author} {\bibfnamefont {D.}~\bibnamefont
  {O'Sullivan}},\ }\bibfield  {title} {\bibinfo {title} {Measures of spatial
  segregation},\ }\href {https://doi.org/10.1111/j.0081-1750.2004.00150.x}
  {\bibfield  {journal} {\bibinfo  {journal} {Sociological Methodology}\
  }\textbf {\bibinfo {volume} {34}},\ \bibinfo {pages} {121} (\bibinfo {year}
  {2004})}\BibitemShut {NoStop}%
\bibitem [{\citenamefont {Roberto}(2015)}]{Roberto2015}%
  \BibitemOpen
  \bibfield  {author} {\bibinfo {author} {\bibfnamefont {E.}~\bibnamefont
  {Roberto}},\ }\bibfield  {title} {\bibinfo {title} {The divergence index: A
  decomposable measure of segregation and inequality},\ }\bibfield  {journal}
  {\bibinfo  {journal} {arXiv}\ }\href
  {https://doi.org/10.48550/ARXIV.1508.01167} {10.48550/ARXIV.1508.01167}
  (\bibinfo {year} {2015}),\ \Eprint {https://arxiv.org/abs/1508.01167}
  {1508.01167} \BibitemShut {NoStop}%
\bibitem [{\citenamefont {Simonyan}\ \emph {et~al.}(2014)\citenamefont
  {Simonyan}, \citenamefont {Vedaldi},\ and\ \citenamefont
  {Zisserman}}]{Simonyan2014}%
  \BibitemOpen
  \bibfield  {author} {\bibinfo {author} {\bibfnamefont {K.}~\bibnamefont
  {Simonyan}}, \bibinfo {author} {\bibfnamefont {A.}~\bibnamefont {Vedaldi}},\
  and\ \bibinfo {author} {\bibfnamefont {A.}~\bibnamefont {Zisserman}},\
  }\href@noop {} {\bibinfo {title} {Deep inside convolutional networks:
  Visualising image classification models and saliency maps}} (\bibinfo {year}
  {2014}),\ \Eprint {https://arxiv.org/abs/1312.6034} {arXiv:1312.6034 [cs.CV]}
  \BibitemShut {NoStop}%
\bibitem [{\citenamefont {Conti}\ \emph {et~al.}(2002)\citenamefont {Conti},
  \citenamefont {Meerson}, \citenamefont {Peleg},\ and\ \citenamefont
  {Sasorov}}]{Conti2002}%
  \BibitemOpen
  \bibfield  {author} {\bibinfo {author} {\bibfnamefont {M.}~\bibnamefont
  {Conti}}, \bibinfo {author} {\bibfnamefont {B.}~\bibnamefont {Meerson}},
  \bibinfo {author} {\bibfnamefont {A.}~\bibnamefont {Peleg}},\ and\ \bibinfo
  {author} {\bibfnamefont {P.~V.}\ \bibnamefont {Sasorov}},\ }\bibfield
  {title} {\bibinfo {title} {Phase ordering with a global conservation law:
  Ostwald ripening and coalescence},\ }\href
  {https://doi.org/10.1103/physreve.65.046117} {\bibfield  {journal} {\bibinfo
  {journal} {Physical Review E}\ }\textbf {\bibinfo {volume} {65}},\ \bibinfo
  {pages} {046117} (\bibinfo {year} {2002})}\BibitemShut {NoStop}%
\bibitem [{\citenamefont {{US Census
  Bureau}}(2023{\natexlab{a}})}]{censusMigration2023}%
  \BibitemOpen
  \bibfield  {author} {\bibinfo {author} {\bibnamefont {{US Census Bureau}}},\
  }\href@noop {} {\bibinfo {title} {Cps historical migration/geographic
  mobility tables}},\ \bibinfo {howpublished}
  {\url{https://www.census.gov/data/tables/time-series/demo/geographic-mobility/historic.html}}
  (\bibinfo {year} {2023}{\natexlab{a}}),\ \bibinfo {note} {table A-6.
  Accessed: 2024-07-22}\BibitemShut {NoStop}%
\bibitem [{\citenamefont {Fisher}(1937)}]{Fisher1937}%
  \BibitemOpen
  \bibfield  {author} {\bibinfo {author} {\bibfnamefont {R.~A.}\ \bibnamefont
  {Fisher}},\ }\bibfield  {title} {\bibinfo {title} {The wave of advance of
  advantageous genes},\ }\href
  {https://doi.org/10.1111/j.1469-1809.1937.tb02153.x} {\bibfield  {journal}
  {\bibinfo  {journal} {Annals of Eugenics}\ }\textbf {\bibinfo {volume} {7}},\
  \bibinfo {pages} {355} (\bibinfo {year} {1937})}\BibitemShut {NoStop}%
\bibitem [{\citenamefont {Hallatschek}\ \emph {et~al.}(2023)\citenamefont
  {Hallatschek}, \citenamefont {Datta}, \citenamefont {Drescher}, \citenamefont
  {Dunkel}, \citenamefont {Elgeti}, \citenamefont {Waclaw},\ and\ \citenamefont
  {Wingreen}}]{Hallatschek2023}%
  \BibitemOpen
  \bibfield  {author} {\bibinfo {author} {\bibfnamefont {O.}~\bibnamefont
  {Hallatschek}}, \bibinfo {author} {\bibfnamefont {S.~S.}\ \bibnamefont
  {Datta}}, \bibinfo {author} {\bibfnamefont {K.}~\bibnamefont {Drescher}},
  \bibinfo {author} {\bibfnamefont {J.}~\bibnamefont {Dunkel}}, \bibinfo
  {author} {\bibfnamefont {J.}~\bibnamefont {Elgeti}}, \bibinfo {author}
  {\bibfnamefont {B.}~\bibnamefont {Waclaw}},\ and\ \bibinfo {author}
  {\bibfnamefont {N.~S.}\ \bibnamefont {Wingreen}},\ }\bibfield  {title}
  {\bibinfo {title} {Proliferating active matter},\ }\href
  {https://doi.org/10.1038/s42254-023-00593-0} {\bibfield  {journal} {\bibinfo
  {journal} {Nature Reviews Physics}\ }\textbf {\bibinfo {volume} {5}},\
  \bibinfo {pages} {407} (\bibinfo {year} {2023})}\BibitemShut {NoStop}%
\bibitem [{\citenamefont {Crow}\ and\ \citenamefont {Kimura}(1970)}]{Crow1970}%
  \BibitemOpen
  \bibfield  {author} {\bibinfo {author} {\bibfnamefont {J.~F.}\ \bibnamefont
  {Crow}}\ and\ \bibinfo {author} {\bibfnamefont {M.}~\bibnamefont {Kimura}},\
  }\href@noop {} {\emph {\bibinfo {title} {An Introduction to Population
  Genetics Theory}}}\ (\bibinfo  {publisher} {Blackburn Press},\ \bibinfo
  {year} {1970})\ p.\ \bibinfo {pages} {608}\BibitemShut {NoStop}%
\bibitem [{\citenamefont {Kimura}(1983)}]{Kimura1983}%
  \BibitemOpen
  \bibfield  {author} {\bibinfo {author} {\bibfnamefont {M.}~\bibnamefont
  {Kimura}},\ }\href@noop {} {\emph {\bibinfo {title} {The neutral theory of
  molecular evolution}}}\ (\bibinfo  {publisher} {Cambridge University Press},\
  \bibinfo {year} {1983})\BibitemShut {NoStop}%
\bibitem [{\citenamefont {Tsimring}\ \emph {et~al.}(1996)\citenamefont
  {Tsimring}, \citenamefont {Levine},\ and\ \citenamefont
  {Kessler}}]{Tsimring1996}%
  \BibitemOpen
  \bibfield  {author} {\bibinfo {author} {\bibfnamefont {L.~S.}\ \bibnamefont
  {Tsimring}}, \bibinfo {author} {\bibfnamefont {H.}~\bibnamefont {Levine}},\
  and\ \bibinfo {author} {\bibfnamefont {D.~A.}\ \bibnamefont {Kessler}},\
  }\bibfield  {title} {\bibinfo {title} {{RNA} virus evolution via a
  fitness-space model},\ }\href {https://doi.org/10.1103/physrevlett.76.4440}
  {\bibfield  {journal} {\bibinfo  {journal} {Physical Review Letters}\
  }\textbf {\bibinfo {volume} {76}},\ \bibinfo {pages} {4440} (\bibinfo {year}
  {1996})}\BibitemShut {NoStop}%
\bibitem [{\citenamefont {Mustonen}\ and\ \citenamefont
  {Lässig}(2009)}]{Mustonen2009}%
  \BibitemOpen
  \bibfield  {author} {\bibinfo {author} {\bibfnamefont {V.}~\bibnamefont
  {Mustonen}}\ and\ \bibinfo {author} {\bibfnamefont {M.}~\bibnamefont
  {Lässig}},\ }\bibfield  {title} {\bibinfo {title} {From fitness landscapes
  to seascapes: non-equilibrium dynamics of selection and adaptation},\ }\href
  {https://doi.org/10.1016/j.tig.2009.01.002} {\bibfield  {journal} {\bibinfo
  {journal} {Trends in Genetics}\ }\textbf {\bibinfo {volume} {25}},\ \bibinfo
  {pages} {111} (\bibinfo {year} {2009})}\BibitemShut {NoStop}%
\bibitem [{\citenamefont {Cremer}\ \emph {et~al.}(2019)\citenamefont {Cremer},
  \citenamefont {Honda}, \citenamefont {Tang}, \citenamefont {Wong-Ng},
  \citenamefont {Vergassola},\ and\ \citenamefont {Hwa}}]{Cremer2019}%
  \BibitemOpen
  \bibfield  {author} {\bibinfo {author} {\bibfnamefont {J.}~\bibnamefont
  {Cremer}}, \bibinfo {author} {\bibfnamefont {T.}~\bibnamefont {Honda}},
  \bibinfo {author} {\bibfnamefont {Y.}~\bibnamefont {Tang}}, \bibinfo {author}
  {\bibfnamefont {J.}~\bibnamefont {Wong-Ng}}, \bibinfo {author} {\bibfnamefont
  {M.}~\bibnamefont {Vergassola}},\ and\ \bibinfo {author} {\bibfnamefont
  {T.}~\bibnamefont {Hwa}},\ }\bibfield  {title} {\bibinfo {title} {Chemotaxis
  as a navigation strategy to boost range expansion},\ }\href
  {https://doi.org/10.1038/s41586-019-1733-y} {\bibfield  {journal} {\bibinfo
  {journal} {Nature}\ }\textbf {\bibinfo {volume} {575}},\ \bibinfo {pages}
  {658} (\bibinfo {year} {2019})}\BibitemShut {NoStop}%
\bibitem [{\citenamefont {Cahn}\ and\ \citenamefont
  {Hilliard}(1958)}]{Cahn1958}%
  \BibitemOpen
  \bibfield  {author} {\bibinfo {author} {\bibfnamefont {J.~W.}\ \bibnamefont
  {Cahn}}\ and\ \bibinfo {author} {\bibfnamefont {J.~E.}\ \bibnamefont
  {Hilliard}},\ }\bibfield  {title} {\bibinfo {title} {Free energy of a
  nonuniform system. i. interfacial free energy},\ }\href
  {https://doi.org/10.1063/1.1744102} {\bibfield  {journal} {\bibinfo
  {journal} {The Journal of Chemical Physics}\ }\textbf {\bibinfo {volume}
  {28}},\ \bibinfo {pages} {258} (\bibinfo {year} {1958})}\BibitemShut
  {NoStop}%
\bibitem [{\citenamefont {Hohenberg}\ and\ \citenamefont
  {Halperin}(1977)}]{Hohenberg1977}%
  \BibitemOpen
  \bibfield  {author} {\bibinfo {author} {\bibfnamefont {P.~C.}\ \bibnamefont
  {Hohenberg}}\ and\ \bibinfo {author} {\bibfnamefont {B.~I.}\ \bibnamefont
  {Halperin}},\ }\bibfield  {title} {\bibinfo {title} {Theory of dynamic
  critical phenomena},\ }\href {https://doi.org/10.1103/revmodphys.49.435}
  {\bibfield  {journal} {\bibinfo  {journal} {Reviews of Modern Physics}\
  }\textbf {\bibinfo {volume} {49}},\ \bibinfo {pages} {435} (\bibinfo {year}
  {1977})}\BibitemShut {NoStop}%
\bibitem [{\citenamefont {Keller}\ and\ \citenamefont
  {Segel}(1971)}]{Keller1971}%
  \BibitemOpen
  \bibfield  {author} {\bibinfo {author} {\bibfnamefont {E.~F.}\ \bibnamefont
  {Keller}}\ and\ \bibinfo {author} {\bibfnamefont {L.~A.}\ \bibnamefont
  {Segel}},\ }\bibfield  {title} {\bibinfo {title} {Model for chemotaxis},\
  }\href {https://doi.org/10.1016/0022-5193(71)90050-6} {\bibfield  {journal}
  {\bibinfo  {journal} {Journal of Theoretical Biology}\ }\textbf {\bibinfo
  {volume} {30}},\ \bibinfo {pages} {225} (\bibinfo {year} {1971})}\BibitemShut
  {NoStop}%
\bibitem [{\citenamefont {Adler}(1975)}]{Adler1975}%
  \BibitemOpen
  \bibfield  {author} {\bibinfo {author} {\bibfnamefont {J.}~\bibnamefont
  {Adler}},\ }\bibfield  {title} {\bibinfo {title} {Chemotaxis in bacteria},\
  }\href {https://doi.org/10.1146/annurev.bi.44.070175.002013} {\bibfield
  {journal} {\bibinfo  {journal} {Annual Review of Biochemistry}\ }\textbf
  {\bibinfo {volume} {44}},\ \bibinfo {pages} {341} (\bibinfo {year}
  {1975})}\BibitemShut {NoStop}%
\bibitem [{\citenamefont {Berg}(1975)}]{Berg1975}%
  \BibitemOpen
  \bibfield  {author} {\bibinfo {author} {\bibfnamefont {H.~C.}\ \bibnamefont
  {Berg}},\ }\bibfield  {title} {\bibinfo {title} {Chemotaxis in bacteria},\
  }\href {https://doi.org/10.1146/annurev.bb.04.060175.001003} {\bibfield
  {journal} {\bibinfo  {journal} {Annual Review of Biophysics and
  Bioengineering}\ }\textbf {\bibinfo {volume} {4}},\ \bibinfo {pages} {119}
  (\bibinfo {year} {1975})}\BibitemShut {NoStop}%
\bibitem [{\citenamefont {Vergassola}\ \emph {et~al.}(2007)\citenamefont
  {Vergassola}, \citenamefont {Villermaux},\ and\ \citenamefont
  {Shraiman}}]{Vergassola2007}%
  \BibitemOpen
  \bibfield  {author} {\bibinfo {author} {\bibfnamefont {M.}~\bibnamefont
  {Vergassola}}, \bibinfo {author} {\bibfnamefont {E.}~\bibnamefont
  {Villermaux}},\ and\ \bibinfo {author} {\bibfnamefont {B.~I.}\ \bibnamefont
  {Shraiman}},\ }\bibfield  {title} {\bibinfo {title} {‘infotaxis’ as a
  strategy for searching without gradients},\ }\href
  {https://doi.org/10.1038/nature05464} {\bibfield  {journal} {\bibinfo
  {journal} {Nature}\ }\textbf {\bibinfo {volume} {445}},\ \bibinfo {pages}
  {406–409} (\bibinfo {year} {2007})}\BibitemShut {NoStop}%
\bibitem [{\citenamefont {Farley}\ \emph {et~al.}(1993)\citenamefont {Farley},
  \citenamefont {Steeh}, \citenamefont {Jackson}, \citenamefont {Krysan},\ and\
  \citenamefont {Reeves}}]{Farley1993}%
  \BibitemOpen
  \bibfield  {author} {\bibinfo {author} {\bibfnamefont {R.}~\bibnamefont
  {Farley}}, \bibinfo {author} {\bibfnamefont {C.}~\bibnamefont {Steeh}},
  \bibinfo {author} {\bibfnamefont {T.}~\bibnamefont {Jackson}}, \bibinfo
  {author} {\bibfnamefont {M.}~\bibnamefont {Krysan}},\ and\ \bibinfo {author}
  {\bibfnamefont {K.}~\bibnamefont {Reeves}},\ }\bibfield  {title} {\bibinfo
  {title} {Continued racial residential segregation in detroit: "chocolate
  city, vanilla suburbs" revisited},\ }\href
  {http://www.jstor.org/stable/24832753} {\bibfield  {journal} {\bibinfo
  {journal} {Journal of Housing Research}\ }\textbf {\bibinfo {volume} {4}},\
  \bibinfo {pages} {1} (\bibinfo {year} {1993})}\BibitemShut {NoStop}%
\bibitem [{\citenamefont {Clark}(2002)}]{Clark2002}%
  \BibitemOpen
  \bibfield  {author} {\bibinfo {author} {\bibfnamefont {W.~A.~V.}\
  \bibnamefont {Clark}},\ }\bibfield  {title} {\bibinfo {title} {Ethnic
  preferences and ethnic perceptions in multi-ethnic settings},\ }\href
  {https://doi.org/10.2747/0272-3638.23.3.237} {\bibfield  {journal} {\bibinfo
  {journal} {Urban Geography}\ }\textbf {\bibinfo {volume} {23}},\ \bibinfo
  {pages} {237} (\bibinfo {year} {2002})}\BibitemShut {NoStop}%
\bibitem [{\citenamefont {Farley}\ \emph {et~al.}(1978)\citenamefont {Farley},
  \citenamefont {Schuman}, \citenamefont {Bianchi}, \citenamefont {Colasanto},\
  and\ \citenamefont {Hatchett}}]{Farley1978}%
  \BibitemOpen
  \bibfield  {author} {\bibinfo {author} {\bibfnamefont {R.}~\bibnamefont
  {Farley}}, \bibinfo {author} {\bibfnamefont {H.}~\bibnamefont {Schuman}},
  \bibinfo {author} {\bibfnamefont {S.}~\bibnamefont {Bianchi}}, \bibinfo
  {author} {\bibfnamefont {D.}~\bibnamefont {Colasanto}},\ and\ \bibinfo
  {author} {\bibfnamefont {S.}~\bibnamefont {Hatchett}},\ }\bibfield  {title}
  {\bibinfo {title} {"chocolate city, vanilla suburbs": Will the trend toward
  racially separate communities continue?},\ }\href
  {https://doi.org/10.1016/0049-089x(78)90017-0} {\bibfield  {journal}
  {\bibinfo  {journal} {Social Science Research}\ }\textbf {\bibinfo {volume}
  {7}},\ \bibinfo {pages} {319} (\bibinfo {year} {1978})}\BibitemShut {NoStop}%
\bibitem [{\citenamefont {Clark}(1991)}]{Clark1991}%
  \BibitemOpen
  \bibfield  {author} {\bibinfo {author} {\bibfnamefont {W.}~\bibnamefont
  {Clark}},\ }\bibfield  {title} {\bibinfo {title} {Residential preferences and
  neighborhood racial segregation: A test of the schelling segregation model},\
  }\href {https://doi.org/10.2307/2061333} {\bibfield  {journal} {\bibinfo
  {journal} {Demography}\ }\textbf {\bibinfo {volume} {28}},\ \bibinfo {pages}
  {1} (\bibinfo {year} {1991})}\BibitemShut {NoStop}%
\bibitem [{\citenamefont {Bobo}\ and\ \citenamefont
  {Zubrinsky}(1996)}]{Bobo1996}%
  \BibitemOpen
  \bibfield  {author} {\bibinfo {author} {\bibfnamefont {L.}~\bibnamefont
  {Bobo}}\ and\ \bibinfo {author} {\bibfnamefont {C.~L.}\ \bibnamefont
  {Zubrinsky}},\ }\bibfield  {title} {\bibinfo {title} {Attitudes on
  residential integration: Perceived status differences, mere in-group
  preference, or racial prejudice?},\ }\href {https://doi.org/10.2307/2580385}
  {\bibfield  {journal} {\bibinfo  {journal} {Social Forces}\ }\textbf
  {\bibinfo {volume} {74}},\ \bibinfo {pages} {883} (\bibinfo {year}
  {1996})}\BibitemShut {NoStop}%
\bibitem [{\citenamefont {Giacomin}\ and\ \citenamefont
  {Lebowitz}(1996)}]{Giacomin1996}%
  \BibitemOpen
  \bibfield  {author} {\bibinfo {author} {\bibfnamefont {G.}~\bibnamefont
  {Giacomin}}\ and\ \bibinfo {author} {\bibfnamefont {J.~L.}\ \bibnamefont
  {Lebowitz}},\ }\bibfield  {title} {\bibinfo {title} {Exact macroscopic
  description of phase segregation in model alloys with long range
  interactions},\ }\href {https://doi.org/10.1103/physrevlett.76.1094}
  {\bibfield  {journal} {\bibinfo  {journal} {Physical Review Letters}\
  }\textbf {\bibinfo {volume} {76}},\ \bibinfo {pages} {1094} (\bibinfo {year}
  {1996})}\BibitemShut {NoStop}%
\bibitem [{\citenamefont {{US Census
  Bureau}}(2023{\natexlab{b}})}]{censusMigrationByRace2020}%
  \BibitemOpen
  \bibfield  {author} {\bibinfo {author} {\bibnamefont {{US Census Bureau}}},\
  }\href@noop {} {\bibinfo {title} {Cps historical migration/geographic
  mobility tables}},\ \bibinfo {howpublished}
  {\url{https://www.census.gov/data/tables/2020/demo/geographic-mobility/cps-2020.html}}
  (\bibinfo {year} {2023}{\natexlab{b}}),\ \bibinfo {note} {table 2. Accessed:
  2025-02-24}\BibitemShut {NoStop}%
\bibitem [{\citenamefont {Desmond}\ and\ \citenamefont
  {Shollenberger}(2015)}]{Desmond2015}%
  \BibitemOpen
  \bibfield  {author} {\bibinfo {author} {\bibfnamefont {M.}~\bibnamefont
  {Desmond}}\ and\ \bibinfo {author} {\bibfnamefont {T.}~\bibnamefont
  {Shollenberger}},\ }\bibfield  {title} {\bibinfo {title} {Forced displacement
  from rental housing: Prevalence and neighborhood consequences},\ }\href
  {https://doi.org/10.1007/s13524-015-0419-9} {\bibfield  {journal} {\bibinfo
  {journal} {Demography}\ }\textbf {\bibinfo {volume} {52}},\ \bibinfo {pages}
  {1751} (\bibinfo {year} {2015})}\BibitemShut {NoStop}%
\bibitem [{\citenamefont {Lemoy}\ \emph {et~al.}(2011)\citenamefont {Lemoy},
  \citenamefont {Bertin},\ and\ \citenamefont {Jensen}}]{Lemoy2011}%
  \BibitemOpen
  \bibfield  {author} {\bibinfo {author} {\bibfnamefont {R.}~\bibnamefont
  {Lemoy}}, \bibinfo {author} {\bibfnamefont {E.}~\bibnamefont {Bertin}},\ and\
  \bibinfo {author} {\bibfnamefont {P.}~\bibnamefont {Jensen}},\ }\bibfield
  {title} {\bibinfo {title} {Socio-economic utility and chemical potential},\
  }\href {https://doi.org/10.1209/0295-5075/93/38002} {\bibfield  {journal}
  {\bibinfo  {journal} {{EPL} (Europhysics Letters)}\ }\textbf {\bibinfo
  {volume} {93}},\ \bibinfo {pages} {38002} (\bibinfo {year}
  {2011})}\BibitemShut {NoStop}%
\bibitem [{\citenamefont {Fruchart}\ \emph {et~al.}(2021)\citenamefont
  {Fruchart}, \citenamefont {Hanai}, \citenamefont {Littlewood},\ and\
  \citenamefont {Vitelli}}]{Fruchart2021}%
  \BibitemOpen
  \bibfield  {author} {\bibinfo {author} {\bibfnamefont {M.}~\bibnamefont
  {Fruchart}}, \bibinfo {author} {\bibfnamefont {R.}~\bibnamefont {Hanai}},
  \bibinfo {author} {\bibfnamefont {P.~B.}\ \bibnamefont {Littlewood}},\ and\
  \bibinfo {author} {\bibfnamefont {V.}~\bibnamefont {Vitelli}},\ }\bibfield
  {title} {\bibinfo {title} {Non-reciprocal phase transitions},\ }\href
  {https://doi.org/10.1038/s41586-021-03375-9} {\bibfield  {journal} {\bibinfo
  {journal} {Nature}\ }\textbf {\bibinfo {volume} {592}},\ \bibinfo {pages}
  {363} (\bibinfo {year} {2021})}\BibitemShut {NoStop}%
\bibitem [{\citenamefont {Dinelli}\ \emph {et~al.}(2023)\citenamefont
  {Dinelli}, \citenamefont {O’Byrne}, \citenamefont {Curatolo}, \citenamefont
  {Zhao}, \citenamefont {Sollich},\ and\ \citenamefont
  {Tailleur}}]{Dinelli2023}%
  \BibitemOpen
  \bibfield  {author} {\bibinfo {author} {\bibfnamefont {A.}~\bibnamefont
  {Dinelli}}, \bibinfo {author} {\bibfnamefont {J.}~\bibnamefont {O’Byrne}},
  \bibinfo {author} {\bibfnamefont {A.}~\bibnamefont {Curatolo}}, \bibinfo
  {author} {\bibfnamefont {Y.}~\bibnamefont {Zhao}}, \bibinfo {author}
  {\bibfnamefont {P.}~\bibnamefont {Sollich}},\ and\ \bibinfo {author}
  {\bibfnamefont {J.}~\bibnamefont {Tailleur}},\ }\bibfield  {title} {\bibinfo
  {title} {Non-reciprocity across scales in active mixtures},\ }\bibfield
  {journal} {\bibinfo  {journal} {Nature Communications}\ }\textbf {\bibinfo
  {volume} {14}},\ \href {https://doi.org/10.1038/s41467-023-42713-5}
  {10.1038/s41467-023-42713-5} (\bibinfo {year} {2023})\BibitemShut {NoStop}%
\bibitem [{\citenamefont {You}\ \emph {et~al.}(2020)\citenamefont {You},
  \citenamefont {Baskaran},\ and\ \citenamefont {Marchetti}}]{You2020}%
  \BibitemOpen
  \bibfield  {author} {\bibinfo {author} {\bibfnamefont {Z.}~\bibnamefont
  {You}}, \bibinfo {author} {\bibfnamefont {A.}~\bibnamefont {Baskaran}},\ and\
  \bibinfo {author} {\bibfnamefont {M.~C.}\ \bibnamefont {Marchetti}},\
  }\bibfield  {title} {\bibinfo {title} {{Nonreciprocity as a generic route to
  traveling states}},\ }\href {https://doi.org/10.1073/pnas.2010318117}
  {\bibfield  {journal} {\bibinfo  {journal} {Proceedings of the National
  Academy of Sciences}\ }\textbf {\bibinfo {volume} {117}},\ \bibinfo {pages}
  {19767} (\bibinfo {year} {2020})}\BibitemShut {NoStop}%
\bibitem [{\citenamefont {Saha}\ \emph {et~al.}(2020)\citenamefont {Saha},
  \citenamefont {Agudo-Canalejo},\ and\ \citenamefont
  {Golestanian}}]{Saha2020}%
  \BibitemOpen
  \bibfield  {author} {\bibinfo {author} {\bibfnamefont {S.}~\bibnamefont
  {Saha}}, \bibinfo {author} {\bibfnamefont {J.}~\bibnamefont
  {Agudo-Canalejo}},\ and\ \bibinfo {author} {\bibfnamefont {R.}~\bibnamefont
  {Golestanian}},\ }\bibfield  {title} {\bibinfo {title} {{Scalar Active
  Mixtures: The Nonreciprocal Cahn-Hilliard Model}},\ }\href
  {https://doi.org/10.1103/PhysRevX.10.041009} {\bibfield  {journal} {\bibinfo
  {journal} {Physical Review X}\ }\textbf {\bibinfo {volume} {10}},\ \bibinfo
  {pages} {041009} (\bibinfo {year} {2020})}\BibitemShut {NoStop}%
\bibitem [{\citenamefont {Ivlev}\ \emph {et~al.}(2015)\citenamefont {Ivlev},
  \citenamefont {Bartnick}, \citenamefont {Heinen}, \citenamefont {Du},
  \citenamefont {Nosenko},\ and\ \citenamefont {Löwen}}]{Ivlev2015}%
  \BibitemOpen
  \bibfield  {author} {\bibinfo {author} {\bibfnamefont {A.~V.}\ \bibnamefont
  {Ivlev}}, \bibinfo {author} {\bibfnamefont {J.}~\bibnamefont {Bartnick}},
  \bibinfo {author} {\bibfnamefont {M.}~\bibnamefont {Heinen}}, \bibinfo
  {author} {\bibfnamefont {C.-R.}\ \bibnamefont {Du}}, \bibinfo {author}
  {\bibfnamefont {V.}~\bibnamefont {Nosenko}},\ and\ \bibinfo {author}
  {\bibfnamefont {H.}~\bibnamefont {Löwen}},\ }\bibfield  {title} {\bibinfo
  {title} {Statistical mechanics where newton’s third law is broken},\ }\href
  {https://doi.org/10.1103/physrevx.5.011035} {\bibfield  {journal} {\bibinfo
  {journal} {Physical Review X}\ }\textbf {\bibinfo {volume} {5}},\ \bibinfo
  {pages} {011035} (\bibinfo {year} {2015})}\BibitemShut {NoStop}%
\bibitem [{\citenamefont {Hargus}\ \emph {et~al.}(2021)\citenamefont {Hargus},
  \citenamefont {Epstein},\ and\ \citenamefont {Mandadapu}}]{Hargus2021}%
  \BibitemOpen
  \bibfield  {author} {\bibinfo {author} {\bibfnamefont {C.}~\bibnamefont
  {Hargus}}, \bibinfo {author} {\bibfnamefont {J.~M.}\ \bibnamefont
  {Epstein}},\ and\ \bibinfo {author} {\bibfnamefont {K.~K.}\ \bibnamefont
  {Mandadapu}},\ }\bibfield  {title} {\bibinfo {title} {Odd diffusivity of
  chiral random motion},\ }\href
  {https://doi.org/10.1103/physrevlett.127.178001} {\bibfield  {journal}
  {\bibinfo  {journal} {Physical Review Letters}\ }\textbf {\bibinfo {volume}
  {127}},\ \bibinfo {pages} {178001} (\bibinfo {year} {2021})}\BibitemShut
  {NoStop}%
\bibitem [{\citenamefont {Baek}\ \emph {et~al.}(2018)\citenamefont {Baek},
  \citenamefont {Solon}, \citenamefont {Xu}, \citenamefont {Nikola},\ and\
  \citenamefont {Kafri}}]{Baek2018}%
  \BibitemOpen
  \bibfield  {author} {\bibinfo {author} {\bibfnamefont {Y.}~\bibnamefont
  {Baek}}, \bibinfo {author} {\bibfnamefont {A.~P.}\ \bibnamefont {Solon}},
  \bibinfo {author} {\bibfnamefont {X.}~\bibnamefont {Xu}}, \bibinfo {author}
  {\bibfnamefont {N.}~\bibnamefont {Nikola}},\ and\ \bibinfo {author}
  {\bibfnamefont {Y.}~\bibnamefont {Kafri}},\ }\bibfield  {title} {\bibinfo
  {title} {Generic long-range interactions between passive bodies in an active
  fluid},\ }\href {https://doi.org/10.1103/physrevlett.120.058002} {\bibfield
  {journal} {\bibinfo  {journal} {Physical Review Letters}\ }\textbf {\bibinfo
  {volume} {120}},\ \bibinfo {pages} {058002} (\bibinfo {year}
  {2018})}\BibitemShut {NoStop}%
\bibitem [{\citenamefont {Frohoff-Hülsmann}\ \emph {et~al.}(2021)\citenamefont
  {Frohoff-Hülsmann}, \citenamefont {Wrembel},\ and\ \citenamefont
  {Thiele}}]{FrohoffHuelsmann2021}%
  \BibitemOpen
  \bibfield  {author} {\bibinfo {author} {\bibfnamefont {T.}~\bibnamefont
  {Frohoff-Hülsmann}}, \bibinfo {author} {\bibfnamefont {J.}~\bibnamefont
  {Wrembel}},\ and\ \bibinfo {author} {\bibfnamefont {U.}~\bibnamefont
  {Thiele}},\ }\bibfield  {title} {\bibinfo {title} {Suppression of coarsening
  and emergence of oscillatory behavior in a cahn-hilliard model with
  nonvariational coupling},\ }\href
  {https://doi.org/10.1103/physreve.103.042602} {\bibfield  {journal} {\bibinfo
   {journal} {Physical Review E}\ }\textbf {\bibinfo {volume} {103}},\ \bibinfo
  {pages} {042602} (\bibinfo {year} {2021})}\BibitemShut {NoStop}%
\bibitem [{\citenamefont {Brauns}\ and\ \citenamefont
  {Marchetti}(2024)}]{Brauns2024}%
  \BibitemOpen
  \bibfield  {author} {\bibinfo {author} {\bibfnamefont {F.}~\bibnamefont
  {Brauns}}\ and\ \bibinfo {author} {\bibfnamefont {M.~C.}\ \bibnamefont
  {Marchetti}},\ }\bibfield  {title} {\bibinfo {title} {Nonreciprocal pattern
  formation of conserved fields},\ }\href
  {https://doi.org/10.1103/physrevx.14.021014} {\bibfield  {journal} {\bibinfo
  {journal} {Physical Review X}\ }\textbf {\bibinfo {volume} {14}},\ \bibinfo
  {pages} {021014} (\bibinfo {year} {2024})}\BibitemShut {NoStop}%
\bibitem [{\citenamefont {Scheibner}\ \emph {et~al.}(2020)\citenamefont
  {Scheibner}, \citenamefont {Souslov}, \citenamefont {Banerjee}, \citenamefont
  {Sur{\'{o}}wka}, \citenamefont {Irvine},\ and\ \citenamefont
  {Vitelli}}]{Scheibner2020}%
  \BibitemOpen
  \bibfield  {author} {\bibinfo {author} {\bibfnamefont {C.}~\bibnamefont
  {Scheibner}}, \bibinfo {author} {\bibfnamefont {A.}~\bibnamefont {Souslov}},
  \bibinfo {author} {\bibfnamefont {D.}~\bibnamefont {Banerjee}}, \bibinfo
  {author} {\bibfnamefont {P.}~\bibnamefont {Sur{\'{o}}wka}}, \bibinfo {author}
  {\bibfnamefont {W.~T.~M.}\ \bibnamefont {Irvine}},\ and\ \bibinfo {author}
  {\bibfnamefont {V.}~\bibnamefont {Vitelli}},\ }\bibfield  {title} {\bibinfo
  {title} {Odd elasticity},\ }\href {https://doi.org/10.1038/s41567-020-0795-y}
  {\bibfield  {journal} {\bibinfo  {journal} {Nature Physics}\ }\textbf
  {\bibinfo {volume} {16}},\ \bibinfo {pages} {475} (\bibinfo {year}
  {2020})}\BibitemShut {NoStop}%
\bibitem [{\citenamefont {Fruchart}\ \emph {et~al.}(2023)\citenamefont
  {Fruchart}, \citenamefont {Scheibner},\ and\ \citenamefont
  {Vitelli}}]{Fruchart2023}%
  \BibitemOpen
  \bibfield  {author} {\bibinfo {author} {\bibfnamefont {M.}~\bibnamefont
  {Fruchart}}, \bibinfo {author} {\bibfnamefont {C.}~\bibnamefont
  {Scheibner}},\ and\ \bibinfo {author} {\bibfnamefont {V.}~\bibnamefont
  {Vitelli}},\ }\bibfield  {title} {\bibinfo {title} {Odd viscosity and odd
  elasticity},\ }\href
  {https://doi.org/10.1146/annurev-conmatphys-040821-125506} {\bibfield
  {journal} {\bibinfo  {journal} {Annual Review of Condensed Matter Physics}\
  }\textbf {\bibinfo {volume} {14}},\ \bibinfo {pages} {471} (\bibinfo {year}
  {2023})}\BibitemShut {NoStop}%
\bibitem [{\citenamefont {Banerjee}\ \emph {et~al.}(2022)\citenamefont
  {Banerjee}, \citenamefont {Mandal}, \citenamefont {Banerjee}, \citenamefont
  {Thutupalli},\ and\ \citenamefont {Rao}}]{Banerjee2022}%
  \BibitemOpen
  \bibfield  {author} {\bibinfo {author} {\bibfnamefont {J.~P.}\ \bibnamefont
  {Banerjee}}, \bibinfo {author} {\bibfnamefont {R.}~\bibnamefont {Mandal}},
  \bibinfo {author} {\bibfnamefont {D.~S.}\ \bibnamefont {Banerjee}}, \bibinfo
  {author} {\bibfnamefont {S.}~\bibnamefont {Thutupalli}},\ and\ \bibinfo
  {author} {\bibfnamefont {M.}~\bibnamefont {Rao}},\ }\bibfield  {title}
  {\bibinfo {title} {Unjamming and emergent nonreciprocity in active ploughing
  through a compressible viscoelastic fluid},\ }\bibfield  {journal} {\bibinfo
  {journal} {Nature Communications}\ }\textbf {\bibinfo {volume} {13}},\ \href
  {https://doi.org/10.1038/s41467-022-31984-z} {10.1038/s41467-022-31984-z}
  (\bibinfo {year} {2022})\BibitemShut {NoStop}%
\bibitem [{\citenamefont {Keim}\ \emph {et~al.}(2019)\citenamefont {Keim},
  \citenamefont {Paulsen}, \citenamefont {Zeravcic}, \citenamefont {Sastry},\
  and\ \citenamefont {Nagel}}]{Keim2019}%
  \BibitemOpen
  \bibfield  {author} {\bibinfo {author} {\bibfnamefont {N.~C.}\ \bibnamefont
  {Keim}}, \bibinfo {author} {\bibfnamefont {J.~D.}\ \bibnamefont {Paulsen}},
  \bibinfo {author} {\bibfnamefont {Z.}~\bibnamefont {Zeravcic}}, \bibinfo
  {author} {\bibfnamefont {S.}~\bibnamefont {Sastry}},\ and\ \bibinfo {author}
  {\bibfnamefont {S.~R.}\ \bibnamefont {Nagel}},\ }\bibfield  {title} {\bibinfo
  {title} {Memory formation in matter},\ }\href
  {https://doi.org/10.1103/revmodphys.91.035002} {\bibfield  {journal}
  {\bibinfo  {journal} {Reviews of Modern Physics}\ }\textbf {\bibinfo {volume}
  {91}},\ \bibinfo {pages} {035002} (\bibinfo {year} {2019})}\BibitemShut
  {NoStop}%
\bibitem [{\citenamefont {Kerr}\ \emph {et~al.}(2002)\citenamefont {Kerr},
  \citenamefont {Riley}, \citenamefont {Feldman},\ and\ \citenamefont
  {Bohannan}}]{Kerr2002}%
  \BibitemOpen
  \bibfield  {author} {\bibinfo {author} {\bibfnamefont {B.}~\bibnamefont
  {Kerr}}, \bibinfo {author} {\bibfnamefont {M.~A.}\ \bibnamefont {Riley}},
  \bibinfo {author} {\bibfnamefont {M.~W.}\ \bibnamefont {Feldman}},\ and\
  \bibinfo {author} {\bibfnamefont {B.~J.~M.}\ \bibnamefont {Bohannan}},\
  }\bibfield  {title} {\bibinfo {title} {Local dispersal promotes biodiversity
  in a real-life game of rock–paper–scissors},\ }\href
  {https://doi.org/10.1038/nature00823} {\bibfield  {journal} {\bibinfo
  {journal} {Nature}\ }\textbf {\bibinfo {volume} {418}},\ \bibinfo {pages}
  {171} (\bibinfo {year} {2002})}\BibitemShut {NoStop}%
\bibitem [{\citenamefont {Velicer}(2003)}]{Velicer2003}%
  \BibitemOpen
  \bibfield  {author} {\bibinfo {author} {\bibfnamefont {G.~J.}\ \bibnamefont
  {Velicer}},\ }\bibfield  {title} {\bibinfo {title} {Social strife in the
  microbial world},\ }\href {https://doi.org/10.1016/s0966-842x(03)00152-5}
  {\bibfield  {journal} {\bibinfo  {journal} {Trends in Microbiology}\ }\textbf
  {\bibinfo {volume} {11}},\ \bibinfo {pages} {330} (\bibinfo {year}
  {2003})}\BibitemShut {NoStop}%
\bibitem [{\citenamefont {Hibbing}\ \emph {et~al.}(2009)\citenamefont
  {Hibbing}, \citenamefont {Fuqua}, \citenamefont {Parsek},\ and\ \citenamefont
  {Peterson}}]{Hibbing2009}%
  \BibitemOpen
  \bibfield  {author} {\bibinfo {author} {\bibfnamefont {M.~E.}\ \bibnamefont
  {Hibbing}}, \bibinfo {author} {\bibfnamefont {C.}~\bibnamefont {Fuqua}},
  \bibinfo {author} {\bibfnamefont {M.~R.}\ \bibnamefont {Parsek}},\ and\
  \bibinfo {author} {\bibfnamefont {S.~B.}\ \bibnamefont {Peterson}},\
  }\bibfield  {title} {\bibinfo {title} {Bacterial competition: surviving and
  thriving in the microbial jungle},\ }\href
  {https://doi.org/10.1038/nrmicro2259} {\bibfield  {journal} {\bibinfo
  {journal} {Nature Reviews Microbiology}\ }\textbf {\bibinfo {volume} {8}},\
  \bibinfo {pages} {15} (\bibinfo {year} {2009})}\BibitemShut {NoStop}%
\bibitem [{\citenamefont {Kelsic}\ \emph {et~al.}(2015)\citenamefont {Kelsic},
  \citenamefont {Zhao}, \citenamefont {Vetsigian},\ and\ \citenamefont
  {Kishony}}]{Kelsic2015}%
  \BibitemOpen
  \bibfield  {author} {\bibinfo {author} {\bibfnamefont {E.~D.}\ \bibnamefont
  {Kelsic}}, \bibinfo {author} {\bibfnamefont {J.}~\bibnamefont {Zhao}},
  \bibinfo {author} {\bibfnamefont {K.}~\bibnamefont {Vetsigian}},\ and\
  \bibinfo {author} {\bibfnamefont {R.}~\bibnamefont {Kishony}},\ }\bibfield
  {title} {\bibinfo {title} {Counteraction of antibiotic production and
  degradation stabilizes microbial communities},\ }\href
  {https://doi.org/10.1038/nature14485} {\bibfield  {journal} {\bibinfo
  {journal} {Nature}\ }\textbf {\bibinfo {volume} {521}},\ \bibinfo {pages}
  {516} (\bibinfo {year} {2015})}\BibitemShut {NoStop}%
\bibitem [{\citenamefont {{Van Dyken}}\ \emph {et~al.}(2013)\citenamefont {{Van
  Dyken}}, \citenamefont {M{\"{u}}ller}, \citenamefont {Mack},\ and\
  \citenamefont {Desai}}]{VanDyken2013}%
  \BibitemOpen
  \bibfield  {author} {\bibinfo {author} {\bibfnamefont {J.~D.}\ \bibnamefont
  {{Van Dyken}}}, \bibinfo {author} {\bibfnamefont {M.~J.}\ \bibnamefont
  {M{\"{u}}ller}}, \bibinfo {author} {\bibfnamefont {K.~M.}\ \bibnamefont
  {Mack}},\ and\ \bibinfo {author} {\bibfnamefont {M.~M.}\ \bibnamefont
  {Desai}},\ }\bibfield  {title} {\bibinfo {title} {{Spatial Population
  Expansion Promotes the Evolution of Cooperation in an Experimental Prisoner's
  Dilemma}},\ }\href {https://doi.org/10.1016/j.cub.2013.04.026} {\bibfield
  {journal} {\bibinfo  {journal} {Current Biology}\ }\textbf {\bibinfo {volume}
  {23}},\ \bibinfo {pages} {919} (\bibinfo {year} {2013})}\BibitemShut
  {NoStop}%
\bibitem [{\citenamefont {Ratzke}\ \emph {et~al.}(2020)\citenamefont {Ratzke},
  \citenamefont {Barrere},\ and\ \citenamefont {Gore}}]{Ratzke2020}%
  \BibitemOpen
  \bibfield  {author} {\bibinfo {author} {\bibfnamefont {C.}~\bibnamefont
  {Ratzke}}, \bibinfo {author} {\bibfnamefont {J.}~\bibnamefont {Barrere}},\
  and\ \bibinfo {author} {\bibfnamefont {J.}~\bibnamefont {Gore}},\ }\bibfield
  {title} {\bibinfo {title} {Strength of species interactions determines
  biodiversity and stability in microbial communities},\ }\href
  {https://doi.org/10.1038/s41559-020-1099-4} {\bibfield  {journal} {\bibinfo
  {journal} {Nature Ecology \& Evolution}\ }\textbf {\bibinfo {volume} {4}},\
  \bibinfo {pages} {376} (\bibinfo {year} {2020})}\BibitemShut {NoStop}%
\bibitem [{\citenamefont {Miller}\ and\ \citenamefont
  {Bassler}(2001)}]{Miller2001}%
  \BibitemOpen
  \bibfield  {author} {\bibinfo {author} {\bibfnamefont {M.~B.}\ \bibnamefont
  {Miller}}\ and\ \bibinfo {author} {\bibfnamefont {B.~L.}\ \bibnamefont
  {Bassler}},\ }\bibfield  {title} {\bibinfo {title} {Quorum sensing in
  bacteria},\ }\href {https://doi.org/10.1146/annurev.micro.55.1.165}
  {\bibfield  {journal} {\bibinfo  {journal} {Annual Review of Microbiology}\
  }\textbf {\bibinfo {volume} {55}},\ \bibinfo {pages} {165–199} (\bibinfo
  {year} {2001})}\BibitemShut {NoStop}%
\bibitem [{\citenamefont {Waters}\ and\ \citenamefont
  {Bassler}(2005)}]{Waters2005}%
  \BibitemOpen
  \bibfield  {author} {\bibinfo {author} {\bibfnamefont {C.~M.}\ \bibnamefont
  {Waters}}\ and\ \bibinfo {author} {\bibfnamefont {B.~L.}\ \bibnamefont
  {Bassler}},\ }\bibfield  {title} {\bibinfo {title} {Quorum sensing:
  Cell-to-cell communication in bacteria},\ }\href
  {https://doi.org/10.1146/annurev.cellbio.21.012704.131001} {\bibfield
  {journal} {\bibinfo  {journal} {Annual Review of Cell and Developmental
  Biology}\ }\textbf {\bibinfo {volume} {21}},\ \bibinfo {pages} {319}
  (\bibinfo {year} {2005})}\BibitemShut {NoStop}%
\bibitem [{\citenamefont {Mills}\ \emph {et~al.}(1967)\citenamefont {Mills},
  \citenamefont {Peterson},\ and\ \citenamefont {Spiegelman}}]{Mills1967}%
  \BibitemOpen
  \bibfield  {author} {\bibinfo {author} {\bibfnamefont {D.~R.}\ \bibnamefont
  {Mills}}, \bibinfo {author} {\bibfnamefont {R.~L.}\ \bibnamefont
  {Peterson}},\ and\ \bibinfo {author} {\bibfnamefont {S.}~\bibnamefont
  {Spiegelman}},\ }\bibfield  {title} {\bibinfo {title} {An extracellular
  darwinian experiment with a self-duplicating nucleic acid molecule.},\ }\href
  {https://doi.org/10.1073/pnas.58.1.217} {\bibfield  {journal} {\bibinfo
  {journal} {Proceedings of the National Academy of Sciences}\ }\textbf
  {\bibinfo {volume} {58}},\ \bibinfo {pages} {217} (\bibinfo {year}
  {1967})}\BibitemShut {NoStop}%
\bibitem [{\citenamefont {Smith}(1985)}]{Smith1985}%
  \BibitemOpen
  \bibfield  {author} {\bibinfo {author} {\bibfnamefont {G.~P.}\ \bibnamefont
  {Smith}},\ }\bibfield  {title} {\bibinfo {title} {Filamentous fusion phage:
  Novel expression vectors that display cloned antigens on the virion
  surface},\ }\href {https://doi.org/10.1126/science.4001944} {\bibfield
  {journal} {\bibinfo  {journal} {Science}\ }\textbf {\bibinfo {volume}
  {228}},\ \bibinfo {pages} {1315} (\bibinfo {year} {1985})}\BibitemShut
  {NoStop}%
\bibitem [{\citenamefont {Chen}\ and\ \citenamefont {Arnold}(1991)}]{Chen1991}%
  \BibitemOpen
  \bibfield  {author} {\bibinfo {author} {\bibfnamefont {K.}~\bibnamefont
  {Chen}}\ and\ \bibinfo {author} {\bibfnamefont {F.~H.}\ \bibnamefont
  {Arnold}},\ }\bibfield  {title} {\bibinfo {title} {Enzyme engineering for
  nonaqueous solvents: Random mutagenesis to enhance activity of subtilisin e
  in polar organic media},\ }\href {https://doi.org/10.1038/nbt1191-1073}
  {\bibfield  {journal} {\bibinfo  {journal} {Bio/Technology}\ }\textbf
  {\bibinfo {volume} {9}},\ \bibinfo {pages} {1073} (\bibinfo {year}
  {1991})}\BibitemShut {NoStop}%
\bibitem [{\citenamefont {Balaban}\ \emph {et~al.}(2004)\citenamefont
  {Balaban}, \citenamefont {Merrin}, \citenamefont {Chait}, \citenamefont
  {Kowalik},\ and\ \citenamefont {Leibler}}]{Balaban2004}%
  \BibitemOpen
  \bibfield  {author} {\bibinfo {author} {\bibfnamefont {N.~Q.}\ \bibnamefont
  {Balaban}}, \bibinfo {author} {\bibfnamefont {J.}~\bibnamefont {Merrin}},
  \bibinfo {author} {\bibfnamefont {R.}~\bibnamefont {Chait}}, \bibinfo
  {author} {\bibfnamefont {L.}~\bibnamefont {Kowalik}},\ and\ \bibinfo {author}
  {\bibfnamefont {S.}~\bibnamefont {Leibler}},\ }\bibfield  {title} {\bibinfo
  {title} {Bacterial persistence as a phenotypic switch},\ }\href
  {https://doi.org/10.1126/science.1099390} {\bibfield  {journal} {\bibinfo
  {journal} {Science}\ }\textbf {\bibinfo {volume} {305}},\ \bibinfo {pages}
  {1622} (\bibinfo {year} {2004})}\BibitemShut {NoStop}%
\bibitem [{\citenamefont {Kussell}\ and\ \citenamefont
  {Leibler}(2005)}]{Kussell2005}%
  \BibitemOpen
  \bibfield  {author} {\bibinfo {author} {\bibfnamefont {E.}~\bibnamefont
  {Kussell}}\ and\ \bibinfo {author} {\bibfnamefont {S.}~\bibnamefont
  {Leibler}},\ }\bibfield  {title} {\bibinfo {title} {Phenotypic diversity,
  population growth, and information in fluctuating environments},\ }\href
  {https://doi.org/10.1126/science.1114383} {\bibfield  {journal} {\bibinfo
  {journal} {Science}\ }\textbf {\bibinfo {volume} {309}},\ \bibinfo {pages}
  {2075} (\bibinfo {year} {2005})}\BibitemShut {NoStop}%
\bibitem [{\citenamefont {Murugan}\ \emph {et~al.}(2021)\citenamefont
  {Murugan}, \citenamefont {Husain}, \citenamefont {Rust}, \citenamefont
  {Hepler}, \citenamefont {Bass}, \citenamefont {Pietsch}, \citenamefont
  {Swain}, \citenamefont {Jena}, \citenamefont {Toettcher}, \citenamefont
  {Chakraborty}, \citenamefont {Sprenger}, \citenamefont {Mora}, \citenamefont
  {Walczak}, \citenamefont {Rivoire}, \citenamefont {Wang}, \citenamefont
  {Wood}, \citenamefont {Skanata}, \citenamefont {Kussell}, \citenamefont
  {Ranganathan}, \citenamefont {Shih},\ and\ \citenamefont
  {Goldenfeld}}]{Murugan2021}%
  \BibitemOpen
  \bibfield  {author} {\bibinfo {author} {\bibfnamefont {A.}~\bibnamefont
  {Murugan}}, \bibinfo {author} {\bibfnamefont {K.}~\bibnamefont {Husain}},
  \bibinfo {author} {\bibfnamefont {M.~J.}\ \bibnamefont {Rust}}, \bibinfo
  {author} {\bibfnamefont {C.}~\bibnamefont {Hepler}}, \bibinfo {author}
  {\bibfnamefont {J.}~\bibnamefont {Bass}}, \bibinfo {author} {\bibfnamefont
  {J.~M.~J.}\ \bibnamefont {Pietsch}}, \bibinfo {author} {\bibfnamefont
  {P.~S.}\ \bibnamefont {Swain}}, \bibinfo {author} {\bibfnamefont {S.~G.}\
  \bibnamefont {Jena}}, \bibinfo {author} {\bibfnamefont {J.~E.}\ \bibnamefont
  {Toettcher}}, \bibinfo {author} {\bibfnamefont {A.~K.}\ \bibnamefont
  {Chakraborty}}, \bibinfo {author} {\bibfnamefont {K.~G.}\ \bibnamefont
  {Sprenger}}, \bibinfo {author} {\bibfnamefont {T.}~\bibnamefont {Mora}},
  \bibinfo {author} {\bibfnamefont {A.~M.}\ \bibnamefont {Walczak}}, \bibinfo
  {author} {\bibfnamefont {O.}~\bibnamefont {Rivoire}}, \bibinfo {author}
  {\bibfnamefont {S.}~\bibnamefont {Wang}}, \bibinfo {author} {\bibfnamefont
  {K.~B.}\ \bibnamefont {Wood}}, \bibinfo {author} {\bibfnamefont
  {A.}~\bibnamefont {Skanata}}, \bibinfo {author} {\bibfnamefont
  {E.}~\bibnamefont {Kussell}}, \bibinfo {author} {\bibfnamefont
  {R.}~\bibnamefont {Ranganathan}}, \bibinfo {author} {\bibfnamefont {H.-Y.}\
  \bibnamefont {Shih}},\ and\ \bibinfo {author} {\bibfnamefont
  {N.}~\bibnamefont {Goldenfeld}},\ }\bibfield  {title} {\bibinfo {title}
  {Roadmap on biology in time varying environments},\ }\href
  {https://doi.org/10.1088/1478-3975/abde8d} {\bibfield  {journal} {\bibinfo
  {journal} {Physical Biology}\ }\textbf {\bibinfo {volume} {18}},\ \bibinfo
  {pages} {041502} (\bibinfo {year} {2021})}\BibitemShut {NoStop}%
\bibitem [{\citenamefont {Marchi}\ \emph {et~al.}(2021)\citenamefont {Marchi},
  \citenamefont {Lässig}, \citenamefont {Walczak},\ and\ \citenamefont
  {Mora}}]{Marchi2021}%
  \BibitemOpen
  \bibfield  {author} {\bibinfo {author} {\bibfnamefont {J.}~\bibnamefont
  {Marchi}}, \bibinfo {author} {\bibfnamefont {M.}~\bibnamefont {Lässig}},
  \bibinfo {author} {\bibfnamefont {A.~M.}\ \bibnamefont {Walczak}},\ and\
  \bibinfo {author} {\bibfnamefont {T.}~\bibnamefont {Mora}},\ }\bibfield
  {title} {\bibinfo {title} {Antigenic waves of virus–immune coevolution},\
  }\bibfield  {journal} {\bibinfo  {journal} {Proceedings of the National
  Academy of Sciences}\ }\textbf {\bibinfo {volume} {118}},\ \href
  {https://doi.org/10.1073/pnas.2103398118} {10.1073/pnas.2103398118} (\bibinfo
  {year} {2021})\BibitemShut {NoStop}%
\bibitem [{\citenamefont {Lässig}\ \emph {et~al.}(2017)\citenamefont
  {Lässig}, \citenamefont {Mustonen},\ and\ \citenamefont
  {Walczak}}]{Lassig2017}%
  \BibitemOpen
  \bibfield  {author} {\bibinfo {author} {\bibfnamefont {M.}~\bibnamefont
  {Lässig}}, \bibinfo {author} {\bibfnamefont {V.}~\bibnamefont {Mustonen}},\
  and\ \bibinfo {author} {\bibfnamefont {A.~M.}\ \bibnamefont {Walczak}},\
  }\bibfield  {title} {\bibinfo {title} {Predicting evolution},\ }\bibfield
  {journal} {\bibinfo  {journal} {Nature Ecology and Evolution}\ }\textbf
  {\bibinfo {volume} {1}},\ \href {https://doi.org/10.1038/s41559-017-0077}
  {10.1038/s41559-017-0077} (\bibinfo {year} {2017})\BibitemShut {NoStop}%
\bibitem [{\citenamefont {{Turing, Alan}}(1952)}]{TuringAlan1952}%
  \BibitemOpen
  \bibfield  {author} {\bibinfo {author} {\bibnamefont {{Turing, Alan}}},\
  }\bibfield  {title} {\bibinfo {title} {The chemical basis of morphogenesis},\
  }\href {https://doi.org/10.1098/rstb.1952.0012} {\bibfield  {journal}
  {\bibinfo  {journal} {Philosophical Transactions of the Royal Society of
  London. Series B, Biological Sciences}\ }\textbf {\bibinfo {volume} {237}},\
  \bibinfo {pages} {37} (\bibinfo {year} {1952})}\BibitemShut {NoStop}%
\bibitem [{\citenamefont {Gilmour}\ \emph {et~al.}(2017)\citenamefont
  {Gilmour}, \citenamefont {Rembold},\ and\ \citenamefont
  {Leptin}}]{Gilmour2017}%
  \BibitemOpen
  \bibfield  {author} {\bibinfo {author} {\bibfnamefont {D.}~\bibnamefont
  {Gilmour}}, \bibinfo {author} {\bibfnamefont {M.}~\bibnamefont {Rembold}},\
  and\ \bibinfo {author} {\bibfnamefont {M.}~\bibnamefont {Leptin}},\
  }\bibfield  {title} {\bibinfo {title} {From morphogen to morphogenesis and
  back},\ }\href {https://doi.org/10.1038/nature21348} {\bibfield  {journal}
  {\bibinfo  {journal} {Nature}\ }\textbf {\bibinfo {volume} {541}},\ \bibinfo
  {pages} {311–320} (\bibinfo {year} {2017})}\BibitemShut {NoStop}%
\bibitem [{\citenamefont {Lam}(1983)}]{Lam1983}%
  \BibitemOpen
  \bibfield  {author} {\bibinfo {author} {\bibfnamefont {N.~S.-N.}\
  \bibnamefont {Lam}},\ }\bibfield  {title} {\bibinfo {title} {Spatial
  interpolation methods: A review},\ }\href
  {https://doi.org/10.1559/152304083783914958} {\bibfield  {journal} {\bibinfo
  {journal} {The American Cartographer}\ }\textbf {\bibinfo {volume} {10}},\
  \bibinfo {pages} {129} (\bibinfo {year} {1983})}\BibitemShut {NoStop}%
\bibitem [{\citenamefont {Guyer}\ \emph {et~al.}(2009)\citenamefont {Guyer},
  \citenamefont {Wheeler},\ and\ \citenamefont {Warren}}]{Guyer2009}%
  \BibitemOpen
  \bibfield  {author} {\bibinfo {author} {\bibfnamefont {J.~E.}\ \bibnamefont
  {Guyer}}, \bibinfo {author} {\bibfnamefont {D.}~\bibnamefont {Wheeler}},\
  and\ \bibinfo {author} {\bibfnamefont {J.~A.}\ \bibnamefont {Warren}},\
  }\bibfield  {title} {\bibinfo {title} {Fipy: Partial differential equations
  with python},\ }\href {https://doi.org/10.1109/mcse.2009.52} {\bibfield
  {journal} {\bibinfo  {journal} {Computing in Science \& Engineering}\
  }\textbf {\bibinfo {volume} {11}},\ \bibinfo {pages} {6} (\bibinfo {year}
  {2009})}\BibitemShut {NoStop}%
\bibitem [{\citenamefont {Tobler}(1979)}]{Tobler1979}%
  \BibitemOpen
  \bibfield  {author} {\bibinfo {author} {\bibfnamefont {W.~R.}\ \bibnamefont
  {Tobler}},\ }\bibfield  {title} {\bibinfo {title} {Smooth pycnophylactic
  interpolation for geographical regions},\ }\href
  {https://doi.org/10.1080/01621459.1979.10481647} {\bibfield  {journal}
  {\bibinfo  {journal} {Journal of the American Statistical Association}\
  }\textbf {\bibinfo {volume} {74}},\ \bibinfo {pages} {519} (\bibinfo {year}
  {1979})}\BibitemShut {NoStop}%
\bibitem [{Note1()}]{Note1}%
  \BibitemOpen
  \bibinfo {note} {Code is available online at \protect \url
  {https://github.com/jcolen/sociohydro}}\BibitemShut {NoStop}%
\bibitem [{\citenamefont {Liu}\ \emph {et~al.}(2022)\citenamefont {Liu},
  \citenamefont {Mao}, \citenamefont {Wu}, \citenamefont {Feichtenhofer},
  \citenamefont {Darrell},\ and\ \citenamefont {Xie}}]{Liu2022}%
  \BibitemOpen
  \bibfield  {author} {\bibinfo {author} {\bibfnamefont {Z.}~\bibnamefont
  {Liu}}, \bibinfo {author} {\bibfnamefont {H.}~\bibnamefont {Mao}}, \bibinfo
  {author} {\bibfnamefont {C.-Y.}\ \bibnamefont {Wu}}, \bibinfo {author}
  {\bibfnamefont {C.}~\bibnamefont {Feichtenhofer}}, \bibinfo {author}
  {\bibfnamefont {T.}~\bibnamefont {Darrell}},\ and\ \bibinfo {author}
  {\bibfnamefont {S.}~\bibnamefont {Xie}},\ }\href
  {https://doi.org/10.48550/ARXIV.2201.03545} {\bibinfo {title} {A convnet for
  the 2020s}} (\bibinfo {year} {2022})\BibitemShut {NoStop}%
\bibitem [{\citenamefont {Verbavatz}\ and\ \citenamefont
  {Barthelemy}(2020)}]{Verbavatz2020}%
  \BibitemOpen
  \bibfield  {author} {\bibinfo {author} {\bibfnamefont {V.}~\bibnamefont
  {Verbavatz}}\ and\ \bibinfo {author} {\bibfnamefont {M.}~\bibnamefont
  {Barthelemy}},\ }\bibfield  {title} {\bibinfo {title} {The growth equation of
  cities},\ }\href {https://doi.org/10.1038/s41586-020-2900-x} {\bibfield
  {journal} {\bibinfo  {journal} {Nature}\ }\textbf {\bibinfo {volume} {587}},\
  \bibinfo {pages} {397} (\bibinfo {year} {2020})}\BibitemShut {NoStop}%
\bibitem [{\citenamefont {Reia}\ \emph {et~al.}(2022)\citenamefont {Reia},
  \citenamefont {Rao}, \citenamefont {Barthelemy},\ and\ \citenamefont
  {Ukkusuri}}]{Reia2022}%
  \BibitemOpen
  \bibfield  {author} {\bibinfo {author} {\bibfnamefont {S.~M.}\ \bibnamefont
  {Reia}}, \bibinfo {author} {\bibfnamefont {P.~S.~C.}\ \bibnamefont {Rao}},
  \bibinfo {author} {\bibfnamefont {M.}~\bibnamefont {Barthelemy}},\ and\
  \bibinfo {author} {\bibfnamefont {S.~V.}\ \bibnamefont {Ukkusuri}},\
  }\bibfield  {title} {\bibinfo {title} {Spatial structure of city population
  growth},\ }\bibfield  {journal} {\bibinfo  {journal} {Nature Communications}\
  }\textbf {\bibinfo {volume} {13}},\ \href
  {https://doi.org/10.1038/s41467-022-33527-y} {10.1038/s41467-022-33527-y}
  (\bibinfo {year} {2022})\BibitemShut {NoStop}%
\bibitem [{\citenamefont {Farley}\ \emph {et~al.}(1997)\citenamefont {Farley},
  \citenamefont {Fielding},\ and\ \citenamefont {Krysan}}]{Farley1997}%
  \BibitemOpen
  \bibfield  {author} {\bibinfo {author} {\bibfnamefont {R.}~\bibnamefont
  {Farley}}, \bibinfo {author} {\bibfnamefont {E.~L.}\ \bibnamefont
  {Fielding}},\ and\ \bibinfo {author} {\bibfnamefont {M.}~\bibnamefont
  {Krysan}},\ }\bibfield  {title} {\bibinfo {title} {The residential
  preferences of blacks and whites: A four-metropolis analysis},\ }\href
  {https://doi.org/10.1080/10511482.1997.9521278} {\bibfield  {journal}
  {\bibinfo  {journal} {Housing Policy Debate}\ }\textbf {\bibinfo {volume}
  {8}},\ \bibinfo {pages} {763} (\bibinfo {year} {1997})}\BibitemShut {NoStop}%
\bibitem [{\citenamefont {Bertin}(2021)}]{Bertin2021}%
  \BibitemOpen
  \bibfield  {author} {\bibinfo {author} {\bibfnamefont {E.}~\bibnamefont
  {Bertin}},\ }\href@noop {} {\emph {\bibinfo {title} {Statistical Physics of
  Complex Systems A Concise Introduction}}}\ (\bibinfo  {publisher} {Springer
  International Publishing AG},\ \bibinfo {year} {2021})\ p.\ \bibinfo {pages}
  {291}\BibitemShut {NoStop}%
\bibitem [{\citenamefont {Glauber}(1963)}]{Glauber1963}%
  \BibitemOpen
  \bibfield  {author} {\bibinfo {author} {\bibfnamefont {R.~J.}\ \bibnamefont
  {Glauber}},\ }\bibfield  {title} {\bibinfo {title} {Time-dependent statistics
  of the ising model},\ }\href {https://doi.org/10.1063/1.1703954} {\bibfield
  {journal} {\bibinfo  {journal} {Journal of Mathematical Physics}\ }\textbf
  {\bibinfo {volume} {4}},\ \bibinfo {pages} {294} (\bibinfo {year}
  {1963})}\BibitemShut {NoStop}%
\bibitem [{\citenamefont {Bouchaud}(2013)}]{Bouchaud2013}%
  \BibitemOpen
  \bibfield  {author} {\bibinfo {author} {\bibfnamefont {J.-P.}\ \bibnamefont
  {Bouchaud}},\ }\bibfield  {title} {\bibinfo {title} {Crises and collective
  socio-economic phenomena: Simple models and challenges},\ }\href
  {https://doi.org/10.1007/s10955-012-0687-3} {\bibfield  {journal} {\bibinfo
  {journal} {Journal of Statistical Physics}\ }\textbf {\bibinfo {volume}
  {151}},\ \bibinfo {pages} {567} (\bibinfo {year} {2013})}\BibitemShut
  {NoStop}%
\bibitem [{\citenamefont {Eyring}(1935)}]{Eyring1935}%
  \BibitemOpen
  \bibfield  {author} {\bibinfo {author} {\bibfnamefont {H.}~\bibnamefont
  {Eyring}},\ }\bibfield  {title} {\bibinfo {title} {The activated complex in
  chemical reactions},\ }\href {https://doi.org/10.1063/1.1749604} {\bibfield
  {journal} {\bibinfo  {journal} {The Journal of Chemical Physics}\ }\textbf
  {\bibinfo {volume} {3}},\ \bibinfo {pages} {107} (\bibinfo {year}
  {1935})}\BibitemShut {NoStop}%
\bibitem [{\citenamefont {Frohoff-Hülsmann}\ \emph {et~al.}(2023)\citenamefont
  {Frohoff-Hülsmann}, \citenamefont {Holl}, \citenamefont {Knobloch},
  \citenamefont {Gurevich},\ and\ \citenamefont
  {Thiele}}]{FrohoffHuelsmann2023b}%
  \BibitemOpen
  \bibfield  {author} {\bibinfo {author} {\bibfnamefont {T.}~\bibnamefont
  {Frohoff-Hülsmann}}, \bibinfo {author} {\bibfnamefont {M.~P.}\ \bibnamefont
  {Holl}}, \bibinfo {author} {\bibfnamefont {E.}~\bibnamefont {Knobloch}},
  \bibinfo {author} {\bibfnamefont {S.~V.}\ \bibnamefont {Gurevich}},\ and\
  \bibinfo {author} {\bibfnamefont {U.}~\bibnamefont {Thiele}},\ }\bibfield
  {title} {\bibinfo {title} {Stationary broken parity states in active matter
  models},\ }\href {https://doi.org/10.1103/physreve.107.064210} {\bibfield
  {journal} {\bibinfo  {journal} {Physical Review E}\ }\textbf {\bibinfo
  {volume} {107}},\ \bibinfo {pages} {064210} (\bibinfo {year}
  {2023})}\BibitemShut {NoStop}%
\bibitem [{\citenamefont {Evans}\ and\ \citenamefont
  {Hanney}(2005)}]{Evans2005}%
  \BibitemOpen
  \bibfield  {author} {\bibinfo {author} {\bibfnamefont {M.~R.}\ \bibnamefont
  {Evans}}\ and\ \bibinfo {author} {\bibfnamefont {T.}~\bibnamefont {Hanney}},\
  }\bibfield  {title} {\bibinfo {title} {Nonequilibrium statistical mechanics
  of the zero-range process and related models},\ }\href
  {https://doi.org/10.1088/0305-4470/38/19/r01} {\bibfield  {journal} {\bibinfo
   {journal} {Journal of Physics A: Mathematical and General}\ }\textbf
  {\bibinfo {volume} {38}},\ \bibinfo {pages} {R195} (\bibinfo {year}
  {2005})}\BibitemShut {NoStop}%
\end{thebibliography}%
